\documentclass[12pt]{article}
\usepackage{epsfig}
\usepackage{axodraw}
\usepackage{amsmath}
\newcommand{\aem}{\alpha_{\mathrm{em}}}
\renewcommand{\d}{\mathrm{d}}
\newcommand{\X}{\mathbf{X}}
\newcommand{\e}{\mathrm{e}}
\newcommand{\f}{\mathrm{f}}
\newcommand{\g}{\mathrm{g}}
\newcommand{\p}{\mathrm{p}}
\newcommand{\q}{\mathrm{q}}
\newcommand{\fbar}{\mathrm{\overline{f}}}
\newcommand{\qbar}{\mathrm{\overline{q}}}
\newcommand{\pbar}{\mathrm{\overline{p}}}
\newcommand{\kT}{k_{\perp}}
\newcommand{\pT}{p_{\perp}}
\newcommand{\pTmin}{p_{\perp,\mathrm{min}}}
\newcommand{\ET}{E_{\perp}}
\newcommand{\shat}{\hat{s}}
\newcommand{\that}{\hat{t}}
\newcommand{\uhat}{\hat{u}}
\newcommand{\sighat}{\hat{\sigma}}
\newcommand{\xhat}{\hat{x}}
\newcommand{\gast}{\gamma^*}

\def\Journal#1#2#3#4{{#1}{\bf #2} (#4) #3}

\def\NPB{{\rm Nucl. Phys.~}{\bf B}}
\def\PLB{{\rm Phys. Lett.~}{\bf B}}
\def\JournalPLB#1#2#3{{\rm Phys. Lett.~}{\bf {#1}B} (#3) #2}

\def\PRD{{\rm Phys. Rev.~}{\bf D}}
\def\ZPC{{\rm Z. Phys.~}{\bf C}}
\def\ZP{\rm Z. Phys.~}
\def\JHEP{\rm J. High Energy Phys.~}
\def\CPC{\rm Computer Phys. Commun.~}
\def\PRP{\rm Phys. Rep.~}
\def\PRV{\rm Phys. Rev.~}
\def\EPJC{{\rm Eur. Phys. J.~}{\bf C}}

\newcommand{\gtrsim}{\raisebox{-0.8mm}%
{\hspace{1mm}$\stackrel{>}{\sim}$\hspace{1mm}}}

{\end{list}}
\newcounter{enumct}

\newlength{\abstwidth}
\newlength{\captivewidth}
\newcommand{\captive}[1]{\rule{5mm}{0mm}%
\begin{minipage}{\captivewidth}%
\caption[small]{#1}\end{minipage}}
 
\begin{document}
 
%set sloppy attitude to line breaks
\sloppy
 
\pagestyle{empty}
 
\begin{flushright}
LU TP 99--11 \\
hep-ph/9907245\\
July 1999
\end{flushright}
 
\vspace{\fill}
 
\begin{center}
{\LARGE\bf Jet Production by Virtual Photons}\\[10mm]
{\Large Christer Friberg\footnote{christer@thep.lu.se }
and Torbj\"orn Sj\"ostrand\footnote{torbjorn@thep.lu.se}}\\ [2mm]
{\it Department of Theoretical Physics,}\\[1mm]
{\it Lund University, Lund, Sweden}
\end{center}
 
\vspace{\fill}
\begin{center}
{\bf Abstract}\\[2ex]
\begin{minipage}{\abstwidth}
The production of jets is studied in collisions of virtual photons,
$\gast\p$ and $\gast\gast$, specifically for applications at HERA and 
LEP2. Photon flux factors are convoluted with matrix elements involving
either direct or resolved photons and, for the latter, with parton
distributions of the photon. Special emphasis is put on the
range of uncertainty in the modeling of the resolved component.
The resulting model is compared with existing data.
\end{minipage}
\end{center}

\vspace{\fill}
 
\clearpage
\pagestyle{plain}
\setcounter{page}{1}

\section{Introduction}

The photon is a complicated object to describe. In the DIS region, i.e. 
when it is very virtual, it can be considered as devoid of any internal
structure, at least to first approximation. In the other extreme, the 
total cross section for real photons is dominated by the resolved 
component of the wave function, where the photon has fluctuated 
into a $\q\qbar$ state. The nature of this resolved component is
still not well understood, especially not the way in which it dies 
out with increasing photon virtuality. This dampening is likely not 
to be a simple function of virtuality, but to depend on the physics
observable being studied, i.e. on the combination of subprocesses
singled out.  

Since our current understanding of QCD does not allow complete 
predictability, one sensible approach is to base ourselves on
QCD-motivated models, where a plausible range of uncertainty can be 
explored. Hopefully comparisons with data may then help constrain
the correct behaviour. The ultimate goal therefore clearly is to
have a testable model for all aspects of the physics of $\gast\p$ and 
$\gast\gast$ collisions. As a stepping stone towards constructing
such a framework, in this paper we explore the physics associated 
with the production of `high-$\pT$' jets in the collision. That is,
we here avoid the processes that only produce activity along the 
$\gast\p$ or $\gast\gast$ collision axis. For resolved photons this
corresponds to the `soft' or `low-$\pT$' events of the hadronic 
physics analogy, for direct ones to the lowest-order DIS process
$\gast\q \to \q$. 

The processes that we will study here instead can be exemplified by
$\gast\gast \to \q\qbar$ (direct), $\gast \g \to \q\qbar$ 
(single-resolved for $\gast\gast$, direct for $\gast\p$) and 
$\g\g \to \q\qbar$ (double-resolved for $\gast\gast$, 
(single-)resolved for $\gast\p$), where the gluons come from the 
parton content of a resolved virtual photon or from the proton.
Note that these are multi-scale processes, at least involving the
virtuality $Q_i^2$ of either photon ($i=1,2$) and the $\pT^2$ of 
the hard subprocess. (In $\gast\gast$ physics the notation $P^2$ 
is often used instead of $Q^2$, especially for the less virtual of
the two photons; here we will use $Q^2$ throughout, however.)
For a resolved photon, the relative transverse momentum $\kT$ of 
the initial $\gast \to \q\qbar$ branching provides a further scale,
at least in our framework. This plethora of scales clearly is a 
challenge to any model builder, but in principle it also offers 
the opportunity to explore QCD in a more differential fashion than 
is normally possible. 

At large photon virtualities, a possible strategy would be to express
the cross sections entirely in terms of processes involving the photon
directly, i.e. to include branchings such as $\gast \to \q'\qbar'$ and
$\q' \to \q'\g$ in the Feynman graphs calculated to describe the
process, so that e.g. the $\gast\p$ process $\g\g \to \q\qbar$ is
calculated as $\gast\g \to \q'\qbar'\q\qbar$. With decreasing
virtuality of the photon, such a fixed-order approach is increasingly
deficient by its lack of the large logarithmic corrections generated
by collinear and soft gluon emission, however.  Furthermore, almost
real photons allow long-lived $\gast \to \q\qbar$ fluctuations, that
then take on the properties of non-perturbative hadronic states,
specifically of vector mesons such as the $\rho^0$.  It is therefore
that an effective description in terms of parton distributions becomes
necessary. Hence the resolved component of the photon, as opposed to
the direct one.

That such a subdivision is more than a technical construct is
excellently illustrated by the $x_{\gamma}^{\mathrm{obs}}$ plots from
HERA \cite{xgobs}. This variable sums up the fraction of the original
photon light-cone momentum carried by the two highest-$\ET$
jets. A clear two-component structure is visible.  The peak close to
$x_{\gamma}^{\mathrm{obs}} = 1$ can be viewed as a smeared footprint
of the direct photon, with all the energy partaking in the hard
interaction, while the broad spectrum at lower
$x_{\gamma}^{\mathrm{obs}}$ is consistent with the resolved photon,
where much of the energy remains in a beam jet. The distinction
between the two is not unique when higher-order effects are included,
but it is always possible to make a functional separation that avoids
double-counting or gaps.

The resolved photon can be further subdivided into low-virtuality
fluctuations, which then are of a nonperturbative character and can be
represented by a set of vector mesons, and high-virtuality ones that
are describable by perturbative $\gast \to \q\qbar$ branchings.  The
former is called the VMD (vector meson dominance) component and the
latter the anomalous one. The parton distributions of the VMD
component are unknown from first principles, and thus have to be based
on reasonable ans\"atze, while the anomalous ones are perturbatively
predictable. This separation is more ambiguous and less well tested
than the one between direct and resolved photons. In principle,
studies on the structure of the beam remnant, e.g. its $\pT$
distribution, should show characteristic patterns. Unfortunately, the
naively expected differences are smeared by higher-order QCD
corrections (especially initial-state radiation), by the possibility
of multiple parton--parton interactions, by hadronization effects, and
so on. (Experimentally, gaps in the detector acceptance, e.g. for the
beam pipe, is a further major worry.) Many of these areas offer
interesting challenges in their own right; e.g. the way in which
multiple interactions die out with virtuality, both that of the photon
itself and that of the $\q\qbar$ pair it fluctuates to.  Models for
one aspect at the time are therefore likely to be inadequate. Instead
we here attempt a combined description of all the relevant physics
topics.

The traditional tool for handling such complex issues is the Monte Carlo
approach. Our starting point is the model for real photons 
\cite{sasevt} and the parton distribution parameterizations of real
and virtual photons \cite{saspdf} already present in the {\sc Pythia}
\cite {pythia} generator. Several further additions and modifications
have been made to model virtual photons, as will be described in the
following. Other generators with an overlapping scope include, 
among others, HERWIG \cite{herwig}, LDC \cite{ldc}, LEPTO \cite{lepto},
PHOJET \cite{phojet} and RAPGAP \cite{rapgap}. The details of the 
approaches are different, 
however, so this gives healthy possibilities to compare and learn.
Another alternative is provided by matrix-element
calculations \cite{nlome}, that do not provide the same complete 
overview but can offer superior descriptions for some purposes.

The plan of this paper is the following. In section 2 the model is
described, with special emphasis on those aspects that are new compared
with the corresponding description for real photons. Thereafter, in
section 3, the dependence of simple observables on model parameters is 
illustrated. Comparisons are shown with some sets of data from $\e\p$ 
and $\e^+\e^-$ colliders, and this is used to constrain partly the
freedom in the model. Based on this experience, some further observables 
are then proposed and studied, to help shed light on the nature of the 
virtual photon. Finally, section 4 contains a summary and outlook.

\section{The Model}

The electromagnetic field surrounding a moving electron can be 
viewed as a flux of photons. Weizs\"acker \cite{weiz} and Williams 
\cite{will} calculated the spectrum of these photons, neglecting 
the photon virtualities and terms involving the longitudinal polarization 
of photons. This approximation is well-known \cite{klaWW} to be a good 
approximation when the scattered lepton is tagged at small scattering 
angles.

In the equivalent photon approximation \cite{EPA}, the cross sections 
for the processes $\e\p\rightarrow\e\X$ and $\e\e\rightarrow\e\e\X$, 
where $\X$ is an arbitrary final state, can then be written as the 
convolutions
\begin{equation}
\d\sigma(\e\p\rightarrow\e\X)=\int \frac{\d \omega}{\omega} \; N(\omega) 
\;\d\sigma(\gamma\p\rightarrow\X)
\label{HERAEPA}
\end{equation}
and 
\begin{equation}
\d\sigma(\e\e\rightarrow\e\e\X)=
\iint \frac{\d \omega_1}{\omega_1}\frac{\d \omega_2}{\omega_2} \;
N(\omega_1) N(\omega_2)
\;\d\sigma(\gamma\gamma\rightarrow\X) ~,
\label{LEPEPA}
\end{equation}
where $\omega$ is the energy of the emitted photon. In this approximation, 
the distribution in photon frequencies $N(\omega)\d \omega / \omega$ is 
obtained by integrating over the photon virtuality $Q^2$. The maximum 
value $Q^2_{\mathrm{max}}$ is usually given by experimental conditions like 
anti-tagging, i.e. that the scattered lepton is not detected if its 
scattering angle is too small.

A better approximation, and the one used in our approach, 
is to keep the $Q^2$ dependence in the photon flux 
$f(y,Q^2)$ (with $y \approx \omega/\omega_{\mathrm{max}}$, see below) 
and in the subprocess cross sections involving the virtual 
photon(s), $\gast\p\rightarrow\X$ and $\gast\gast\rightarrow\X$,
and to sum over the transverse and longitudinal photon polarizations. 
Equations~(\ref{HERAEPA}) and~(\ref{LEPEPA}) then modify to
\begin{equation}
\d\sigma(\e\p\rightarrow\e\X)=\sum_{\xi=\mathrm{T,L}}
\iint \d y \, \d Q^2 
\;f_{\gamma/\e}^{\xi}(y,Q^2) 
\;\d\sigma(\gast_{\xi}\p\rightarrow\X)
\label{PYHERAEPA}
\end{equation}
and 
\begin{equation}
\d\sigma(\e\e\rightarrow\e\e\X)=\!\!\!\sum_{\xi_1,\xi_2=\mathrm{T,L}}
\iiiint \!\d y_1 \, \d Q_1^2 \, \d y_2 \, \d Q_2^2 
\;f_{\gamma/\e}^{\xi_1}(y_1,Q_1^2) f_{\gamma/\e}^{\xi_2}(y_2,Q_2^2)
\;\d\sigma(\gast_{\xi_1}\gast_{\xi_2}\rightarrow\X)\;.
\label{PYLEPEPA}
\end{equation}
For $\e\p$ events, this factorized ansatz is perfectly general, so long 
as azimuthal distributions in the final state are not studied in detail.
In $\e^+\e^-$ events, it is not a good approximation when the 
virtualities $Q_1^2$ and $Q_2^2$ of both photons become of the order of 
the squared invariant mass $W^2$ of the colliding photons 
\cite{GS}. In this region the cross section have terms that depend 
on the relative azimuthal angle of the scattered leptons, and 
the transverse and longitudinal polarizations are non-trivially mixed. 
However, these terms are of order $Q_1^2Q_2^2/W^2$ and can 
be neglected whenever at least one of the photons has low virtuality 
compared to $W^2$. 

\subsection{The Photon Flux}

When $Q^2/W^2$ is small, one can derive \cite{gammaflux,EPA,GS}
\begin{eqnarray}
f_{\gamma/l}^{\mathrm{T}}(y,Q^2) & = & \frac{\aem}{2\pi} 
\left( \frac{(1+(1-y)^2}{y} \frac{1}{Q^2}-\frac{2m_{l}^2y}{Q^4} \right)\;,\\
f_{\gamma/l}^{\mathrm{L}}(y,Q^2) & = & \frac{\aem}{2\pi} 
\frac{2(1-y)}{y} \frac{1}{Q^2}\;, 
\label{LLogflux}
\end{eqnarray}
where $l=\e^{\pm},~\mu^{\pm}$ or~$\tau^{\pm}$. 
In $f_{\gamma/l}^{\mathrm{T}}$ the second term, proportional to 
$m_{l}^2/Q^4$, is not leading log and is therefore often 
omitted. Clearly it is irrelevant at large $Q^2$, but around the lower 
cut-off $Q^2_{\mathrm{min}}$ it significantly dampens the small-$y$ rise 
of the first term. Overall, under realistic conditions, it reduces 
event rates by 5--10\% \cite{GS,Qfourred}.

\begin{figure}[t]
\begin{center}       
\begin{picture}(200,200)(0,0)
\ArrowLine(10,150)(80,150)\Text(35,160)[]{$k_1$}
\ArrowLine(80,150)(160,190)\Text(115,180)[]{$k'_1$}
\ArrowLine(10,50)(80,50)\Text(35,40)[]{$k_2$}
\ArrowLine(80,50)(160,10)\Text(115,20)[]{$k'_2$}
\Photon(80,150)(130,110){5}{6}\Text(100,120)[]{$q_1$}
\Photon(80,50)(130,90){5}{6}\Text(100,80)[]{$q_2$}
\GOval(135,100)(20,10)(0){0.5}
\LongArrow(143,110)(180,130)
\LongArrow(144,105)(180,115)
\LongArrow(145,100)(180,100)\Text(190,100)[]{$\X$}
\LongArrow(144,95)(180,85)
\LongArrow(143,90)(180,70)
\end{picture}     
\end{center}
\captive{Schematic figure of $\gamma\gamma$ processes, illustrating
the notation.
\label{fig:twogammkinem}}
\end{figure}

The $y$ variable is defined as the light-cone fraction the photon takes 
of the incoming lepton momentum. For instance, for $l^+l^-$ events, 
Fig.~\ref{fig:twogammkinem},
\begin{equation}
y_i = \frac{q_i k_j}{k_i k_j} ~, \qquad j=2 (1)~\mathrm{for}~i=1 (2) ~.  
\end{equation} 

Alternatively, the energy fraction the photon takes in the rest frame 
of the collision can be used,
\begin{equation}
x_i = \frac{q_i (k_1 + k_2)}{k_i (k_1 + k_2)} ~, \qquad i=1,2 ~.  
\end{equation} 
The two are simply related,
\begin{equation}
y_i = x_i + \frac{Q_i^2}{s} ~,
\end{equation}
with $s=(k_1 + k_2)^2$. (Here and in the following formulae we have
omitted the lepton and hadron mass terms when it is not of importance 
for the argumentation.) 
Since the Jacobian $\d(y_i, Q_i^2) / \d(x_i, Q_i^2) = 1$, either variable 
would be an equally valid choice for covering the phase space. 
Small $x_i$ values will be of less interest for us, since they lead 
to small $W^2$ and hence no high-$\pT$ jet production, so 
$y_i/x_i \approx 1$ except in the high-$Q^2$ tail, and often the two 
are used interchangeably. Unless special $Q^2$ cuts are imposed, 
cross sections obtained with 
$f_{\gamma/l}^{\mathrm{T,L}}(x,Q^2) \, \d x$ rather than
$f_{\gamma/l}^{\mathrm{T,L}}(y,Q^2) \, \d y$ differ only at the 
per mil level. For comparisons with experimental cuts, 
it is sometimes relevant to know which of the two is being used in 
an analysis.

In the $\e\p$ kinematics, the $x$ and $y$ definitions give that
\begin{equation}
W^2 = x s = y s - Q^2 ~.
\end{equation}
The $W^2$ expression for $l^+l^-$ is more complicated, especially 
because of the dependence on the relative azimuthal angle of the 
scattered leptons,
$\varphi_{12} = \varphi_1 - \varphi_2$:
\begin{eqnarray}
W^2 & = & x_1 x_2 s + \frac{2 Q_1^2 Q_2^2}{s} - 
2 \sqrt{1 - x_1 - \frac{Q_1^2}{s}} \sqrt{1 - x_2 - \frac{Q_2^2}{s}}
Q_1 Q_2 \cos\varphi_{12} \nonumber \\
& = & y_1 y_2 s - Q_1^2 - Q_2^2 + \frac{Q_1^2 Q_2^2}{s} - 
2 \sqrt{1 - y_1} \sqrt{1 - y_2} Q_1 Q_2 \cos\varphi_{12} ~.
\label{W2gaga}
\end{eqnarray}

The lepton scattering angle $\theta_i$ is related to $Q_i^2$ as
\begin{equation}
Q_i^2 = \frac{x_i^2}{1-x_i} m_i^2 + (1-x_i) \left(
s - \frac{2}{(1-x_i)^2} m_i^2 - 2 m_j^2 \right) \sin^2(\theta_i/2) ~, 
\end{equation}
with $m_i^2 = k_i^2 = {k'}^2_i$ and terms of $O(m^4)$ neglected.
The kinematical limits thus are
\begin{eqnarray}
 (Q_i^2)_{\mathrm{min}} & \approx & \frac{x_i^2}{1 - x_i} m_i^2 ~, \\
 (Q_i^2)_{\mathrm{max}} & \approx & (1 - x_i) s ~,
\end{eqnarray} 
unless experimental conditions reduce the $\theta_i$ ranges.

In summary, we will allow the possibility of experimental cuts in the
$x_i$, $y_i$, $Q_i^2$, $\theta_i$ and $W^2$ variables. Within the 
allowed region, the phase space is Monte Carlo sampled according to
$\prod_i (\d Q_i^2/Q_i^2) \, (\d x_i / x_i) \, \d \varphi_i$, with the 
remaining flux factors combined with the cross section factors to give 
the event weight used for eventual acceptance or rejection.

\subsection{Photon Processes}
\label{ME}

The hard-scattering processes are classified according to whether 
one or both photons are resolved. For completeness we here quote
some of the less familiar cross sections.

For the direct process $\gast \gast \to \f\fbar$, 
$\f$ some fermion, the cross sections are \cite{Baier}
\begin{eqnarray}
\frac{\d\sigma_{\mathrm{TT}}}{\d\that} & = & 
2 \pi \alpha_{\mathrm{em}}^2 N_{\f} e_{\f}^4 \,
\frac{\that \uhat - Q_1^2 Q_2^2}{\lambda^2 \that^2 \uhat^2}
\left\{ (\that^2 + \uhat^2) \left[ 1 - 2 F (1 - F) \right] -
2 Q_1^2 Q_2^2 F^2 \right\} ~, 
\label{gamma*Tgamma*T}\\
\frac{\d\sigma_{\mathrm{TL}}}{\d\that} & = & 
8 \pi \alpha_{\mathrm{em}}^2 N_{\f} e_{\f}^4 \,
\frac{Q_2^2 \shat}{\lambda^4 \that^2 \uhat^2}
\left\{ 2 (\that \uhat - Q_1^2 Q_2^2) 
\left[ \that \uhat + \frac{Q_1^2 \shat (\that - \uhat)^2}{\lambda^2} \right]
+ Q_1^4 (\that - \uhat)^2 \right\} ~,  
\label{gamma*Tgamma*L}\\
\frac{\d\sigma_{\mathrm{LT}}}{\d\that} & = & 
\frac{\d\sigma_{\mathrm{TL}}}{\d\that} ( Q_1^2 \leftrightarrow Q_2^2 ) ~,   
\label{gamma*Lgamma*T}\\ 
\frac{\d\sigma_{\mathrm{LL}}}{\d\that} & = & 
32 \pi \alpha_{\mathrm{em}}^2 N_{\f} e_{\f}^4 \,
\frac{Q_1^2 Q_2^2 \shat^2}{\lambda^6 \that^2 \uhat^2} 
(\that \uhat - Q_1^2 Q_2^2) (\that - \uhat)^2 ~,
\label{gamma*Lgamma*L}
\end{eqnarray}  
with $\shat = W^2$,
\begin{eqnarray}
\lambda & = & \sqrt{(\shat + Q_1^2 + Q_2^2)^2 - 4 Q_1^2 Q_2^2} ~, \\
F & = & \frac{ \shat (\shat + Q_1^2 + Q_2^2)}{\lambda^2} ~,
\end{eqnarray}  
$e_{\f}$ the electrical charge and $N_{\f}$ the colour factor, 
3 for a quark and 1 for a lepton. Remember that 
$\shat + \that + \uhat = - Q_1^2 - Q_2^2$, neglecting the fermion mass.
Note that the cross section for a longitudinal photon vanishes as
$Q_i^2$ in the limit $Q_i^2 \to 0$.

For a resolved photon, the photon virtuality scale is included in the 
arguments of the parton distribution but, in the spirit of the parton 
model, the virtuality of the parton inside the photon is not included 
in the matrix elements. Neither is the possibility of the partons being
in longitudinally polarized photons (see below, however). The same 
subprocess cross sections can therefore be used for direct $\gast \p$ 
processes and for single-resolved $\gast\gast$ ones. For 
$\gast \q \to \g \q$ one obtains 
\cite{siggap}
\begin{eqnarray} 
\frac{\d\sighat_{\mathrm{T}}}{\d\that} & = & 
\frac{8}{3} \pi \alpha_{\mathrm{s}} \alpha_{\mathrm{em}} e_{\q}^2 \,
\frac{1}{(\shat + Q_1^2)^2} \left\{
\frac{\shat^2 + \uhat^2 - 2 Q_1^2 \that}{- \shat \uhat} -
\frac{2 Q_1^2 \that}{(\shat +Q_1^2)^2} \right\} ~, 
\label{gamma*Tq} \\
\frac{\d\sighat_{\mathrm{L}}}{\d\that} & = & 
\frac{8}{3} \pi \alpha_{\mathrm{s}} \alpha_{\mathrm{em}} e_{\q}^2 \,
\frac{-4 Q_1^2 \that}{(\shat + Q_1^2)^4} ~,
\label{gamma*Lq}
\end{eqnarray} 
and for $\gast \g \to \q \qbar$
\begin{eqnarray} 
\frac{\d\sighat_{\mathrm{T}}}{\d\that} & = & 
\pi \alpha_{\mathrm{s}} \alpha_{\mathrm{em}} e_{\q}^2 \,
\frac{1}{(\shat + Q_1^2)^2} 
\frac{\that^2 + \uhat^2}{\that \uhat} \left[ 1 - 
\frac{2 Q_1^2 \shat}{(\shat + Q_1^2)^2} \right] ~, 
\label{gamma*Tg}\\ 
\frac{\d\sighat_{\mathrm{L}}}{\d\that} & = & 
\pi \alpha_{\mathrm{s}} \alpha_{\mathrm{em}} e_{\q}^2 \,
\frac{8 Q_1^2 \shat}{(\shat + Q_1^2)^4} ~. 
\label{gamma*Lg}
\end{eqnarray} 
Convolution with parton distributions gives
\begin{eqnarray}
\d\sigma(\gast\p \to \X) & = & \iint \d\xhat_2 \, \d\that \,
f_i^{\p}(\xhat_2,\mu^2) \, 
\frac{\d\sighat}{\d\that}(\shat = \xhat_2 W^2) ~, \\
\d\sigma(\gast\gast \to \X) & = & \iint \d\xhat_2 \, \d\that \,
f_i^{\gast}(\xhat_2,\mu^2,Q_2^2) \, 
\frac{\d\sighat}{\d\that}(\shat = \xhat_2 W^2) ~, 
\end{eqnarray}
where $\mu^2$ is the scale of the hard-scattering subprocess.

Finally we come to resolved processes in $\gast\p$ and doubly-resolved 
ones in $\gast\gast$. There are six basic QCD cross sections, 
$\q\q' \to \q\q'$, $\q\qbar \to \q'\qbar'$, $\q\qbar \to \g\g$,
$\q\g \to \q\g$, $\g\g \to \g\g$ and $\g\g \to \q\qbar$. Since again
parton virtualities are not included, these are the expressions 
familiar from $\p\p$ physics \cite{sigpp} and are not 
listed here. Again a convolution with parton distributions is
necessary,
\begin{eqnarray}
\hspace{-3mm} \d\sigma(\gast\p \to \X) \hspace{-2mm} & = & 
\hspace{-3mm} \iiint \d\xhat_1 \, \d\xhat_2 \, 
\d\that \, f_i^{\gast}(\xhat_1,\mu^2,Q_1^2) \, 
f_j^{\p}(\xhat_2,\mu^2) \, 
\frac{\d\sighat}{\d\that}(\shat = \xhat_1 \xhat_2 W^2) ~, \\
\hspace{-3mm} \d\sigma(\gast\gast \to \X) \hspace{-2mm} & = &  
\hspace{-3mm} \iiint \d\xhat_1 \, \d\xhat_2 \, 
\d\that \, f_i^{\gast}(\xhat_1,\mu^2,Q_1^2) \,  
f_j^{\gast}(\xhat_2,\mu^2,Q_2^2) \, 
\frac{\d\sighat}{\d\that}(\shat = \xhat_1 \xhat_2 W^2) ~.
\end{eqnarray}
In line with the neglect of parton masses, also the $\shat$ 
expression for $\gast\gast$ is simpler than its $W^2$ analogue in 
eq.~(\ref{W2gaga}). When initial-state radiation is subsequently
included, both transverse momenta and spacelike parton virtualities  
are generated, but in such a way that the relation 
$\shat = \xhat_1 \xhat_2 W^2$ is maintained \cite{backwards}.
Differences between alternative $\shat$ definitions being subleading,
they are beyond our standard QCD accuracy.

\subsection{Parton Distributions}
\label{pdf}

One major element of model dependence enters via the choice of parton
distributions for a resolved virtual photon. These distributions
contain a hadronic component that is not perturbatively calculable. It
is therefore necessary to parameterize the solution with input from
experimental data, which mainly is available for (almost) real
photons. In the following we will use the SaS distributions
\cite{saspdf}, which are the ones best suited for our
formalism. Another set of distributions is provided by GRS
\cite{grspdf}, while a simpler recipe for suppression factors relative
to real photons has been proposed by DG \cite{dgpdf}.

The SaS distributions for a real photon can be written as
\begin{equation}
f_a^{\gamma}(x,\mu^2) =
\sum_V \frac{4\pi\aem}{f_V^2} f_a^{\gamma,V}(x,\mu^2; Q_0^2)
+ \frac{\aem}{2\pi} \, \sum_{\q} 2 e_{\q}^2 \,
\int_{Q_0^2}^{\mu^2} \frac{{\d} k^2}{k^2} \,
f_a^{\gamma,\q\qbar}(x,\mu^2;k^2) ~.
\label{decomp}
\end{equation}
Here the sum is over a set of vector mesons
$V = \rho^0, \omega, \phi, \mathrm{J}/\psi$ according to a 
vector-meson-dominance ansatz for low-virtuality fluctuations of
the photon, with experimentally determined couplings $4\pi\aem/f_V^2$.
The higher-virtuality, perturbative, fluctuations are 
represented by an integral over the virtuality $k^2$ and a sum over
quark species. We will refer to the first part as the VMD one and 
the second as the anomalous one. Each component $f^{\gamma,V}$ and 
$f^{\gamma,\q\qbar}$ obeys a unit momentum sum rule, and also obeys 
normal QCD evolution equations. The $f^{\gamma,V}(x,Q_0^2)$ have to 
be determined by a tune to $F_2^{\gamma}(x,\mu^2)$ data, while 
$f^{\gamma,\q\qbar}$ evolve from the boundary condition
\begin{equation}
f_a^{\gamma,\q\qbar}(x,k^2;k^2) =  
\frac{3}{2} \, \left( x^2 + (1-x)^2 \right) \,
( \delta_{a\q} + \delta_{a\qbar} )  ~.
\label{fxinit}
\end{equation}
The $\mu^2$ dependence enter both via the evolution of each component
and via the upper limit of the $\d k^2$ integral. It is the latter 
dependence that generates the so-called anomalous term of the 
photon distribution evolution equations, from which the terminology 
has been taken over for the related event class. 

From the above ansatz, the extension to a virtual photon is given by
the introduction of a dipole dampening factor for each component,
\begin{eqnarray}
f_a^{\gast}(x,\mu^2,Q^2)
& = & \sum_V \frac{4\pi\aem}{f_V^2} \left(
\frac{m_V^2}{m_V^2 + Q^2} \right)^2 \,
f_a^{\gamma,V}(x,\mu^2;\tilde{Q}_0^2)
\nonumber \\
& + & \frac{\aem}{2\pi} \, \sum_{\q} 2 e_{\q}^2 \,
\int_{Q_0^2}^{\mu^2} \frac{{\d} k^2}{k^2} \, \left(
\frac{k^2}{k^2 + Q^2} \right)^2 \, f_a^{\gamma,\q\qbar}(x,\mu^2;k^2)~.
\label{decompvirt}
\end{eqnarray}
Thus, with increasing $Q^2$, the VMD components die away faster than
the anomalous ones, and within the latter the low-$k^2$ ones faster
than the high-$k^2$ ones. In the VMD component, the effective
evolution range is reduced by the introduction of a 
$\tilde{Q}_0^2 = Q_{\mathrm{int}}^2 =Q_0 Q_{\mathrm{eff}} > Q_0^2$, 
with
\begin{equation}
Q_{\mathrm{eff}}^2 = \mu^2 \, \frac{Q_0^2 + Q^2}{\mu^2 + Q^2} \,
\exp \left\{ \frac{Q^2 (\mu^2 - Q_0^2)}{(\mu^2 + Q^2)(Q_0^2 + Q^2)}
\right\} ~.
\end{equation}
As a technical trick, the handling of the $k^2$ integral is
made more tractable by replacing the dipole factor by a
$k^2$-independent multiplicative factor and an increased lower
limit $Q_{\mathrm{int}}^2$ of the integral, 
in such a way that both the momentum sum
and the average evolution range is unchanged. Finally, correction
factors are introduced to ensure that $f_a^{\gast}(x,\mu^2,Q^2) \to 0$
for $\mu^2 \to Q^2$: in the region $Q^2 > \mu^2$ a fixed-order
perturbative description is more appropriate than the leading-log
description in terms of a resolved photon.  We then arrive at the
so-called modified $P_{\mathrm{int}}$ scheme, which is the one used
here.

Since the probed real photon is purely transverse, the above ansatz
does not address the issue of parton distributions of the longitudinal
virtual photons. One could imagine an ansatz based on longitudinally
polarized vector mesons, and branchings $\gast_{\mathrm{L}} \to
\q\qbar$, but currently no parameterization exists along these
lines. We will therefore content ourselves by exploring alternatives 
based on applying simple multiplicative factors $R$ to the results 
obtained for a resolved transverse photon. As usual, processes involving 
longitudinal photons should vanish in the limit $Q^2 \to 0$. To study two 
extremes, the region with a linear rise in $Q^2$ is defined either by 
$Q^2<\mu^2$ or by $Q^2<m_{\rho}^2$, where the former represents the 
perturbative and the latter some non-perturbative scale. Also the 
high-$Q^2$ limit is not well constrained; we will compare two different 
alternatives, one with an asymptotic fall-off like $1/Q^2$ and another 
which approaches a constant ratio, both with respect to the transverse 
resolved photon. (Since we put $f_a^{\gast}(x,\mu^2,Q^2) = 0$ for 
$Q^2>\mu^2$, the $R$ value will actually not be used for large $Q^2$, 
so the choice is not so crucial.) We therefore study the alternative 
ans\"atze
\begin{eqnarray}
R_1(y,Q^2,\mu^2) & = & 1 + a \frac{4 \mu^2 Q^2}{(\mu^2 + Q^2)^2}
\frac{f_{\gamma/l}^{\mathrm{L}}(y,Q^2)}
{f_{\gamma/l}^{\mathrm{T}}(y,Q^2)}~,\label{Rfact1}\\
R_2(y,Q^2,\mu^2) & = & 1 + a \frac{4 Q^2}{(\mu^2 + Q^2)}
\frac{f_{\gamma/l}^{\mathrm{L}}(y,Q^2)}
{f_{\gamma/l}^{\mathrm{T}}(y,Q^2)}~,\label{Rfact2}\\
R_3(y,Q^2,\mu^2) & = & 1 + a \frac{4 Q^2}{(m_{\rho}^2 + Q^2)}
\frac{f_{\gamma/l}^{\mathrm{L}}(y,Q^2)}
{f_{\gamma/l}^{\mathrm{T}}(y,Q^2)}\label{Rfact3}
\end{eqnarray}
with $a=1$ as main contrast to the default $a=0$. The $y$ dependence 
compensates for the difference in photon flux between transverse
and longitudinal photons. $R_1$ and $R_2$ have the same onset at low $Q^2$
but different asymptotic behaviour; for the former case the longitudinal 
part vanishes and for the latter it approaches a constant (w.r.t. the 
transverse case). For the third case the onset is governed by a 
non-perturbative parameter $m_{\rho}$ and it has the same asymptotic limit 
as $R_2$. In a more sophisticated treatment, presumably also the $k^2$ scale 
of the $\gast \to \q\qbar$ fluctuation would enter. In double-resolved 
$\gast\gast$ events one $R$ factor is applied for each side.

Another ambiguity is the choice of $\mu^2$ scale in parton distributions.
For a process such as $\gamma\gamma \to \q\qbar$, with real photons, 
conventional wisdom is that $\mu^2 = -\that$ is the proper scale in the 
limit $\that \to 0$, where $t$-channel graphs dominate the cross 
sections, and $\mu^2 = -\uhat$ in the limit $\uhat \to 0$. 
The combination $\mu^2 = \that\uhat/\shat = \pT^2$ interpolates between
these limits and thus is a traditional choice, sometimes multiplied by 
some constant factor. When the incoming (or outgoing) photons/partons are 
not massless, $\that\uhat/\shat \neq \pT^2$. A possible generalization 
for a direct virtual photon, $Q_1^2 \neq 0$, is
\begin{equation}
\mu^2 = - \frac{\that\uhat}{\that+\uhat} 
      = \frac{\that\uhat}{\shat+Q_1^2} 
      = \pT^2 \, \frac{\shat+Q_1^2}{\shat}   ~.
\label{mutwosimple}
\end{equation}
For a corresponding $t$-channel graph involving the quark from a resolved 
photon, in principle the same relation should hold, with the quark 
virtuality $\hat{Q}_1^2$ substituting for the photon $Q_1^2$ one.
Inside {\sc Pythia}, however, all incoming partons are assumed massless in 
the selection of hard-scattering kinematics. When, later on, the
$\gast \to \q\qbar$ branching is included, the $\shat$ and the rest-frame 
scattering angle are 
left unaffected. Thus also $\pT^2$ is unchanged, while $\that$ and $\uhat$ 
are not. Using the corrected values, one obtains the same final expression
as in eq.~(\ref{mutwosimple}), with $\hat{Q}_1^2$ instead of $Q_1^2$.
Kinematics provides the constraint $\hat{Q}_1^2 > x_1 Q_1^2$ at the 
$\gast \to \q\qbar$ branching. The lower bound is a rather conservative 
estimate of the actual value of $\hat{Q}_1^2$, however, and more typically
one obtains $\hat{Q}_1^2 \approx Q_1^2$ from the $\gast \to \q\qbar$ 
branching. (Further branchings are included in the standard QCD 
parton-shower description. By analogy with the description of other QCD 
processes, we here disregard these shower virtualities.)
A scale choice like in eq.~(\ref{mutwosimple}) therefore should be a 
sensible one both for direct and resolved photons.

When both incoming photons are virtual, the relation between $\mu^2$
and $\pT^2$ becomes more complicated, but in the limit that terms of order
$Q_1^2 Q_2^2/\shat^2$ are neglected, they simplify to
\begin{equation}
\mu^2 = \pT^2 \, \frac{\shat+Q_1^2+Q_2^2}{\shat} ~.
\end{equation}
This expression does not guarantee that $\mu^2 > Q_1^2 + Q_2^2$.
Sometimes such an inequality is assumed \cite{LeifHannes}, so in order
to cover a broader range of scale choices, below we will be comparing six 
different alternatives. In (almost) increasing order these are
\begin{eqnarray}
\mu_1^2 & = & \pT^2 ~,\label{mu1}\\
\mu_2^2 & = & \pT^2 \, \frac{\shat+ x_1 Q_1^2 + x_2 Q_2^2}{\shat}  ~,
\label{mu2}\\
\mu_3^2 & = & \pT^2 \, \frac{\shat+ Q_1^2 + Q_2^2}{\shat} ~,\label{mu3}\\
\mu_4^2 & = & \pT^2 + \frac{Q_1^2 + Q_2^2}{2} ~,\label{mu4}\\
\mu_5^2 & = & \pT^2 + Q_1^2 + Q_2^2 ~,\label{mu5}\\
\mu_6^2 & = & 2 \mu_3^2~.\label{mu6}
\end{eqnarray}  
Only the fifth alternative ensures $f_a^{\gast}(x,\mu^2,Q^2) > 0$ for
arbitrarily large $Q^2$; in all other alternatives the resolved
contribution (at fixed $\pT$) vanish above some $Q^2$ scale. The last 
alternative exploits the well-known freedom of including some multiplicative
factor in any (leading-order) scale choice.
When nothing is mentioned explicitly below, the choice $\mu_3^2$ is used.
We should note that the expressions in the program also contain a dependence 
on final-state masses, e.g. for the production of massive quarks, but this
is left out here since it is not a topic studied in this paper. (It will 
enter briefly in the following, however.) 

\subsection{Other Model Aspects}

The issues discussed above are the main ones that distinguish the
description of processes involving virtual photons from those induced 
by real photons or by hadrons in general. In common for the tree is 
the need to consider the buildup of more complicated partonic
configurations from the lowest-order `skeletons' defined above, 
(\textit{i}) by parton showers, (\textit{ii}) by multiple 
parton--parton interactions and beam remnants, where applicable,
and (\textit{iii}) by the subsequent transformation of these partons 
into the observable hadrons. The latter, hadronization stage can be 
described by the standard string fragmentation framework 
\cite{AGIS}, followed by the decays of unstable primary hadrons, 
and is not further discussed here. In the following we comment 
on the shower, multiple-interaction and beam-remnant aspects.

The parton-shower description is conveniently subdivided into 
initial- and final-state radiation. For wide-angle emissions such a
classification is not unambiguous, and it is necessary to consider
interference effects in order to describe the data \cite{cdfcoher}.
A direct photon is not associated with any initial-state 
QCD radiation.

In the hadronic environment, initial-state radiation normally means an
evolution of a spacelike branch of partons, from an initially
vanishing virtuality up to the scale of the hard process. The
`backwards evolution' strategy \cite{backwards} allows this cascade to
be reconstructed in reverse order. That is, from a parton $b$ coming
in to the hard scattering, the branching $a \to b c$ that produced $b$
is first reconstructed, thereafter the branching that produced $a$,
and so on, down to the lower cutoff scale $Q_0^{\mathrm{sh}}$, in
practice of the order of 1~GeV.  Inclusively, the effect of parton-shower
histories is already taken into account by the $\mu^2$ dependence of the
parton distributions used to select the hard scattering, so what this
procedure does is to associate an exclusive set of 
initial-state-radiation partons to each hard scattering.

For a real photon, the VMD part is assumed to behave like a hadron,
while the lower parton-shower cut-off $Q_0^{\mathrm{sh}}$ has to be 
considered further for the anomalous component. In the spirit of the 
ansatz for parton distributions, eq.~(\ref{decomp}), the $k^2$ value of 
the $\gamma \to \q\qbar$ branching is distributed like $\d k^2/k^2$.
In principle, for a given physical process, there would be a further
`trigger bias' effect to this distribution: at large (small) $x$ values
a large (small) $k^2$ would be favoured since it would be associated with
a small (large) evolution range. As a first approximation, this bias is 
neglected here, i.e. $k^2$ is picked flat in a logarithmic scale 
between $Q_0^2$ and $\mu^2$. When this $k$ is larger than the default
value of $Q_0^{\mathrm{sh}}$, the shower cut-off is increased to instead be
given by $k$. Thus there is no shower evolution below the $k$ scale of 
the $\gamma \to \q\qbar$ branching. Thereby the amount of collinear 
emission along the incoming photon direction is reduced for 
high-virtuality fluctuations. 

As a technical aside, we note that the default value $Q_0 = 0.6$~GeV is 
smaller than $Q_0^{\mathrm{sh}} = 1$~GeV. First, it should be made
clear that $k^2$ and the $Q^2$ virtuality variable of the spacelike shower 
evolution are not defined in precisely the same way and therefore very well 
can differ by factors of order unity, so that also the lower cut-offs may 
differ. However, also disregarding this, there is no reason why the two
cut-offs should agree.
Whereas $Q_0$ is severely constrained e.g. by $F_2^{\gamma}$ data, 
$Q_0^{\mathrm{sh}}$ is just a crude estimate of where emitted partons in the 
cascade are so collinear that their effect may be safely neglected. That is, 
we do not exclude the possibility of parton emissions in the range between
$Q_0$ and $Q_0^{\mathrm{sh}}$, but also do not believe that their inclusion
or not is essential for a description of final-state properties. The effects 
of an increased shower cut-off only show up for $\gamma \to \q\qbar$ 
branchings with $k \gtrsim 2$~GeV, in terms of a reduced parton emission
rate along the incoming photon direction. 
    
The generalization to a virtual photon is now straightforward, given the
extension of $f_a^{\gamma}$ to $f_a^{\gast}$, cf. eqs.~(\ref{decomp}) and
(\ref{decompvirt}). For the VMD piece, the shower cut-off is selected to
be the larger of $Q_{\mathrm{int}}$ and $Q_0^{\mathrm{sh}}$. For the anomalous
piece, a  $k^2$ is picked flat in a logarithmic scale between 
$Q_{\mathrm{int}}^2$ and $\mu^2$, and the shower cut-off is the larger of 
this $k$ and $Q_0^{\mathrm{sh}}$.

Final-state radiation from the scattered partons of the hard interaction 
follow the same pattern as in hadron--hadron collisions, and so need not
be discussed here. Also time-like partons on the `side branches' of an
initial-state cascade can undergo final-state evolution, as part of the
conventional shower description. The new aspect concerns the $\qbar$
`beam remnant' parton of an anomalous $\gast \to \q\qbar$ branching, 
where the $\q$ is the initiator of the spacelike cascade. This parton can
undergo a shower evolution from the $\gast \to \q\qbar$ branching scale
$k$ down to the timelike shower cut-off $m_0^{\mathrm{sh}} \approx 1$~GeV, 
provided $k$ is above the cut-off. For the VMD part of the photon, the beam 
remnant is assumed not to radiate.

In general, a beam remnant contains the flavours `left behind' when one
parton initiates the spacelike shower that leads up to the hard interaction.
Momentum conservation gives its kinematics. This includes both a longitudinal
momentum fraction $1-x$, if the initiator takes $x$, and a transverse  
`primordial $\kT$' recoil. For the VMD components of the photon this 
transverse momentum should be of a nonperturbative character and small,
$\sim 0.5$~GeV, although there are indications of much larger values in
data \cite{largekT}. A Gaussian ansatz is used to pick a $\kT$ vector.
For the anomalous part of the photon, we associate its $\kT$ with the $k$ 
scale that already figured prominently above. 

Multiple parton--parton interactions \cite{multint,sasevt} could be viewed
as a further sophistication of the beam-remnant description. The basic idea
is that hadronic states contain many partons and that therefore several
perturbative interactions may occur in a hadron--hadron collision. Usually
(at most) one of these give rise to visible high-$\pT$ jets, while the
rest only add to the underlying event activity. This additional component  
is vitally needed in order to reproduce the observed multiplicity and 
$\ET$ flow in high-energy $\p\pbar$ collisions, at least if 
string hadronization is supposed to be universal, i.e. have its parameters
constrained by $\e^+\e^-$ data. The main unknown parameter in this approach
is $\pTmin^\mathrm{MI} \approx 2$~GeV, the scale below which 
perturbation theory is assumed not to be applied anymore. Physically, this 
scale can be viewed as related to an effective colour screening phenomenon: 
gluons with a wavelength larger than $1/\pTmin^\mathrm{MI}$ do not 
resolve the colour structure inside the hadron wave function and therefore 
decouple. Of course, this screening should set in gradually, so 
$\pTmin^\mathrm{MI}$ is only an effective parameter. To first 
approximation, the average number of interactions per event is the ratio of 
the jet cross section above $\pTmin^\mathrm{MI}$ to the total 
inelastic cross section. By now there exist convincing evidence in favour 
of multiple interactions also in the HERA data \cite{multintHERA}. 

Almost by definition, direct and single-resolved processes do not lead to 
multiple interactions. Furthermore, within the spectrum of resolved components
of a real photon, one would expect a close-to hadronlike behaviour for the VMD
part and then a gradual fading-away of multiple interactions as the $k^2$
scale of an anomalous fluctuation is increased. This may partly be seen as a 
consequence of that the number of partons (above some fixed $x_0$) decrease
with increasing $k^2$. Furthermore, the physical size of an anomalous
fluctuation should scale like $1/k$, so the screening argument above 
would lead to a scaling-up of $\pTmin^\mathrm{MI}$ roughly by a factor
$k/Q_0$. This in part would be compensated by an expected decrease in the
total cross section of an anomalous fluctuation, about like $1/k^2$ based
on the size scaling argument, but in total there should still be a rather 
rapid fall-off with $k$. As a first approximation, we follow the route adopted 
in \cite{sasevt}, namely to allow multiple interactions for VMD states to the 
same extent as for hadrons, and not at all for the anomalous states.

When generalizing to virtual photons, the same arguments as used above for 
the anomalous component would suggest a fall-off of multiple interactions
in the VMD component with increasing $Q^2$ scale. As a first guess, we
have here chosen to scale up $\pTmin^\mathrm{MI}$ like 
$\sqrt{1 + Q^2/m_V^2}$, while the normalizing total cross section is scaled 
down like $m_V^4/(m_V^2+Q^2)^2$ in accordance with the dipole ansatz.

\section{Results}

\subsection{Basic Distributions}

\subsubsection{$\gast\p$ Partonic Cross Sections}

\begin{figure} [!htb]
   \begin{center}
   \mbox{\psfig{figure=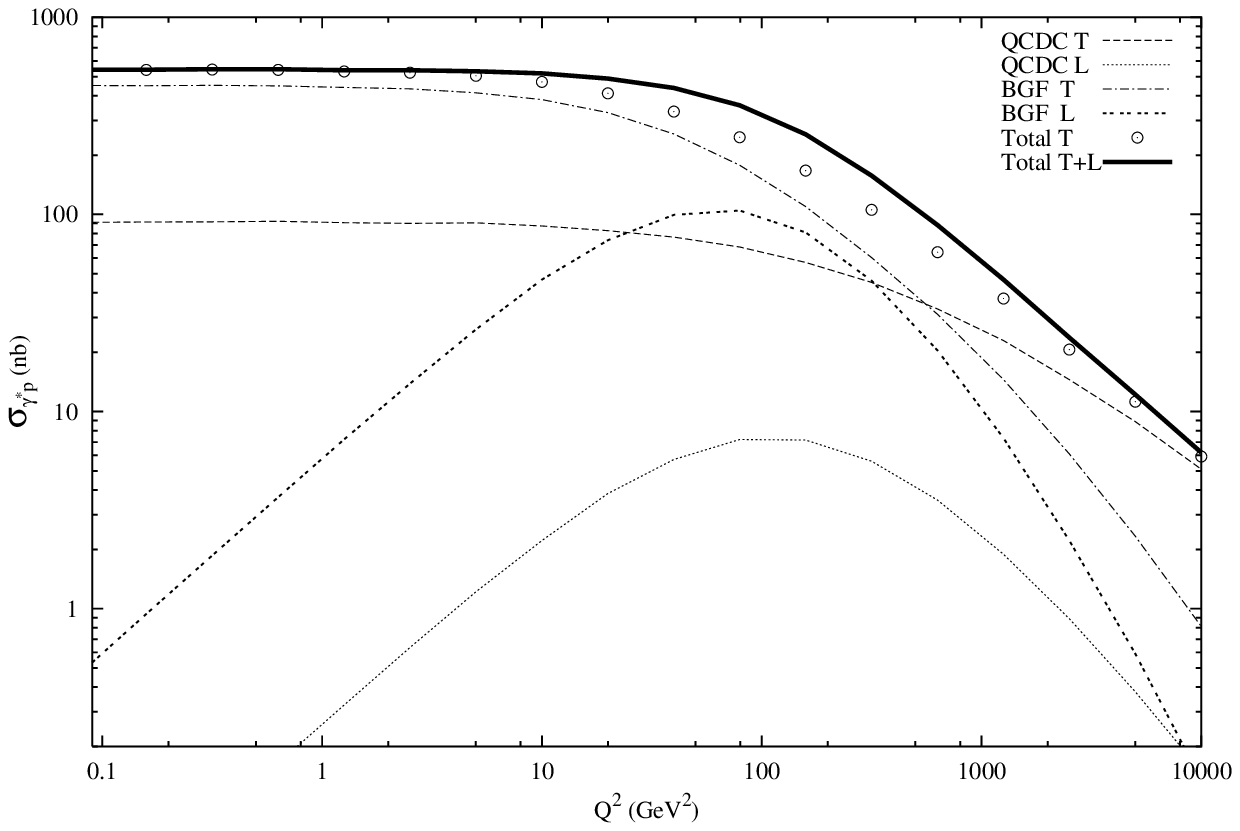,height=90mm}}
   \end{center}
\captive{Contributions to the $\sigma_{\gast\p}$ parton cross section from 
the subprocesses in equations~(\ref{gamma*Tq})--(\ref{gamma*Lg}), 
$\pT^{\mathrm{parton}}>5$~GeV and $W_{\gast\p}=200$~GeV. The circles and
the solid line is the sum of the transverse (T) and the transverse plus the 
longitudinal (T+L) direct processes respectively.
\label{fig:gamma*p}}
\end{figure}

The contribution to the $\sigma_{\gast\p}$ parton cross section from the 
different direct components in equations~(\ref{gamma*Tq})--(\ref{gamma*Lg}), 
the boson--gluon fusion and the QCD Compton processes, 
are shown in Fig.~\ref{fig:gamma*p} as a function of the photon virtuality, 
$Q^2$. The partons produced in the $2 \rightarrow 2$ hard scattering 
subprocess are restricted to have transverse momenta $\pT$ larger than 
5~GeV and the hadronic centre of mass energy $W=\sqrt{s_{\gast\p}}$ is equal 
to 200~GeV.

In the limit $Q^2\rightarrow0$ the cross section vanishes for the 
processes involving a longitudinal photon. In the same 
limit the boson--gluon fusion process with a transverse photon, 
$\gast_{\mathrm{T}} \g \to \q \qbar$, is more important than the QCD Compton
counterpart, $\gast_{\mathrm{T}} \q \to \g \q$; this kinematic region permits 
the gluon/quark from the proton to have a small fraction $x$ of the proton 
momentum, $x=(\shat+Q^2)/(W^2+Q^2)$. 
At large $Q^2$, the opposite is true; $x$ is large and the QCD Compton 
process, involving an initial quark from the proton, is more important than 
the boson--gluon fusion process. Both processes are suppressed by the 
$1/(\shat+Q^2)^2$ factor that provides the main dampening at large $Q^2$.
The longitudinal processes, $\gast_{\mathrm{L}} \q \to \g \q$ and 
$\gast_{\mathrm{L}} \g \to \q \qbar$ have maximal cross sections when 
$Q^2 \sim \shat$, since $\shat \geq 4 \pTmin$ 
this occurs for $Q^2$ around 100~GeV$^2$. 

The proton structure function used to produce Fig.~\ref{fig:gamma*p} was 
GRV~94~LO, which is the default in {\sc Pythia}. A different choice of 
structure function would only change the details but not the conclusions 
made from these results. In the following the GRV~94~LO will be used as the 
proton parton distribution except where otherwise stated.

When studying jet production with low photon virtualities it is likely 
that the photon is resolved. The photon fluctuates then into a quark-antiquark 
pair that develops into a multiparton state and finally one of these partons 
scatters off a parton from the proton. The virtual photon parton distributions 
used for modeling the resolved components are the SaS~1D and~2D distributions. 
The dipole factors in eq.~(\ref{decompvirt}) will dampen the resolved 
components with increasing photon virtualities. 

\begin{figure} [!htb]
  \begin{center}
   \mbox{\psfig{figure=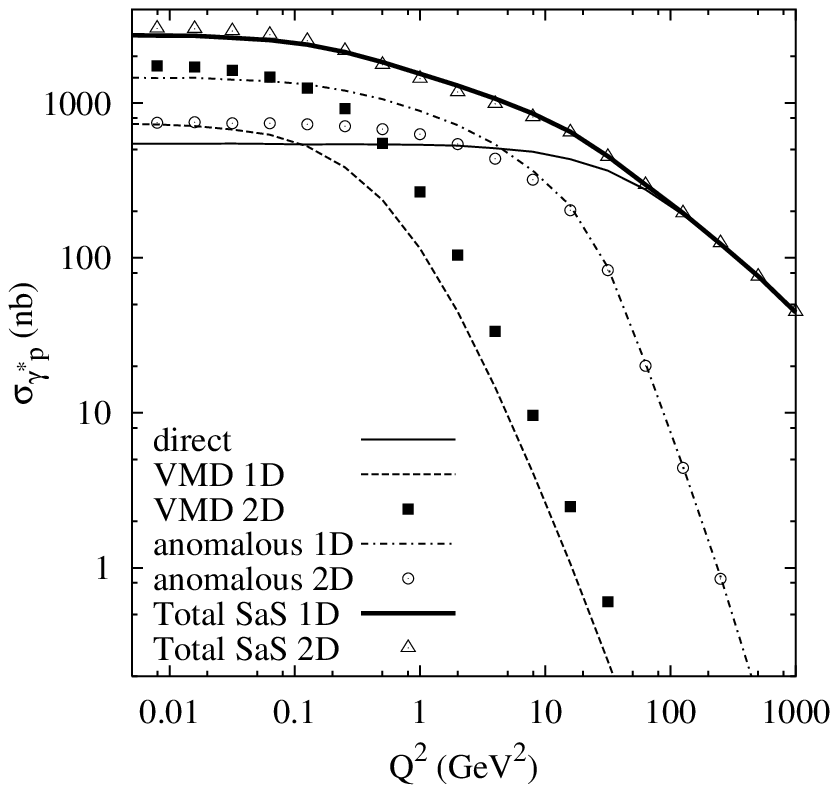,width=78mm}\hspace{-0.5cm}
	\psfig{figure=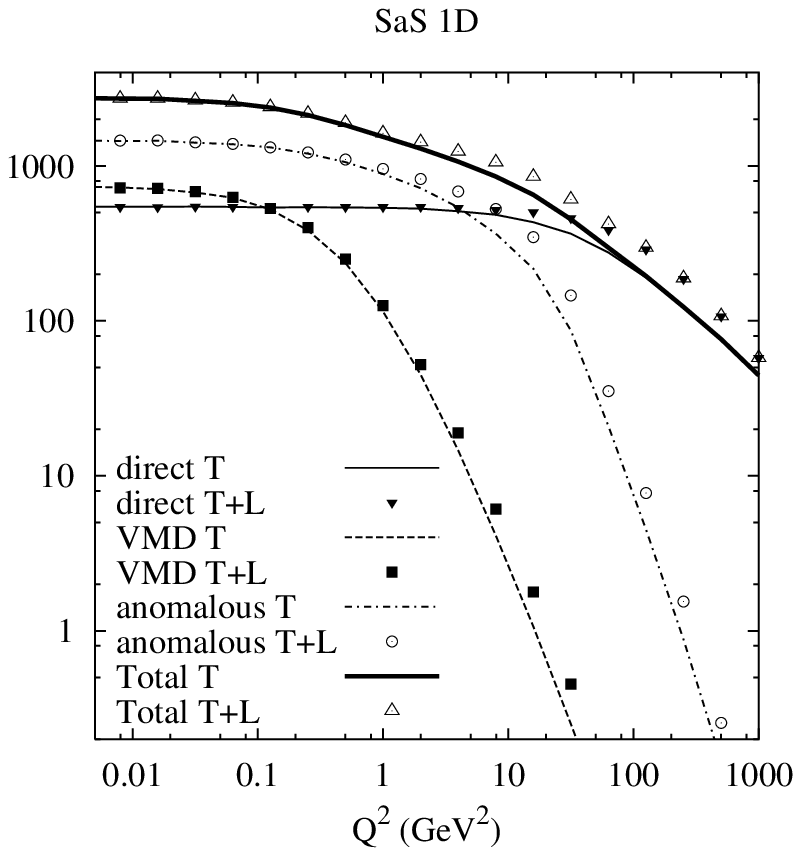,width=78mm}}
  \end{center}
\captive{Contributions to the $\sigma_{\gast\p}$ parton cross section from 
events involving a direct, a VMD and an anomalous photon, 
$\pT^{\mathrm{parton}}>5$~GeV and $\sqrt{s_{\gast\p}}=200$~GeV. To the 
left, the SaS~1D and SaS~2D parton distributions was used, to the right SaS~1D 
was used and a longitudinal resolved photon is modeled with $a=1$ and $y=0.44$ 
(see text).
\label{fig:res-p}}
\end{figure}

The relative importance of the direct components compared to the resolved 
components --- the VMD and anomalous fluctuations --- are shown in 
Fig.~\ref{fig:res-p}. The direct contribution is the sum of the transverse 
QCD Compton and boson--gluon fusion processes studied in 
Fig.~\ref{fig:gamma*p}. 
As expected, the VMD components die away faster than the anomalous ones
which in turn die away faster than the sum of the direct components. Here we
also see the difference between the SaS~1D and SaS~2D distributions, with 
SaS~2D having a higher cut-off for the separation between the VMD component 
and the anomalous component. This is reflected by the larger contribution from
the SaS~2D VMD component compared to the SaS~1D one, and vice versa for the 
anomalous component. 

Longitudinal resolved components are simulated with an extra factor of 
$R_1(y,Q^2,\mu^2)$, eq.~(\ref{Rfact1}), for the transverse resolved components.
 In Fig.~\ref{fig:res-p} an $a$-parameter equal to~1 has been used together
with the SaS~1D photon parton distribution. The $y$ value is set equal to 
0.44 with HERA energies in mind. A noticeable difference is seen in the total 
jet cross section for $Q^2$ around 10~GeV$^2$, since, in this ansatz, the 
maximal contribution for the longitudinal component is obtained at 
$Q^2 = \mu^2$, and the direct and the anomalous component is of the same order 
in this region. With $a=1$ the sum of the resolved components can be almost a 
factor of two larger than with $a=0$, the case with pure transverse resolved 
photons. There is nothing special with the choice $a=1$ except that it models 
a longitudinal resolved component that is of the same order of magnitude as 
the transverse component at intermediate $Q^2$ values; this is the same 
behaviour as seen for the direct components.

\subsubsection{$\gast\gast$ Partonic Cross Sections}

\begin{figure} [!htb]
   \begin{center}
   \mbox{\psfig{figure=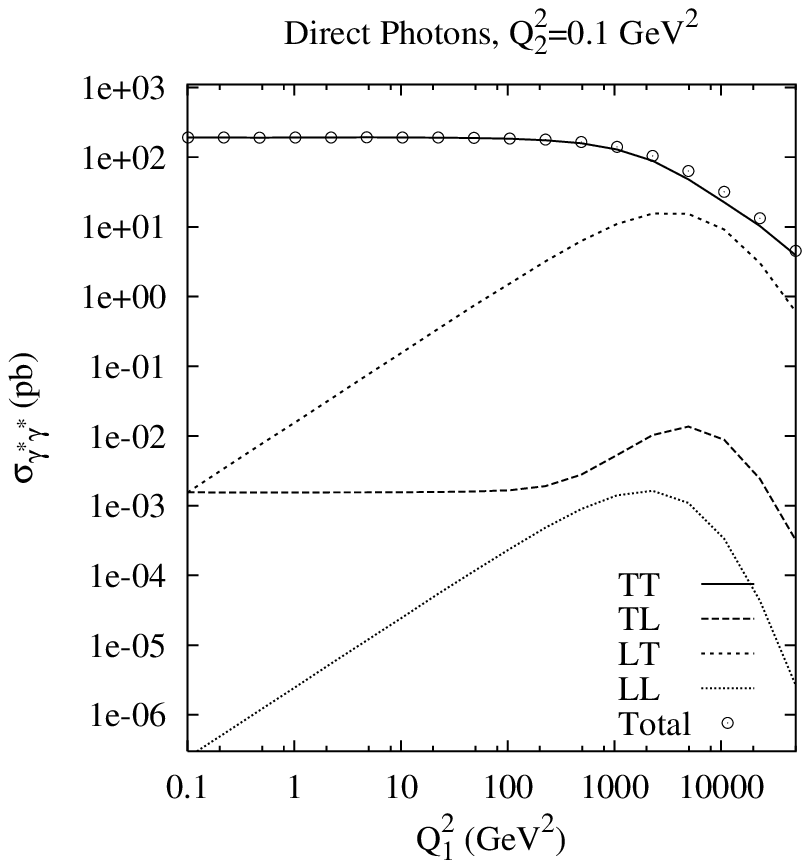,width=78mm}\hspace{-0.5cm}
	 \psfig{figure=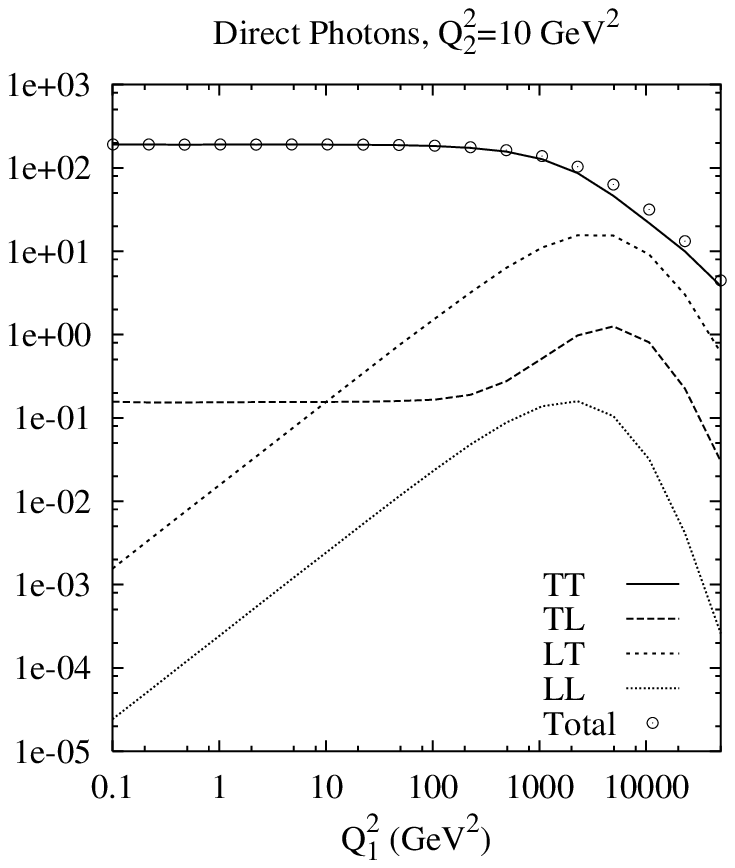,width=78mm}}
   \end{center}
\captive{Contributions to the $\sigma_{\gast\gast}$ parton cross section from 
the subprocesses in equations~(\ref{gamma*Tgamma*T})--(\ref{gamma*Lgamma*L}), 
$\pT^{\mathrm{parton}}>5$~GeV and $\sqrt{s_{\gast\gast}}=100$~GeV. 
The $Q_1^2$ of the first photon is varied while the second photon is kept at 
the fixed virtualities $Q_2^2=0.1~\mathrm{GeV}^2$ or 
$Q_2^2=10~\mathrm{GeV}^2$, respectively.
\label{fig:dirdir}}
\end{figure}

In Fig.~\ref{fig:dirdir} the processes 
$\gast_{\mathrm{T,L}} \gast_{\mathrm{T,L}} \to \q \qbar$, 
equations~(\ref{gamma*Tgamma*T})--(\ref{gamma*Lgamma*L}), have been generated 
at a centre of mass energy $W=\sqrt{s_{\gast\gast}}$ of 100~GeV. For 
simplicity, one of the photons was kept at a fixed virtuality of 0.1~GeV$^2$ 
or 10~GeV$^2$, respectively. The transverse momenta for the partons in the 
hard scattering was restricted to be larger than 5~GeV.

Clearly, the dominant process is $\gast_T \gast_T \to \q \qbar$ but, as the 
virtuality $Q_1^2$ increases, the contribution from the longitudinal 
processes becomes more and more important. From 
eq.~(\ref{gamma*Tgamma*T})--(\ref{gamma*Lgamma*L}) we notice that the 
$\gast_T \gast_L$, $\gast_L \gast_T$ and $\gast_L \gast_L$ direct cross 
sections have a maximum at $Q_1^2 \sim \shat = s_{\gast\gast}$ 
(for $Q_2^2 \ll \shat$). The $\gast_L \gast_L$ process is bounded to never be 
larger than twice the $\gast_T \gast_L$ or $\gast_L \gast_T$ process. 
The $\gast_L \gast_T$ and $\gast_L \gast_L$ processes
clearly go to zero as $Q_1^2 \rightarrow 0$, the $\gast_T \gast_L$ process 
remains finite and approaches the expression for the $\gast_L \g$ matrix 
element, eq.~(\ref{gamma*Lg}), with appropriate factors exchanged (and 
$Q_2^2 \rightarrow Q_1^2$). The $Q_2^2$ proportionality for the 
$\gast_T \gast_L$ and $\gast_L \gast_L$ matrix elements gives the factor of 
100 difference between the two cases in Fig.~\ref{fig:dirdir}. As for the case
with direct processes in $\gast \p$, the 
$\gast_{\xi_1} \gast_{\xi_2} \to \q \qbar$ processes ($\xi_i=$T, L) are 
suppressed by the $1/\lambda^2$ factor for high $Q_i^2$. 

\begin{figure} [!htb]
   \begin{center}
   \mbox{\psfig{figure=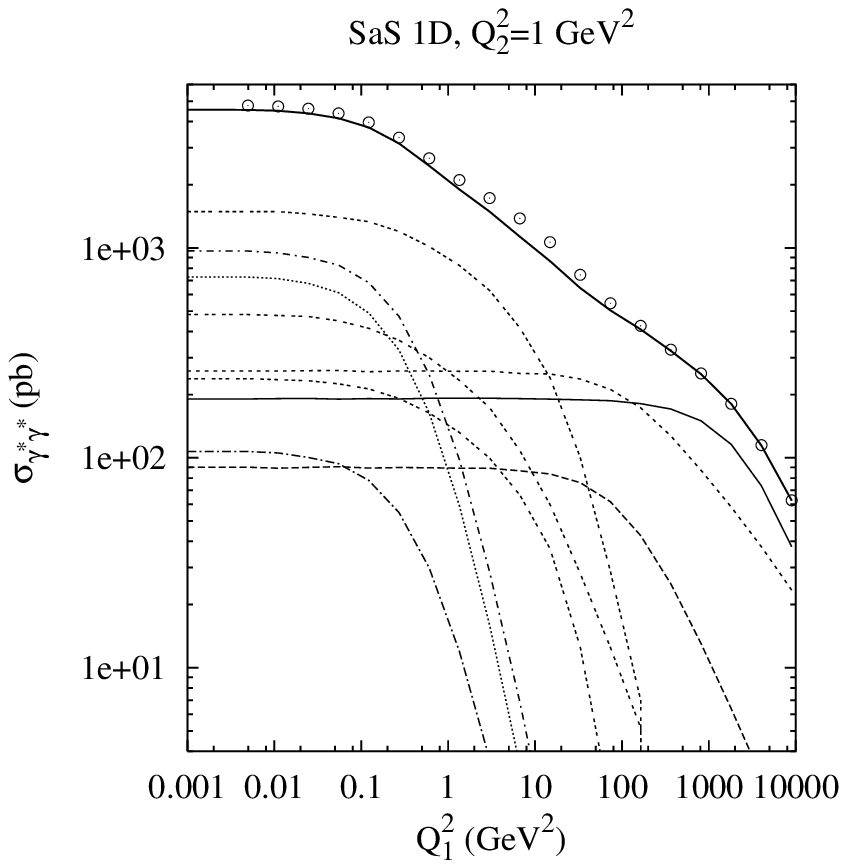,width=78mm}\hspace{-0.5cm}
	 \psfig{figure=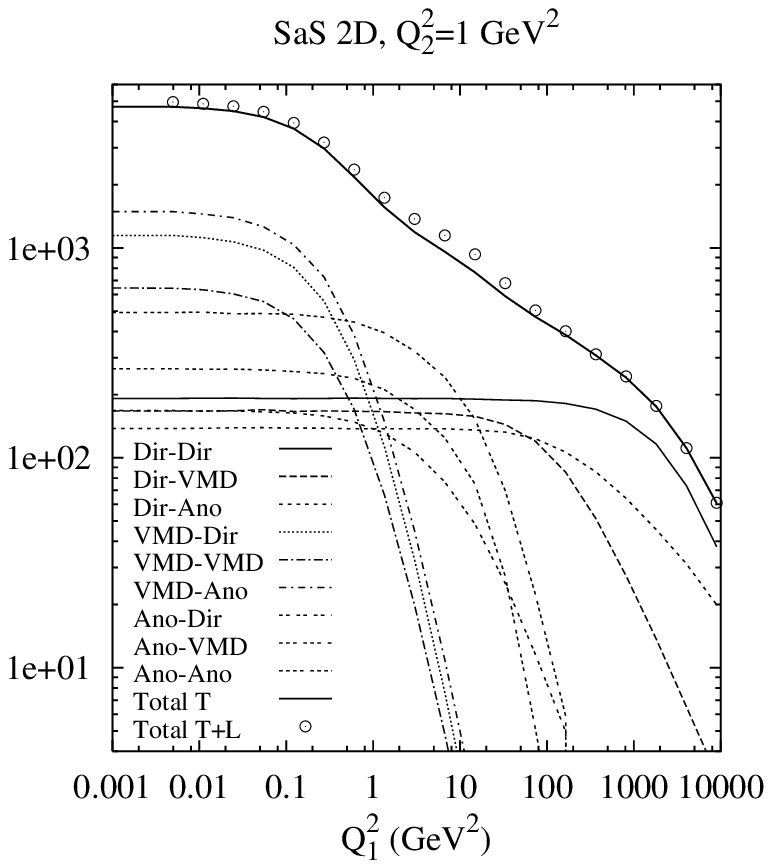,width=78mm}}
   \end{center}
\captive{Contributions to the $\sigma_{\gast\gast}$ parton cross section from 
the nine different event classes, $\pT^{\mathrm{parton}}>5$~GeV and 
$\sqrt{s_{\gast\gast}}=100$~GeV. The $Q_1^2$ variation is for the first 
photon; the second photon is kept at fixed virtuality $Q_2^2=1~\mathrm{GeV}^2$.
\label{fig:gamma*gamma*}}
\end{figure}

With the possibility of having resolved photons, nine event classes is 
obtained. They are illustrated in Fig.~\ref{fig:gamma*gamma*} for the SaS~1D 
and SaS~2D parton distributions, with the second photon having the fixed 
virtuality $Q_2^2=1$~GeV$^2$. Generally, as $Q_1^2$ increases, the VMD 
components (of the first photon) drops first, then the anomalous components 
and at very high $Q_1^2$ remains the direct components (direct--direct, 
direct--VMD and direct--anomalous). Particularly, with the choice 
$Q_2^2=1$~GeV$^2$, the anomalous--anomalous events dominates at low $Q_1^2$ 
for the SaS~1D case, whereas for the SaS~2D case, the VMD components dominate. 
In this ansatz, the $Q_\mathrm{int}$ scheme, the parton distribution for the 
anomalous component $f_a^{\gast}(x,\mu^2,Q^2)=0$ when $\mu^2<Q^2$. 

With the $y$--values taken equal to 0.5 
($s_{\gast\gast} \simeq y_1 y_2 s_{\e^+\e^-}$, 
$\sqrt{s_{\e^+\e^-}} \simeq 200$~GeV) and $a=1$ for $R_1(y,Q^2,\mu^2)$, the 
longitudinally resolved photon contribution can be obtained as the difference 
between the {\em Total T+L} and the {\em Total T} in 
Fig.~\ref{fig:gamma*gamma*}. Again, we see that the contribution is most 
pronounced where $\pTmin^2 \simeq Q^2$. For this special
case, it happens to be where the anomalous component (of the first photon)
is comparable with the direct one. 

\subsubsection{$x_{\gamma}$ Distributions}
\label{xgsec}

\begin{figure} [!htb]
   \begin{center}      
      \mbox{\psfig{figure=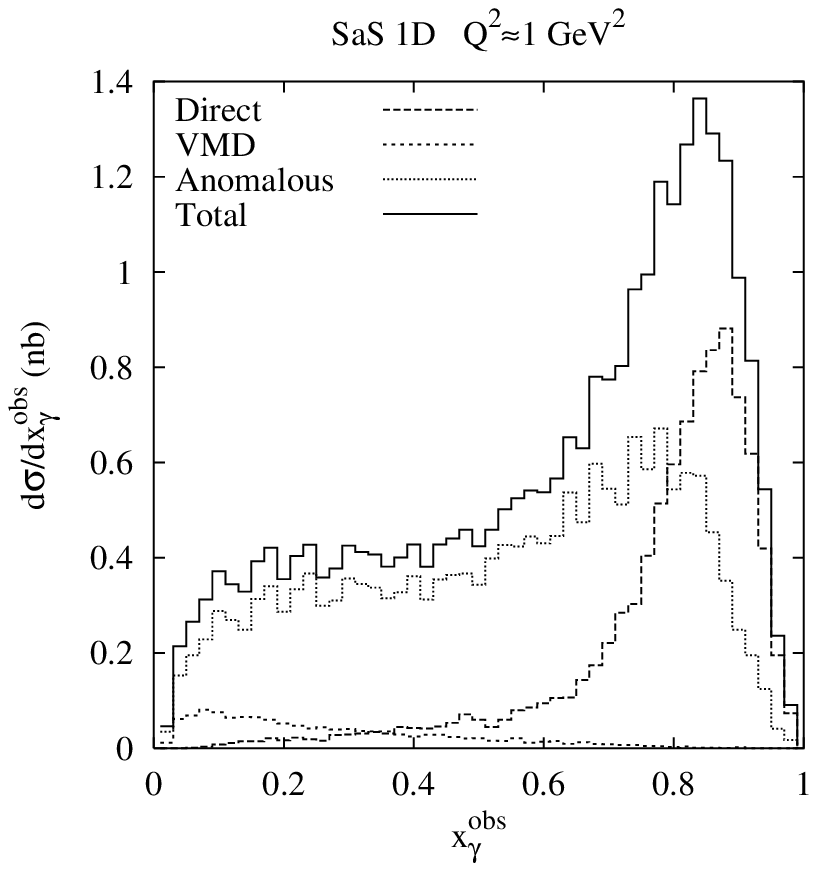,width=78mm}\hspace{-0.5cm}
	    \psfig{figure=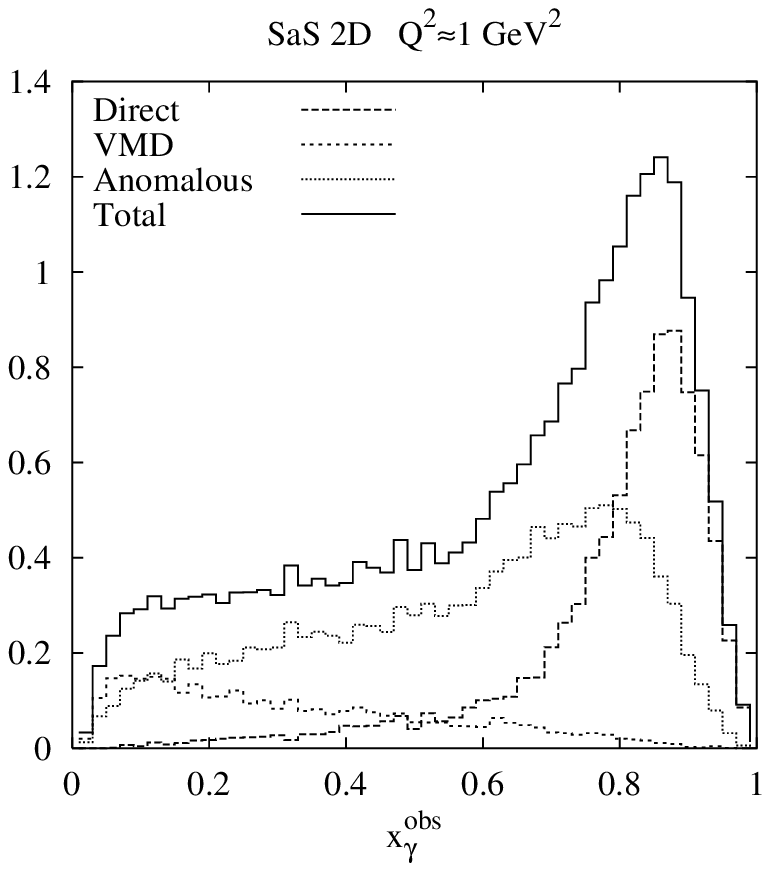,width=78mm}}
      \mbox{\psfig{figure=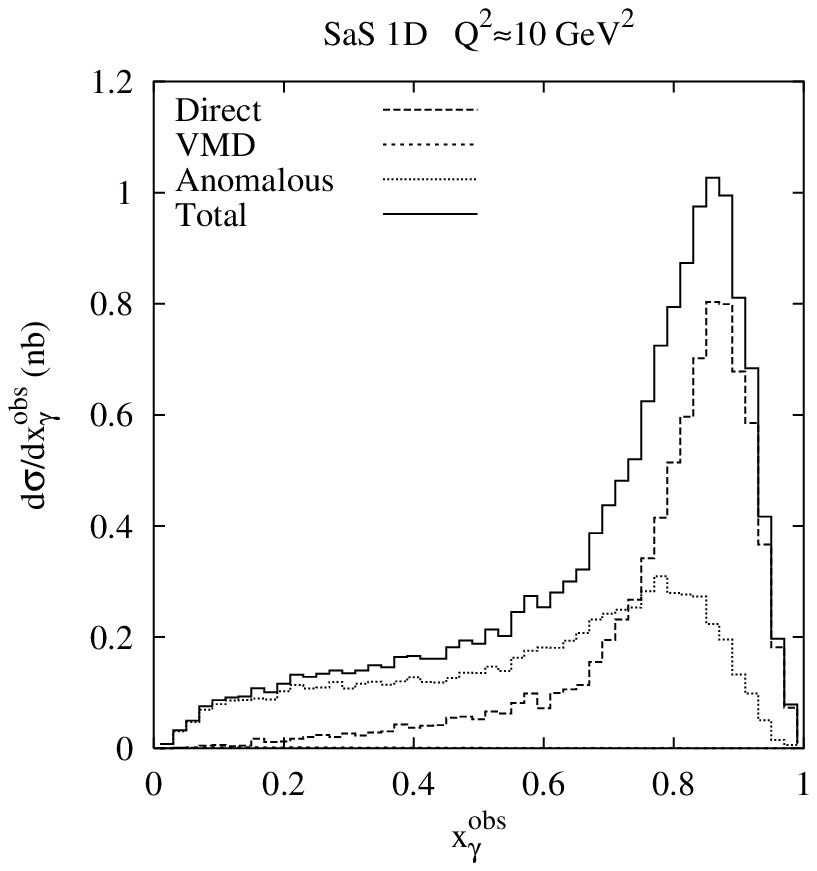,width=78mm}\hspace{-0.5cm}
	    \psfig{figure=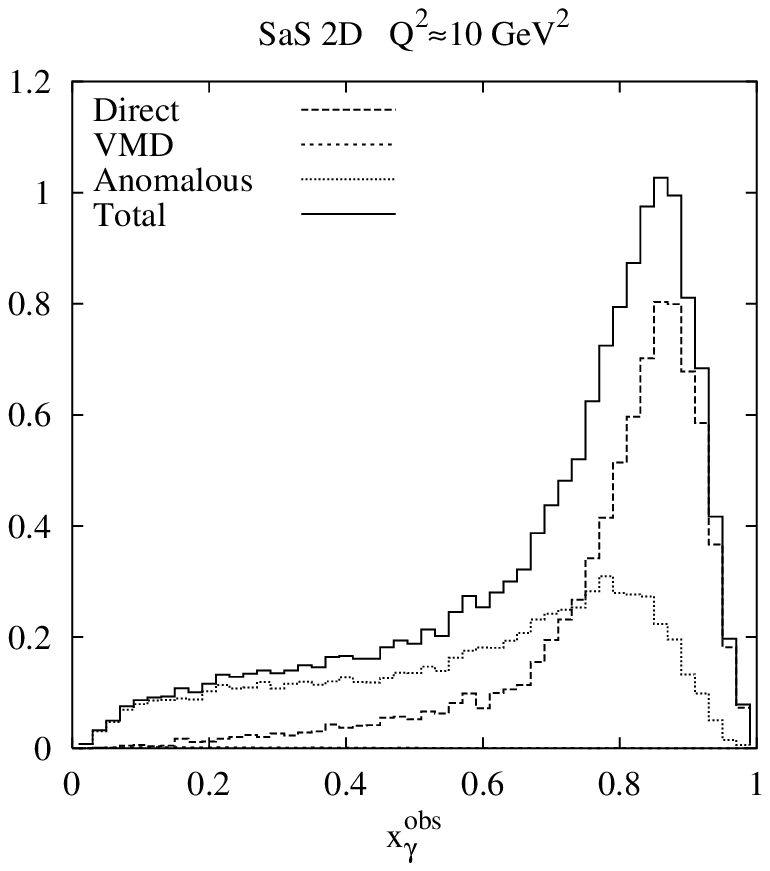,width=78mm}}
   \end{center}
\captive{The $x_{\gamma}^{\mathrm{obs}}$ distribution for the different
photon components. $\pT^{\mathrm{parton}}>3$~GeV, 
$\ET^\mathrm{jet}>6$~GeV, $0.01<y<0.99$ and 
$\sqrt{s_{\e\p}} \simeq 300$~GeV.
\label{fig:xgamma}} 
\end{figure}

In $\e\p$ collisions the variable $x_{\gamma}^{\mathrm{obs}}$ is defined 
as~\cite{xgobs}
\begin{equation}
x_{\gamma}^{\mathrm{obs}}=
\frac{E_{\perp,1}\mathrm{e}^{-\eta_1}+
      E_{\perp,2}\mathrm{e}^{-\eta_2}}{2E_{\gamma}}
\label{xgamma}
\end{equation}
where $E_{\perp,i}$ is the transverse energy and $\eta_i$ is the rapidity 
of the two jets with the highest transverse energy. $E_{\gamma}$ is the 
energy of the photon. $x_{\gamma}^{\mathrm{obs}}$ is then the fraction
of the photon energy (or better, light-cone momentum) that goes into the 
production of the two highest transverse energy jets. In 
Fig.~\ref{fig:xgamma} the $x_{\gamma}^{\mathrm{obs}}$ distribution is shown 
for the different photon components. A cone jet algorithm with cone radius 
$R=1$ is used for jet finding. The partons in the hard scattering are 
restricted to have transverse momenta larger than 3~GeV and the jets to have
a transverse energy larger than 6~GeV. Note that  
$\pTmin^\mathrm{parton}<\pTmin^\mathrm{jet}$ has been used since migration 
from $\pT<\pTmin^\mathrm{jet}$ is likely to happen due to various effects; 
this is discussed further in section~\ref{migration}. The 
$x_{\gamma}^{\mathrm{obs}}$ distribution was generated for two different 
$Q^2$ intervals; 0.9--1.1~GeV$^2$ and 9--11~GeV$^2$. In both cases, an 
$x_{\gamma}^{\mathrm{obs}}$ value of 0.7--0.8 will separate most of the 
direct events from the resolved events. In the low-$Q^2$ case, the SaS~2D 
parton distribution gives less resolved events at large 
$x_{\gamma}^{\mathrm{obs}}$ as compared to SaS~1D. This is because the VMD 
component, which is more important in SaS~2D, is dampened faster with 
increasing $Q^2$ than the anomalous one. For the high-$Q^2$ case, this
difference is not there, since the two anomalous components are approaching 
each other at high $Q^2$, and the VMD pieces are vanishingly small.

\begin{figure} [!htb]
   \begin{center}
   \mbox{\psfig{figure=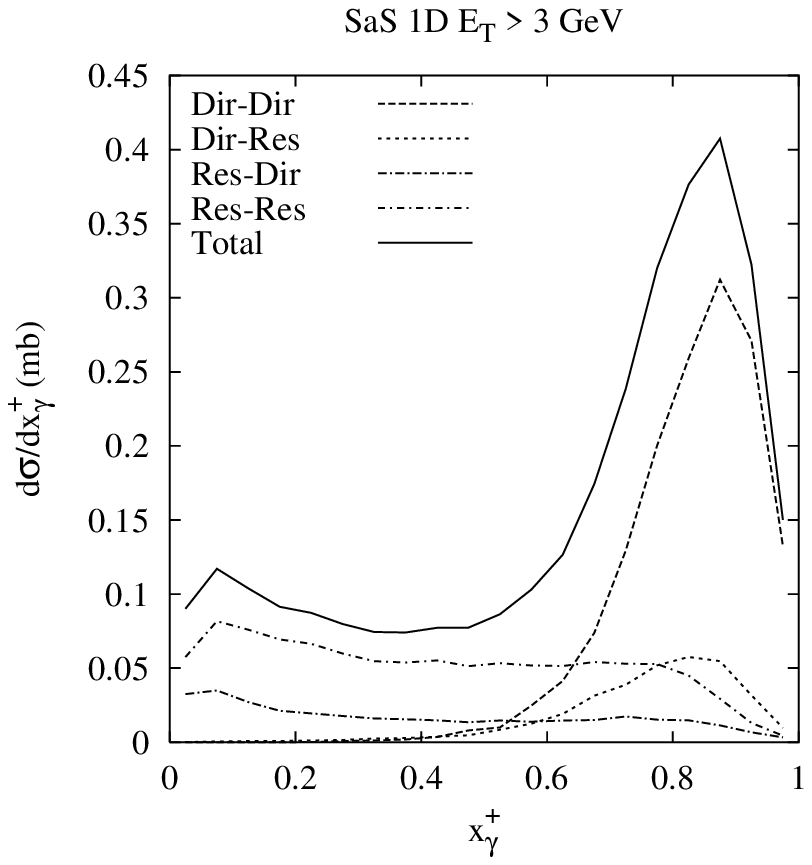,width=78mm}\hspace{-0.5cm}
	\psfig{figure=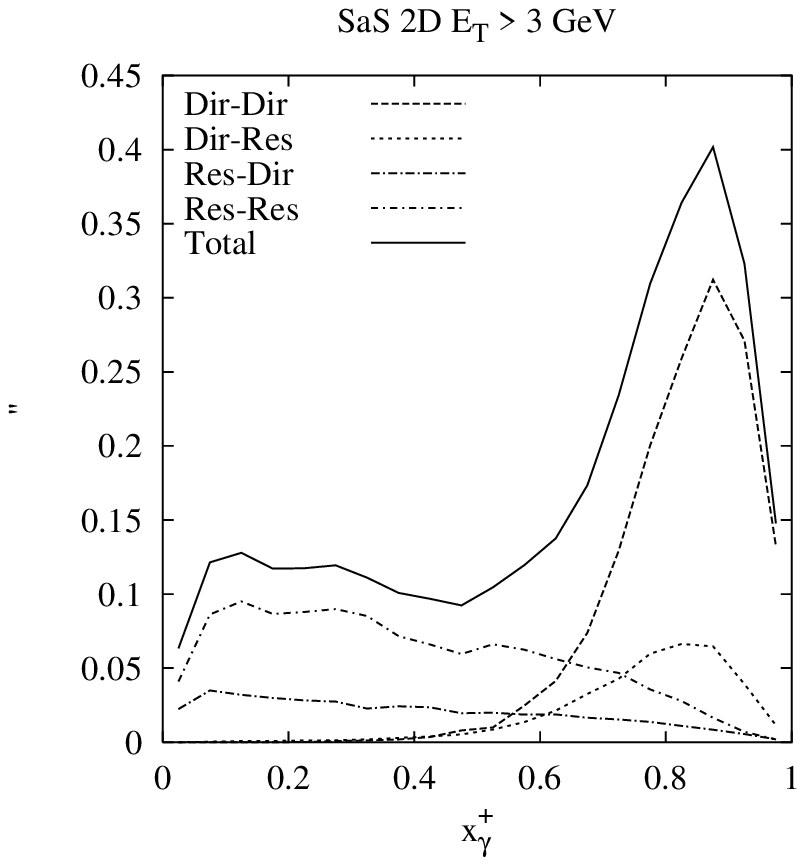,width=78mm}}
   \end{center}
\captive{The $x_{\gamma}^+$ distribution for the direct and resolved
photon components. $\pT^{\mathrm{parton}}=1.5$~GeV, 
$\ET^{\mathrm{jet}}>3$~GeV, $0.01<y_{1,2}<0.99$, $Q_{1,2}^2<0.8$~GeV$^2$ 
and $\sqrt{s_{\e^+\e^-}} = 133$~GeV. 
\label{fig:xgammapm}} 
\end{figure}

In $\e^+\e^-$ collisions the variables $x_{\gamma}^+$ and $x_{\gamma}^-$ are 
defined as~\cite{xgpm} 
\begin{equation}
x_{\gamma}^{\pm}=
\frac{\sum_{\mathrm{jets}}(E \pm p_z)} 
     {\sum_{\mathrm{particles}}(E \pm p_z)}
\label{xgammapm}
\end{equation}
where $p_z$ is the momentum component along the $z$--axis of the $\e^+\e^-$ 
collision and $E$ is the energy of the jets or particles. The sum over 
jets runs over all jets in the event but is limited to the two hardest 
jets for dijet studies. The different direct and resolved components are 
shown in Fig.~\ref{fig:xgammapm} for $Q_i^2 < 0.8~\mathrm{GeV}^2$; each 
combination of components will not be shown here. For these symmetric cuts
the $x_{\gamma}^{\pm}$ distributions are identical, the direct--resolved 
events for one case corresponds to the resolved--direct for the other. 
As expected, direct events are concentrated at large $x_{\gamma}^{\pm}$, 
direct--resolved (resolved--direct) are compatible with direct events in 
$\e\p$, and double--resolved are concentrated at small $x_{\gamma}^{\pm}$ 
--- suggesting a cut at low $x_{\gamma}^+$ and $x_{\gamma}^-$ to separate 
double--resolved from single--resolved events. 

The SaS~2D distribution gives, with $Q_{1,2}^2<0.8$~GeV$^2$, a large 
double--resolved contribution, dominated by the VMD--VMD events. Similarly, 
direct--VMD events dominate the single--resolved contribution. The effective
parameter $\pTmin^{\mathrm{MI}}$ sets the amount of multiple interactions 
for the VMD component. The results shown in this section are sensitive 
to the choice of $\pTmin^{\mathrm{MI}}$, here chosen to 1.4~GeV, which is 
the default in {\sc Pythia}. 

\subsubsection{Scale Choice in Parton Distributions}

To show the dependence on the choice of scale $\mu^2$ in parton distributions
we will study $\e\p$ events, with direct and resolved events treated 
separately. Three extreme choices of scales $\mu_1^2=\pT^2$, 
$\mu_5^2=\pT^2+Q^2$ and $\mu_6^2 \approx 2\pT^2$ are compared. 
For simplicity, we stay with the $\pT$ distribution of the hard scattering 
$2 \rightarrow 2$ subprocess, i.e. avoid hadronization, multiple 
interactions, initial- and final-state bremsstrahlung and jet clustering 
effects. 

\begin{figure} [!htb]
   \begin{center}          
      \mbox{\psfig{figure=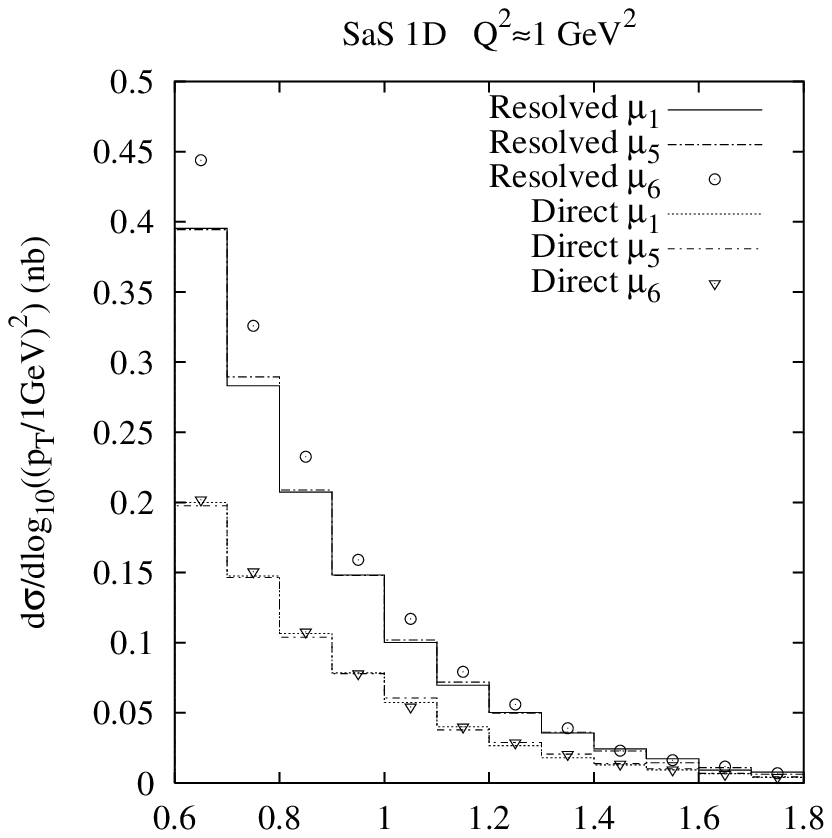,width=78mm}\hspace{-0.5cm}
	    \psfig{figure=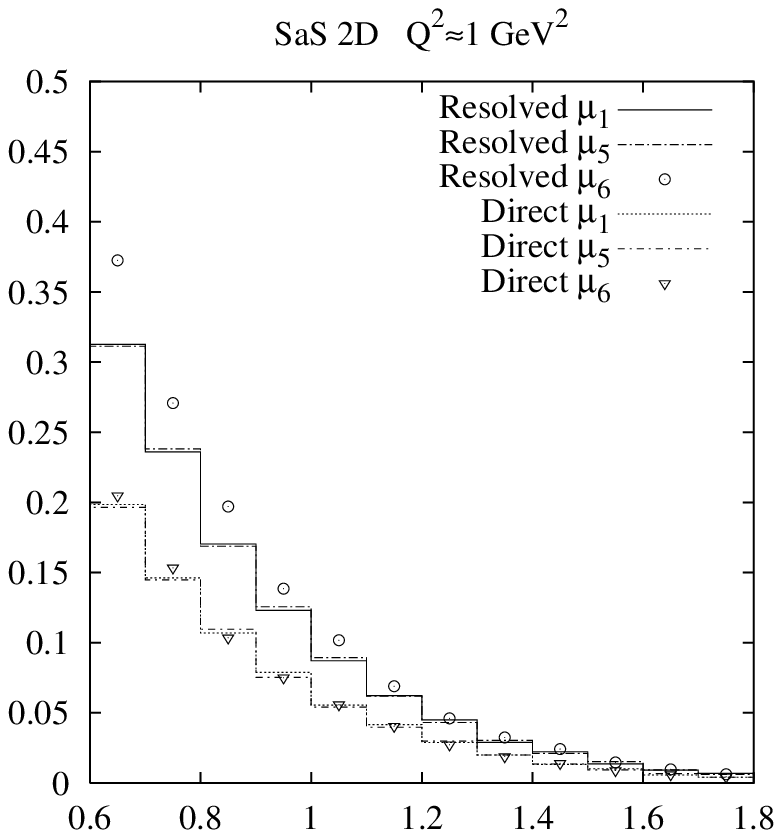,width=78mm}}  
      \mbox{\psfig{figure=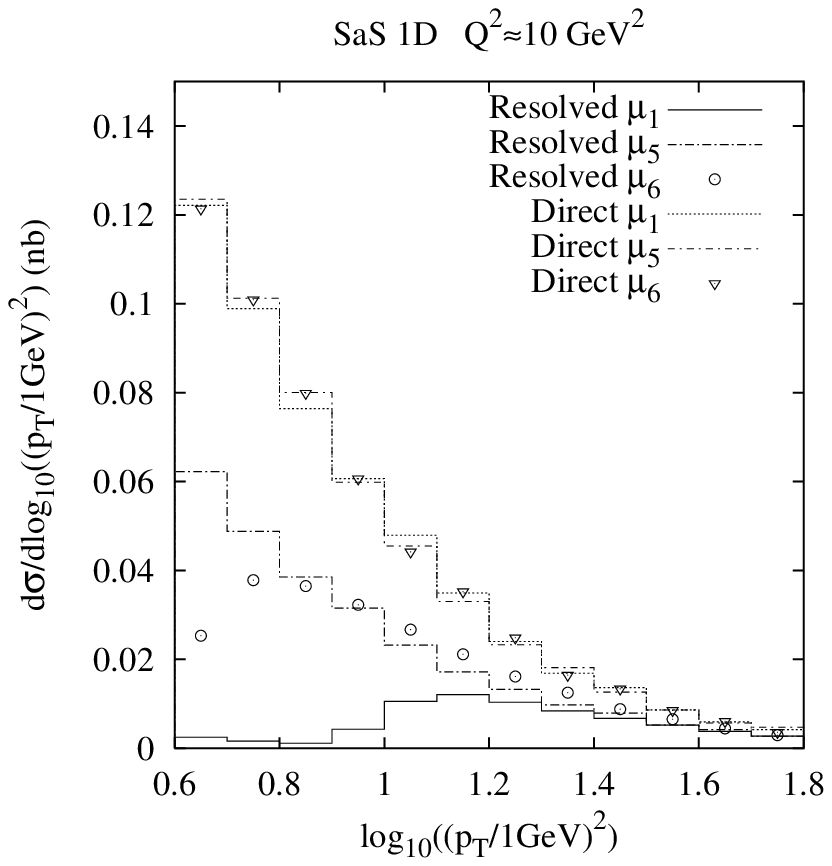,width=78mm}\hspace{-0.5cm}
	    \psfig{figure=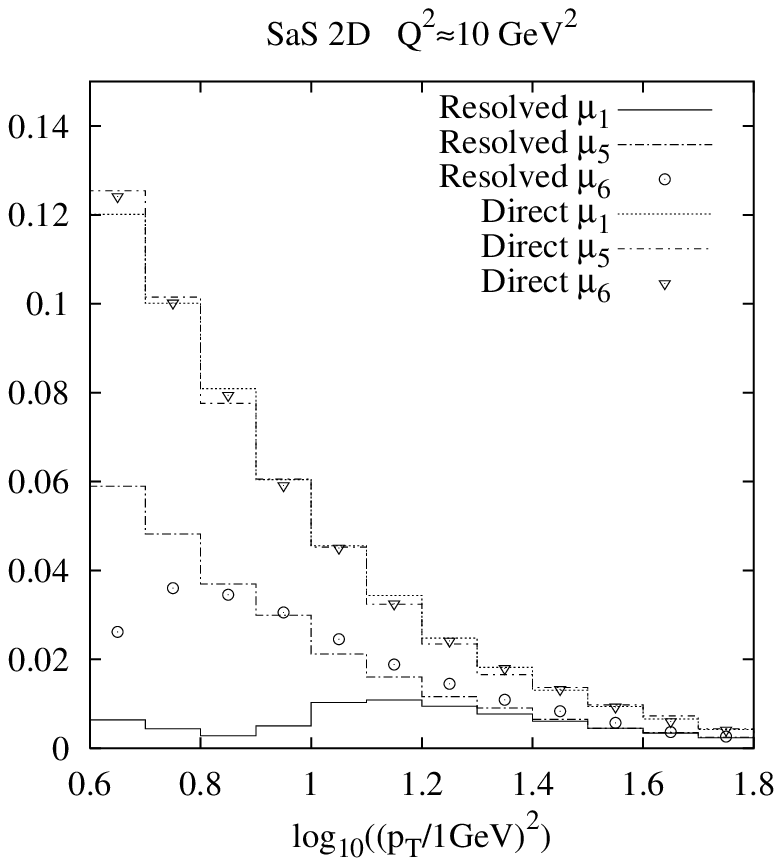,width=78mm}}
   \end{center}
\captive{The $\d\sigma_{\e\p}/\d \log_{10} (\pT/(1\mathrm{GeV}))^2$ cross 
section for different $\mu_i$ scales with $Q^2 \approx 1~\mathrm{GeV}^2$ and 
$Q^2 \approx 10~\mathrm{GeV}^2$, $\sqrt{s_{\e\p}} \simeq 300$~GeV.
\label{fig:mu2}} 
\end{figure}

In Fig.~\ref{fig:mu2}, the 
$\d\sigma_{\e\p}/\d \log_{10} (\pT/(1\mathrm{GeV}))^2$
distribution for the different scale choices are compared at two different 
photon virtualities, $Q^2 \approx 1~\mathrm{GeV}^2$ and 
$Q^2 \approx 10~\mathrm{GeV}^2$. For the low-$Q^2$ case, their differences are 
most pronounced for resolved events and the results are in decreasing order in 
accordance with the scale choice; $\mu_6$, $\mu_5$ and $\mu_1$. Here the 
anomalous component is the dominant one, explaining the difference between 
the SaS~1D and the SaS~2D distributions. For direct events, only a rather 
mild scale-breaking of the proton distributions enters. Since small-$x$ values 
dominate, where distributions increase with $\mu^2$, the $\mu_6$ scale should 
give the largest result but the difference from the other two are within 
errors. 

For $Q^2 \approx 10~\mathrm{GeV}^2$, direct events dominate. The resolved 
results can be divided into two regions;
$\pT^2 < Q^2$ and $\pT^2 > Q^2$. Remember that the $\mu_5$ scale imply 
that the resolved component will not vanish even when $\pT^2 < Q^2$ as can 
be seen in Fig.~\ref{fig:mu2}. With the $\mu_6$ scale, the evolution range 
decreases with $\pT^2$ and finally suppresses resolved photons. The $\mu_1$ 
scale does not allow resolved photons when $\pT^2 < Q^2$. (The tiny tail of
such events comes from charm production, where actually 
$\mu_1^2=\pT^2+m_c^2$ is used rather than only $\pT^2$.) The results with 
the $\mu_5$ and $\mu_6$ scales are, of course, in agreement when 
$\pT^2 \approx Q^2$ and in the tail of the $\pT$ distribution the $\mu_6$ 
scale gives the bigger cross section. At large $Q^2$ values, the difference 
between the two photon parton distributions are reduced. For direct events 
the $\mu_5$ scale is larger than the $\mu_6$ one at low $\pT^2$ but smaller 
at high $\pT^2$; a corresponding pattern is expected to be seen in the cross 
section (again, small-$x$ values dominate). However, differences are small.

\subsection{Comparisons with Data}
\label{migration}

In this section the model is compared with data. We will not make a detailed 
analysis of experimental results but use it to point out model dependences
and to constrain some model parameters. Where applicable, the 
HzTool~\cite{hz} routines will be used for the comparison with data.

$2 \rightarrow 2$ parton interactions normally give rise to 2--jet events.
In leading--order QCD, the jets are balanced in transverse momenta in the 
centre of mass frame of the $\gast\p / \gast\gast$ subsystem. Various effects,
such as primordial $k_{\perp}$, initial- and final-state bremsstrahlung, etc.,
tend to spoil this picture. This increases the $\d\sigma/\d \pT$ 
spectrum at any fixed $\pT$, since jets can be boosted up from lower 
$\pT$. It is here important to remember that $\d\sigma/\d \pT$ is 
dropping steeply with increasing $\pT$, since the matrix elements 
described in section~\ref{ME} are divergent for $\pT \rightarrow 0$. 
Even a symmetric smearing of jet $\pT$, i.e. with shifts 
$+\delta \pT$ and $-\delta \pT$ equally probable, would thus 
increase $\d\sigma/\d \pT$: the jets shifted upwards would migrate into 
a region less populated than those shifted downwards, and therefore 
proportionately have a larger impact. Additionally, the smearing is not 
symmetric, but normally shifts one jet upwards while the other may be shifted 
in either direction. (This holds for primordial $k_{\perp}$ and initial-state 
radiation, while final-state radiation tends to shift both jets downwards.)
In the generation procedure, the divergences are avoided by the introduction 
of a lower cut, $\pTmin^{\mathrm{parton}}$, below which parton
scatterings are not considered. In order to study jets above some 
$\pTmin^\mathrm{jet}$, a 
$\pTmin^\mathrm{parton}<\pTmin^\mathrm{jet}$ 
is required. Stable results are obtained when a hard scattering with 
$\pT^\mathrm{parton}=\pTmin^\mathrm{parton}$ 
for the incoming partons do not produce any jets with $\pT$ above 
$\pTmin^\mathrm{jet}$. Typically 
$\pTmin^\mathrm{parton}=\frac{1}{2} 
 \pTmin^\mathrm{jet}$ or less is required.

\subsubsection{Dijets at HERA}

Dijet angular distributions in photoproduction and deep inelastic scattering
have been studied at HERA \cite{dijetZEUS}. In the center of mass system of 
the two highest transverse energy jets (which are restricted to have a 
transverse energy above a certain value) the angle $\theta^*$ between the 
jets and the beam axis is expected to be distributed differently depending on 
whether or not the photon is resolved. The leading--order direct QCD graphs, 
boson--gluon fusion and QCD Compton, have spin--1/2 quark propagators leading 
to an angular dependence proportional to 
$1/\that \propto (1-|\cos{\theta^*|})^{-1}$. (The jets are generally not 
distinguishable so the absolute value of $\cos{\theta^*}$ cannot be 
measured.) On the contrary, the resolved processes are dominated by 
{\it Rutherford type} scatterings involving a t--channel gluon: 
$\q\q' \to \q\q'$, $\q\g \to \q\g$ and $\g\g \to \g\g$. These thus give an 
angular dependence proportional to 
$1/\that^2 \propto (1-|\cos{\theta^*}|)^{-2}$ i.e. the dijet angular 
distribution is expected to rise steeper with $|\cos{\theta^*}|$ for resolved 
processes than for direct ones. 

\begin{figure} [!htb]
   \begin{center}          
      \mbox{\psfig{figure=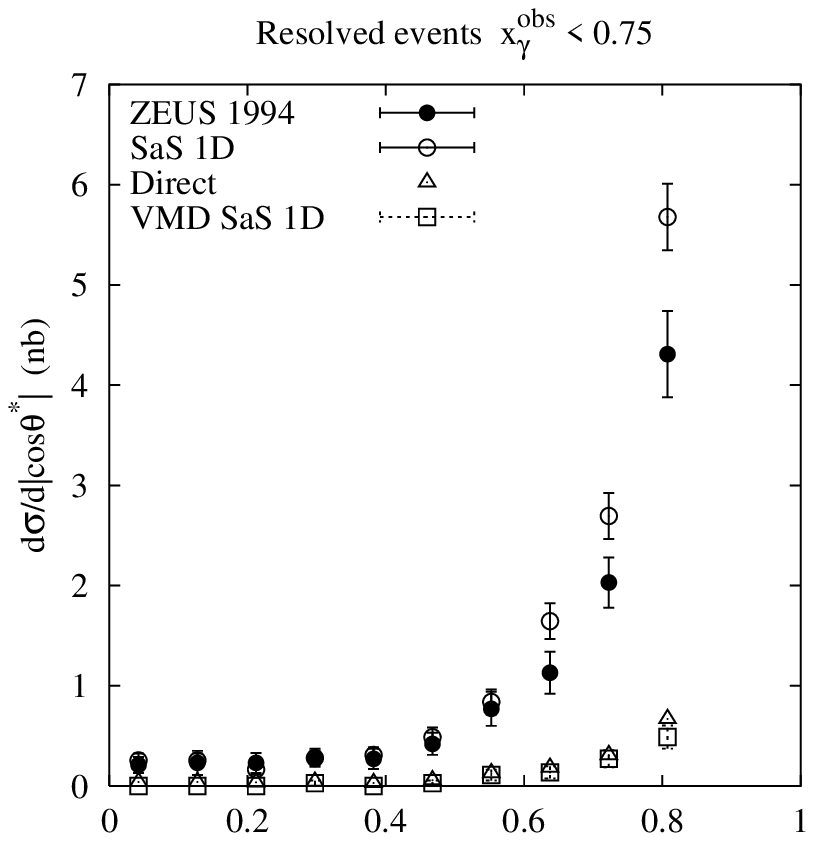,width=78mm}\hspace{-0.5cm}
            \psfig{figure=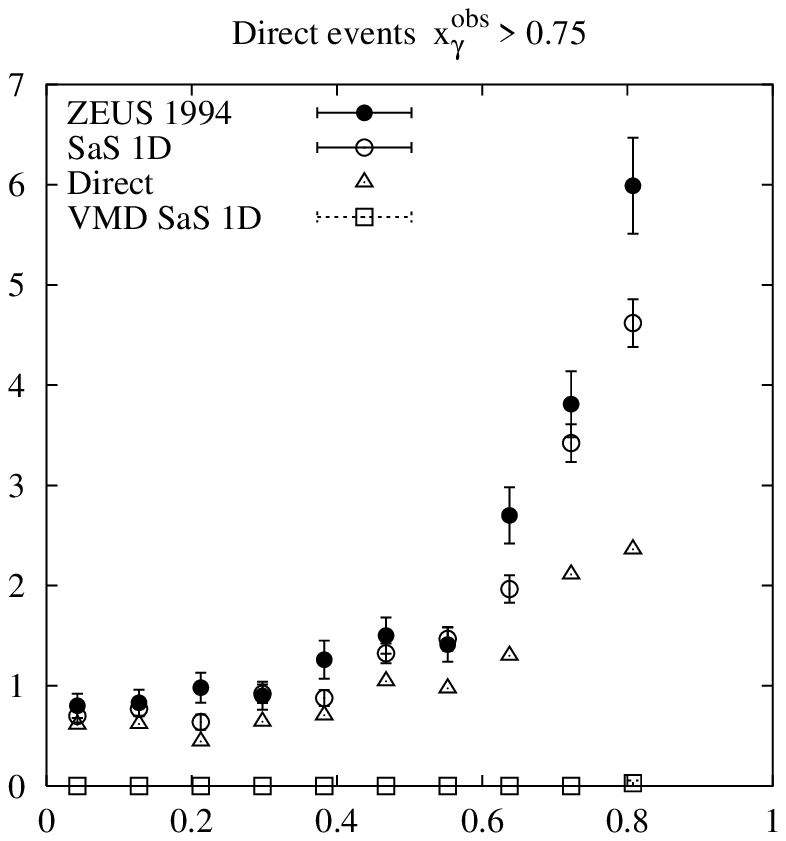,width=78mm}} 
      \mbox{ }
      \mbox{\psfig{figure=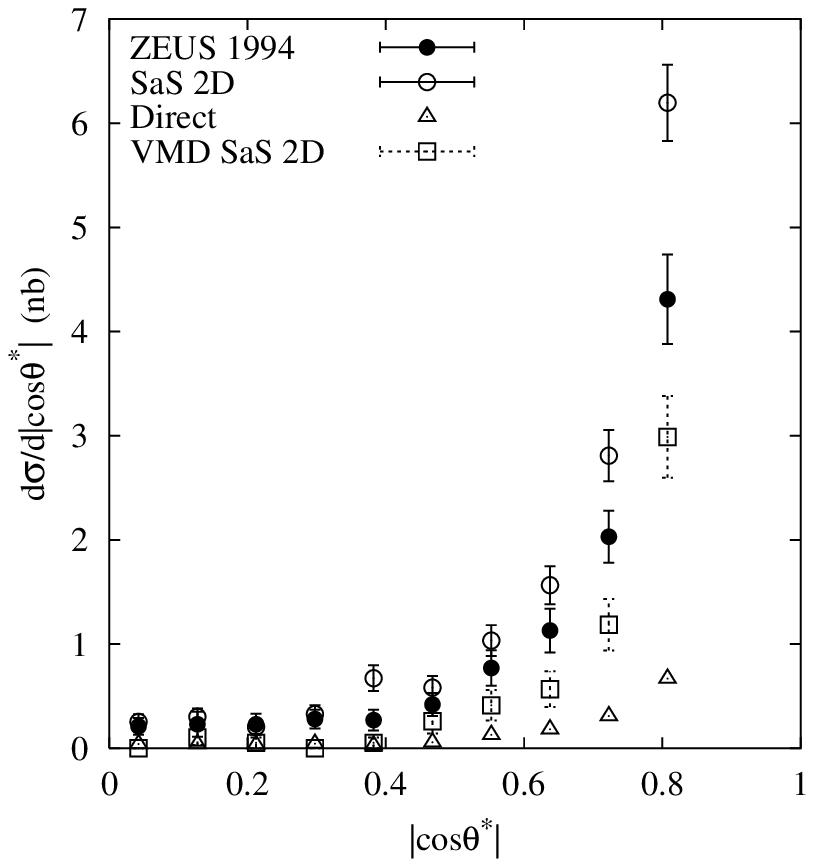,width=78mm}\hspace{-0.5cm}
            \psfig{figure=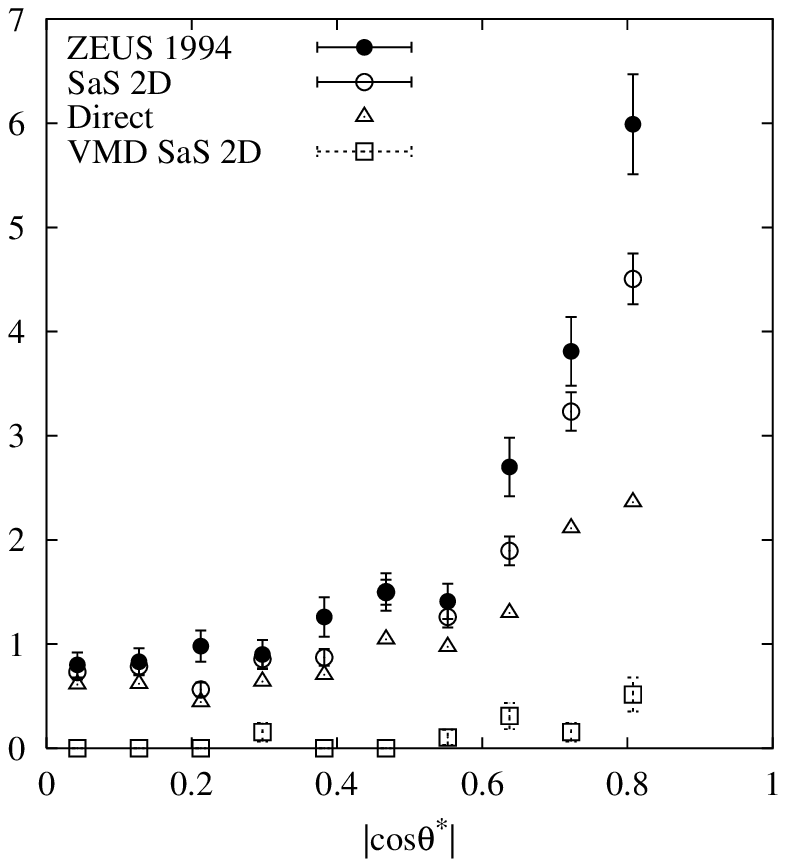,width=78mm}}
   \end{center}
\captive{The dijet angular cross section $\d\sigma/\d |\cos{\theta^*}|$ for  
resolved ($x_{\gamma}^{\mathrm{obs}}<0.75$) and 
direct ($x_{\gamma}^{\mathrm{obs}}>0.75$) processes. 
$\theta^*$ is the jet scattering
angle w.r.t. the beam axis measured in the dijet centre of mass system. 
$Q^2<4$~GeV$^2$ and $0.25<y<0.8$. The SaS~1D and SaS~2D curves give the total 
contribution from the resolved and direct components; the VMD and direct 
components are also shown separately. The anomalous contribution can be 
obtained by subtracting the direct and the VMD contribution from the full one.
\label{dirres}}
\end{figure}

As discussed in the introduction and in section~\ref{xgsec}, with a cut in 
$x_{\gamma}^{\mathrm{obs}}$ it is possible to make a separation between 
events that are likely to be direct from those that are likely to be resolved 
events. An example of this is shown in Fig.~\ref{dirres} where direct and 
resolved events generated by {\sc Pythia} is compared with results presented 
by the ZEUS Collaboration~\cite{dijetZEUS}, with a good agreement in the 
shape of the distributions. The plots were produced by using the HzTool 
package. 

In the analysis of the ZEUS Collaboration a cut 
$x_{\gamma}^{\mathrm{obs}}<0.75$ was used to classify the photon as resolved
while for $x_{\gamma}^{\mathrm{obs}}>0.75$ it was classified as a direct 
photon. Of the events classified as having a direct photon a significant 
contribution comes from events with an anomalous fluctuation of the photon. 
This is well understood since, in this component, the quark (antiquark) often 
carries a large fraction of the original photon momentum. The two jets are 
then likely to be produced at small scattering angles $\theta^*$ as described 
by the matrix elements with gluon exchange. Using the SaS~1D parton 
distribution, the anomalous component is important for either classification, 
whereas the VMD part is highly suppressed and negligible by the 
$x_{\gamma}^{\mathrm{obs}}>0.75$ cut. For SaS~2D, the VMD contribution is 
small for events classified as direct but of equal importance as the anomalous 
part for $x_{\gamma}^{\mathrm{obs}}<0.75$. The choice of scale $\mu_i$ is not 
crucial at this low $Q^2$ values, the $\mu_3$ scale is used and the results 
with $\mu_5$ are within errors. 

The possibility for the incoming electron to emit extra photons, that may go
undetected, is not taken into account by the model. This effect decreases the 
lepton energy; hence the photon energy calculated here would be overestimated. 
A simple estimation of these effects give corrections to the photon energy at 
the percent level. The consequence would be that the given 
$x_{\gamma}^\mathrm{obs}$ distribution will be shifted towards higher 
values. An inclusion of such effects would shuffle some of our resolved 
events into the direct category, thus improving the description of the 
relative amounts. This could be solved by using lepton-inside-lepton 
structure functions, though the implementation of it is postponed for the 
future.

From these observations it is tempting to conclude that a cut in 
$x_{\gamma}^{\mathrm{obs}}$ does not so much separate events with the resolved 
anomalous component (perturbatively calculable fluctuation) of the photon from 
the direct component, but rather separates it from the VMD component 
(non-perturbative fluctuation) of the photon. This statement is true to 
leading order but the anomalous fluctuation contain pieces that a 
next-to-leading-order calculation of the direct component would give. 
Hence, in this sense also anomalous events  can be considered as a 
contribution from the direct component making the $x_{\gamma}^{\mathrm{obs}}$ 
well suited to separate the direct from the resolved component. Nevertheless, 
the classification of the photon into different components used throughout in 
this paper is based on the leading order description, viewing the anomalous 
fluctuation as a resolved photon.

\subsubsection{Inclusive $\e\p$ Jet Cross Sections}

\begin{figure} [!htb]
   \begin{center}       
     \mbox{\psfig{figure=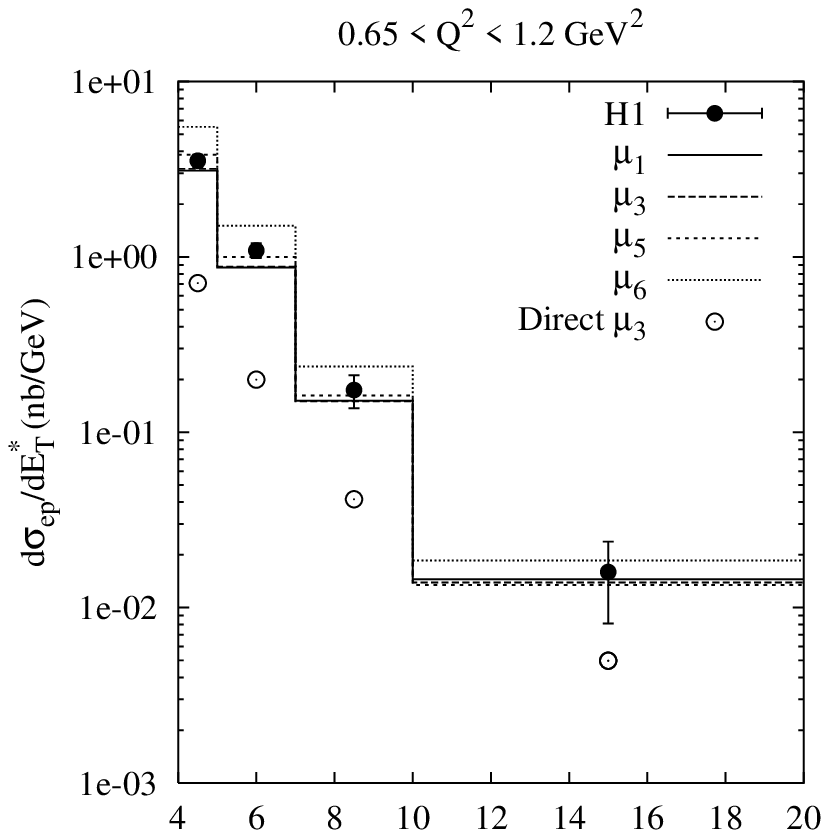,width=78mm}\hspace{-0.5cm}
	   \psfig{figure=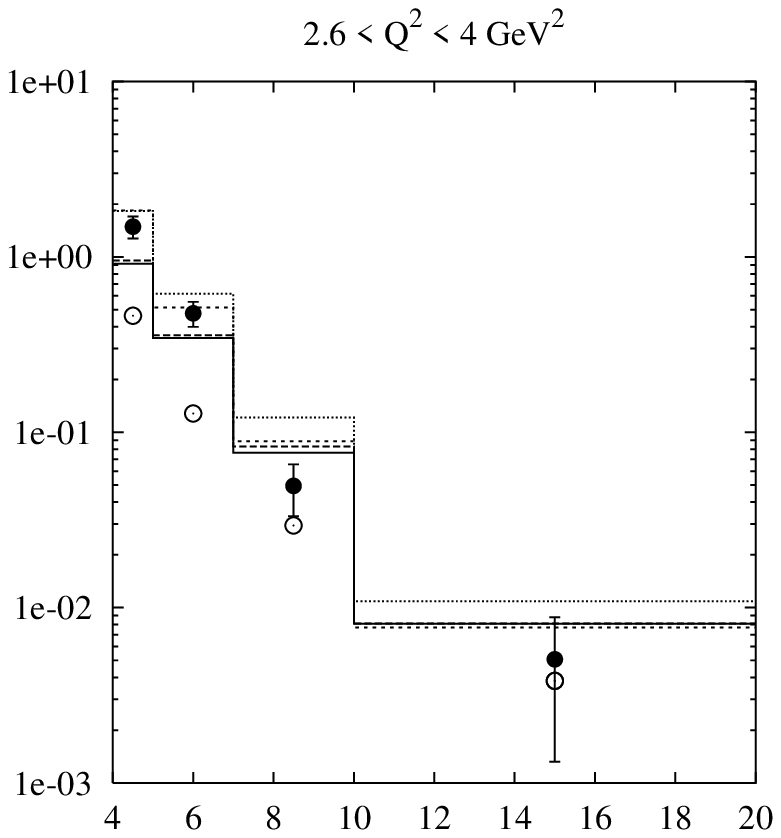,width=78mm}}
     \mbox{ }
     \mbox{\psfig{figure=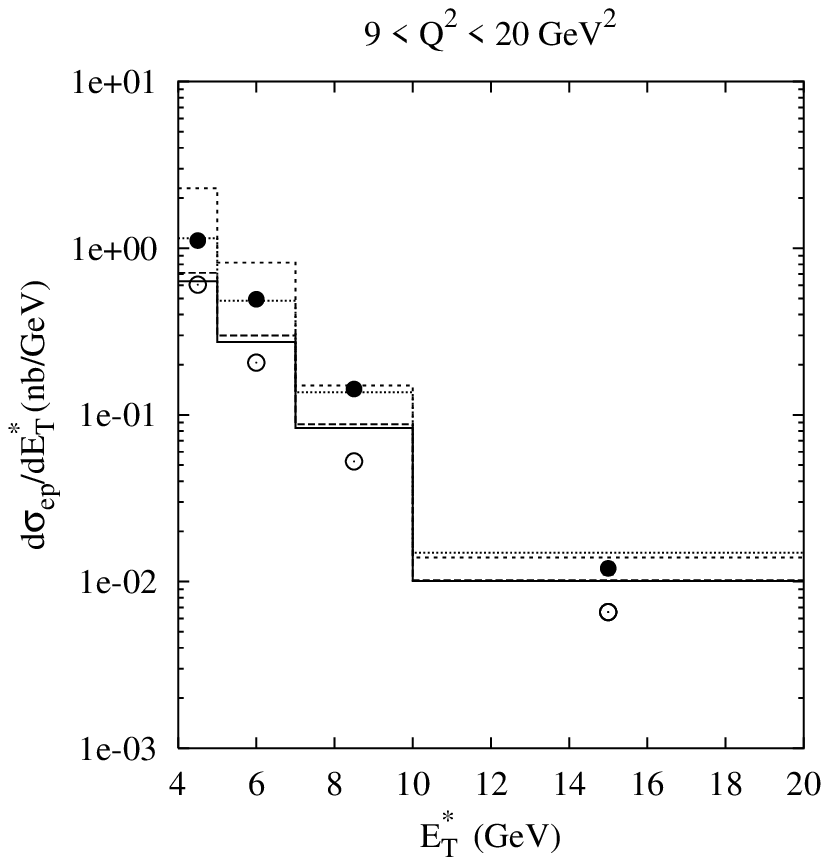,width=78mm}\hspace{-0.5cm}
	   \psfig{figure=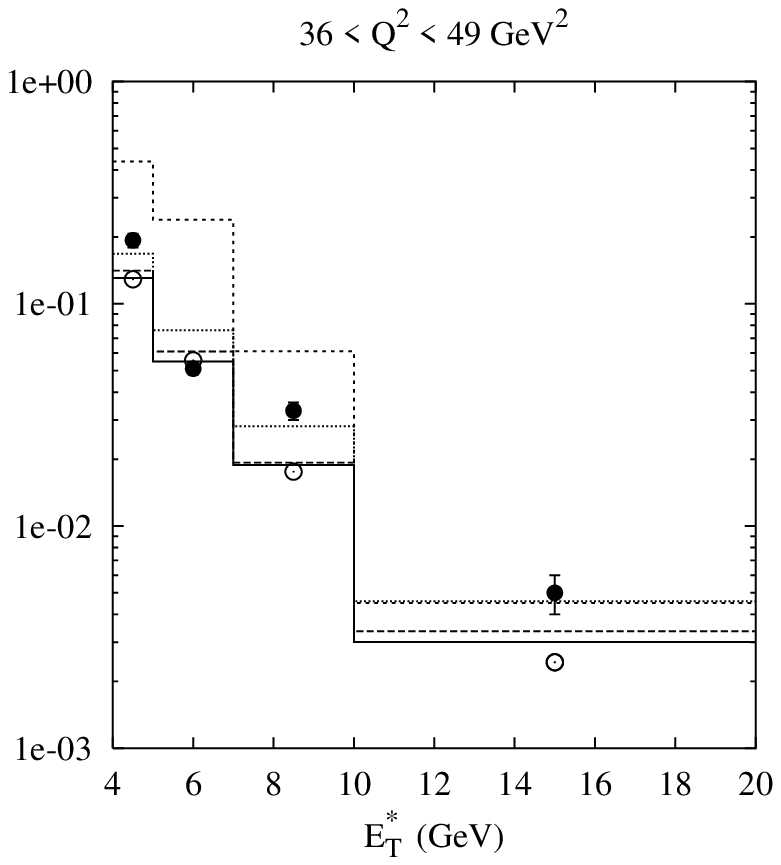,width=78mm}}
   \end{center}
\captive{The differential jet cross section $\d\sigma_{\e\p}/\d E^*_{\perp}$ 
for jets with $-2.5<\eta^*<-0.5$ and $0.3<y<0.6$.
\label{fig:ET}}
\end{figure}

\begin{figure} [!htb]
   \begin{center}       
     \mbox{\psfig{figure=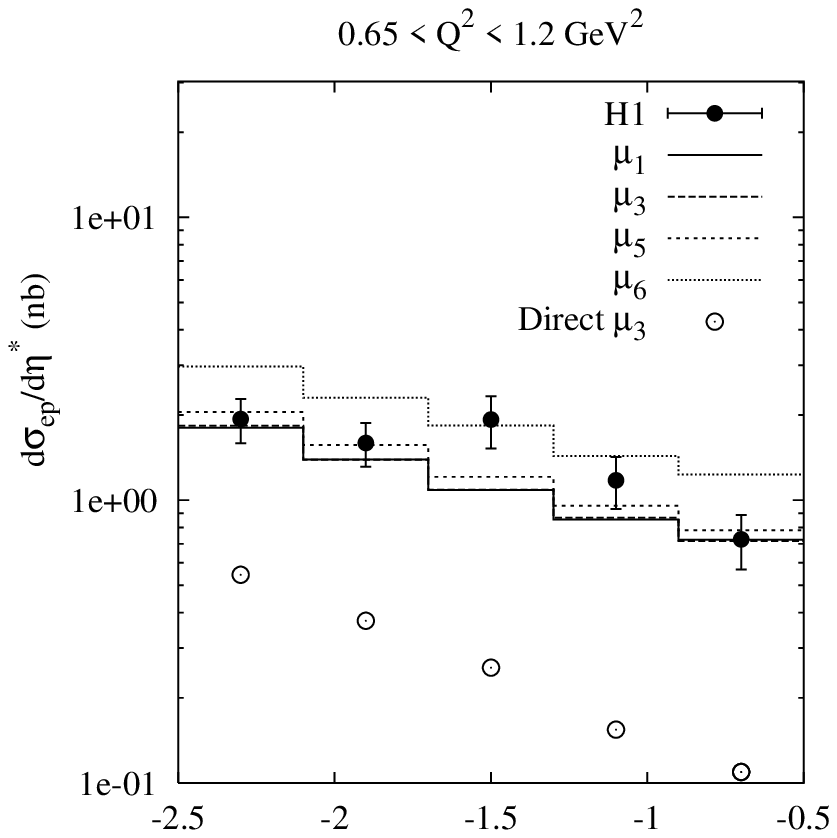,width=78mm}\hspace{-0.5cm}
	   \psfig{figure=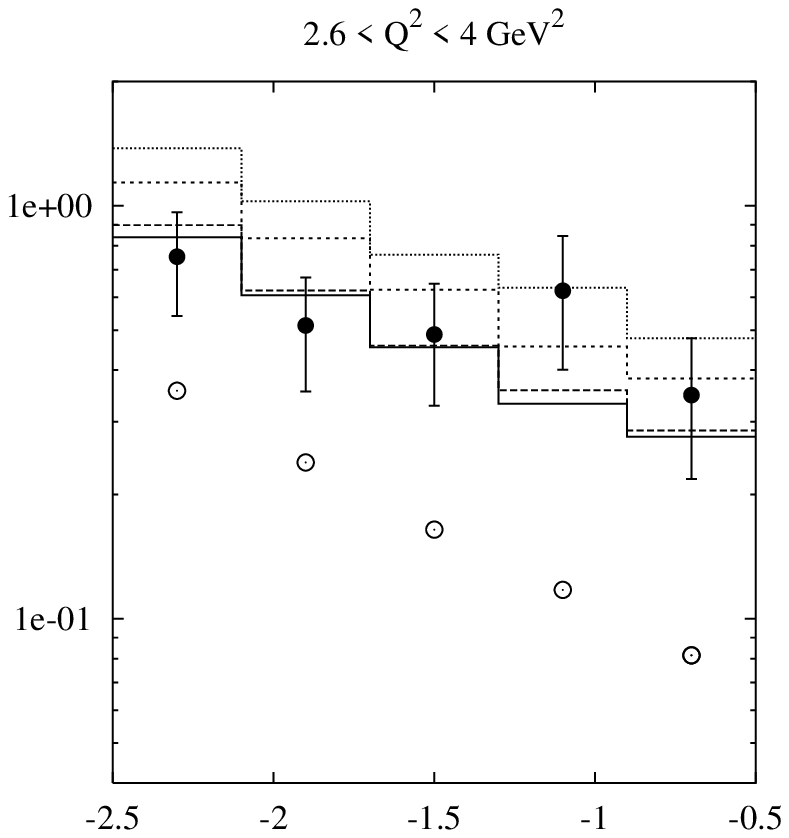,width=78mm}}
     \mbox{}
     \mbox{\psfig{figure=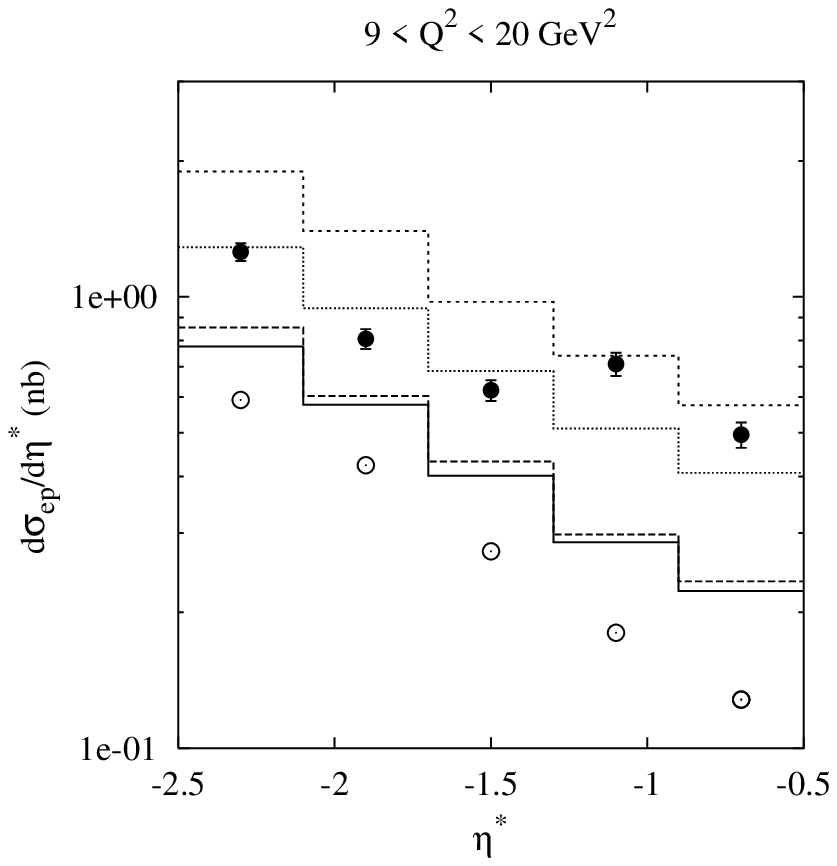,width=78mm}\hspace{-0.5cm}
	   \psfig{figure=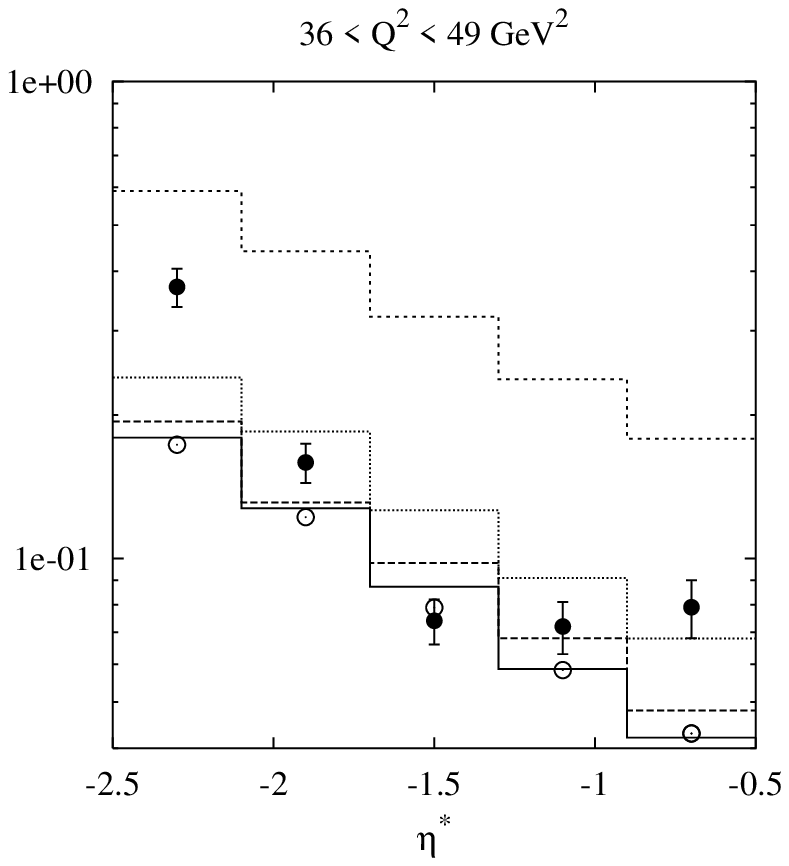,width=78mm}}
   \end{center}
\captive{The differential jet cross section $\d\sigma_{\e\p}/\d \eta^*$ for 
jets with $E^*_{\perp}>5~\mathrm{GeV}$ and $0.3<y<0.6$.
\label{fig:eta}}
\end{figure}

\begin{figure} [!htb]
   \begin{center}    
     \mbox{\psfig{figure=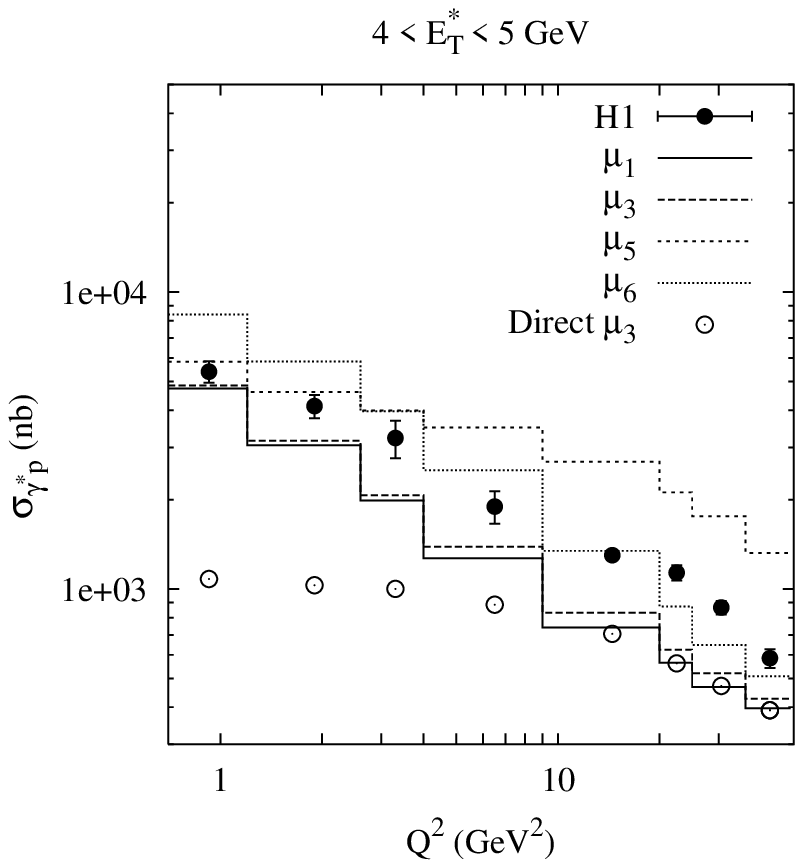,width=78mm}\hspace{-0.5cm}
	   \psfig{figure=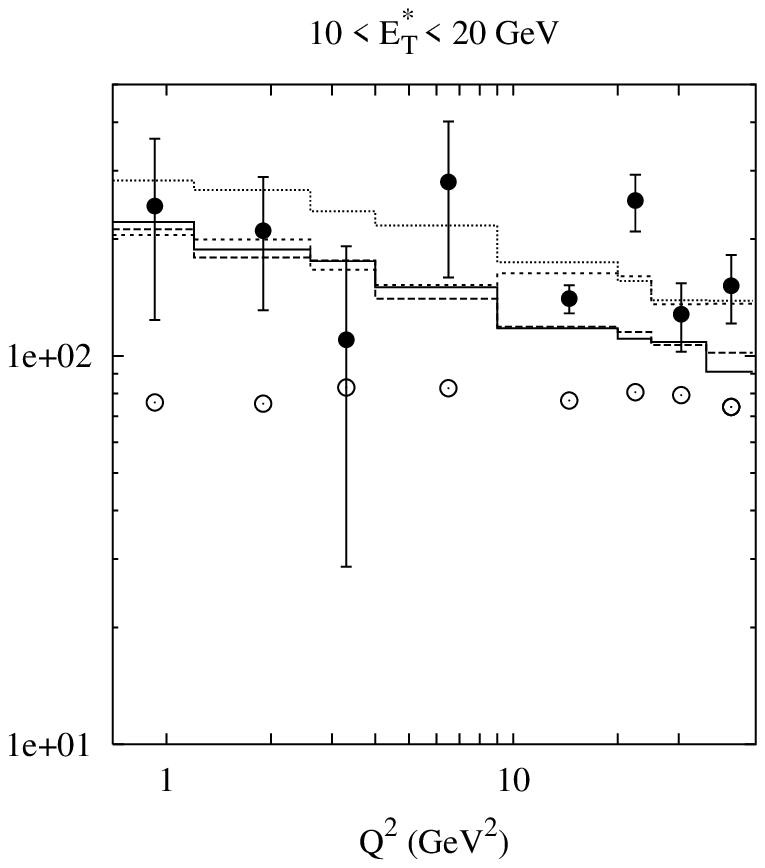,width=78mm}}   
   \end{center}
\captive{The inclusive $\gast\p$ jet cross section $\sigma_{\gast\p}$ for 
jets with $-2.5<\eta^*<-0.5$ and $0.3<y<0.6$.
\label{fig:inc_gp}}
\end{figure}

Inclusive $\e\p$ jet cross sections have been measured by the H1 collaboration 
\cite{lowQ2H1} in the kinematical range $0<Q^2<49~\mathrm{GeV}^2$ and 
$0.3<y<0.6$. The differential jet cross sections 
$\d\sigma_{\e\p}/\d E^*_{\perp}$, $\d\sigma_{\e\p}/\d \eta^*$ and the 
inclusive $\gast\p$ jet cross section in 
Fig.~\ref{fig:ET}, \ref{fig:eta} and~\ref{fig:inc_gp} respectively, 
were produced with the HzTool  package. The $E^*_{\perp}$ and 
$\eta^*$ are calculated in the $\gast\p$ centre of mass frame where the 
incident proton direction corresponds to positive $\eta^*$. 

For $\d\sigma_{\e\p}/\d E^*_{\perp}$ and $\d\sigma_{\e\p}/\d \eta^*$ 
data is available in nine different $Q^2$ bins, four of them are shown here
with similar results for the intermediate bins. The SaS~1D parton 
distribution together with a few different $\mu_i$ scales; $\mu_1$, $\mu_3$, 
$\mu_5$ and $\mu_6$, are used to model the resolved photon component. 
The other choices of scales, $\mu_2$ and $\mu_4$, interpolates between 
these results.

In the highest $Q^2$ bin the direct component is the dominant contribution; 
the virtuality of the photon is for most events of the order of or larger 
than the transverse momenta squared, $Q^2\gtrsim \pT^2$. However, the
resolved component is not negligible and all the scales $\mu_i$, except 
$\mu_1$, depend on the photon virtuality. This gives a larger resolved 
component in this region as compared to the the conventional choice, 
$\mu_1=\pT$. In the low-$Q^2$ bin the $\mu_i$ scales do not differ 
much from $\pT$, i.e. the results are not sensitive to the scale 
choice $\mu_i$. The exception is $\mu_6$, which there overshoots the data. 
The $\mu_4^2=\pT^2+Q^2/2$ scale (not shown) gives nice agreement with 
data for all different $Q^2$ bins. 

Since the VMD part dies out quickly with increasing photon virtuality,
multiple interactions will only be visible at low $Q^2$ (multiple 
interactions for the anomalous component is not in the model so far). 
The anomalous component dominates over the VMD component already at 
1~GeV$^2$. Therefore, multiple interactions for the VMD component can safely 
be neglected for the distributions shown in this section. 

\begin{figure} [!htb]
   \begin{center}       
     \mbox{\psfig{figure=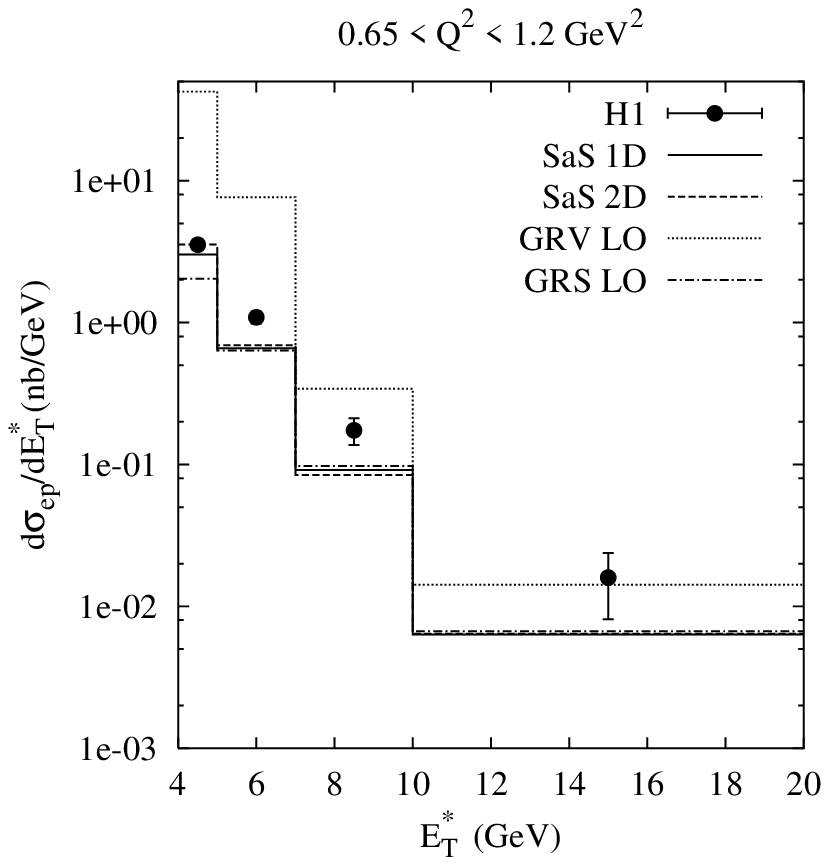,width=78mm}\hspace{-0.5cm}
	   \psfig{figure=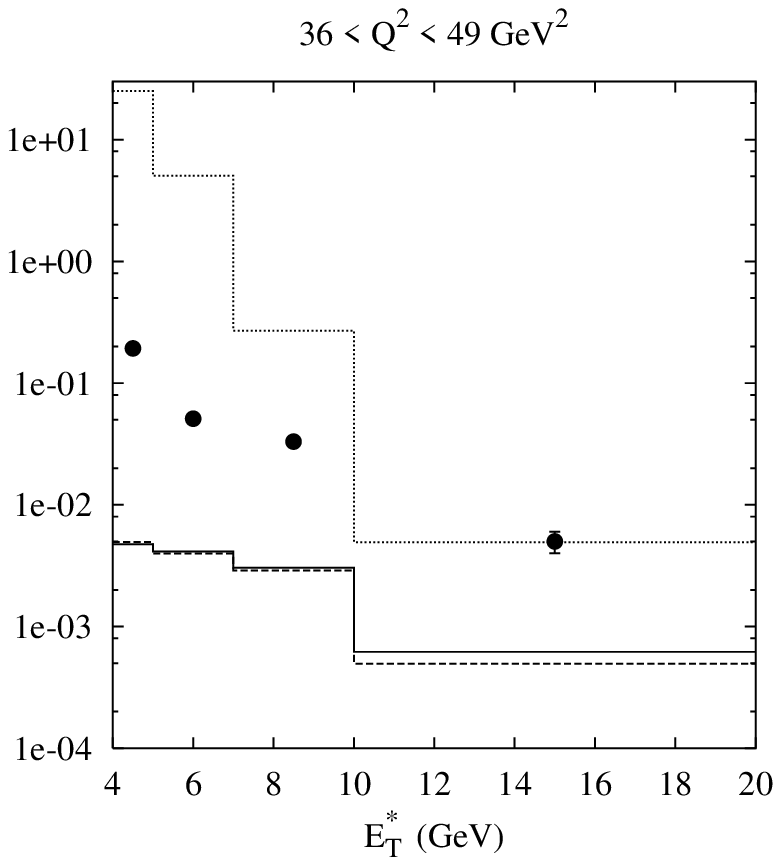,width=78mm}}
     \mbox{ }
     \mbox{\psfig{figure=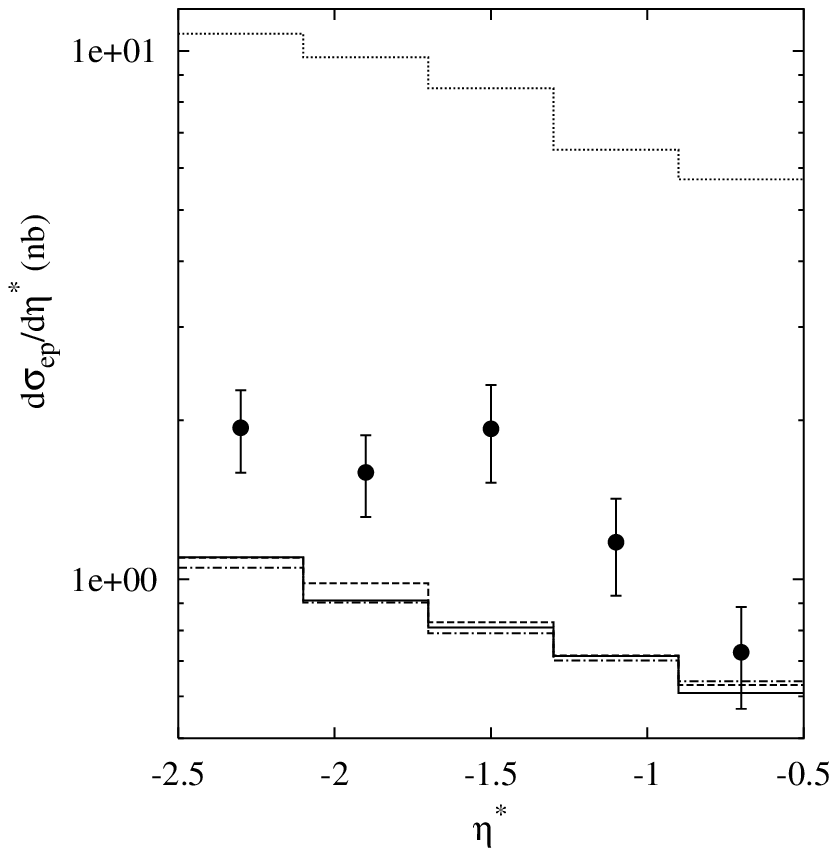,width=78mm}\hspace{-0.5cm}
	   \psfig{figure=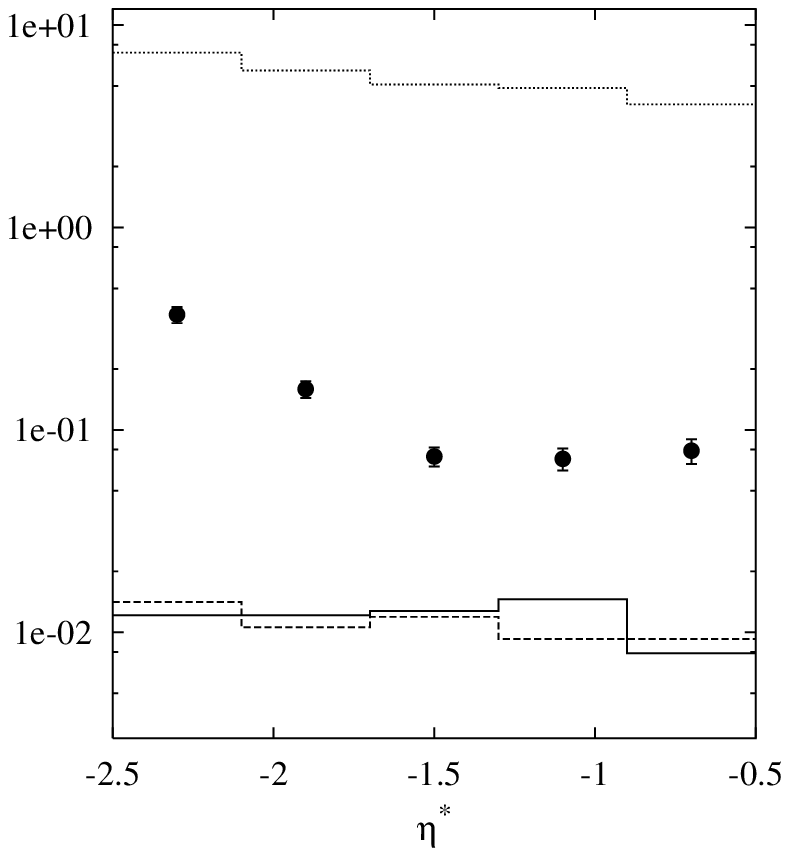,width=78mm}}
   \end{center}
\captive{The differential jet cross sections $\d\sigma_{\e\p}/\d E^*_{\perp}$
and  $\d\sigma_{\e\p}/\d \eta^*$ for jets with $-2.5<\eta^*<-0.5$ and 
$0.3<y<0.6$. Only resolved events are used to show the sensitivity to the 
photon parton distribution.
\label{fig:ET_eta-res}}
\end{figure}

Changing the photon parton distribution from SaS~1D to SaS~2D will give a 
slightly lower result for the low-$Q^2$ bins, Fig.~\ref{fig:ET_eta-res}. 
The GRS~LO~\cite{grspdf} parton distribution gives a similar result as the 
other two. It has a restricted regime of validity, $Q^2 \ll \mu^2$ 
(implemented as $5Q^2 < \mu^2$) and $Q^2 < 10~\mathrm{GeV}^2$. Therefore, 
it is absent in the high-$Q^2$ bin, and some regions of phase space are cut 
out in the low-$\ET^*$ bins and in the $\eta^*$ distributions  
($5Q^2 \gtrsim (\ET^*)^2$). One could imagine larger differences for 
virtual-photon parton distributions that from the onset are more different, 
so the good SaS/GRS agreement is somewhat fortuitous. Using a parton 
distribution for a real photon cannot describe the $Q^2$ dampening in the 
distributions shown in this section. As an example of this the 
GRV~LO~\cite{grvpdf} distribution has been used; the change here is solely 
from the photon flux. 

\begin{figure} [!htb]
   \begin{center}       
     \mbox{\psfig{figure=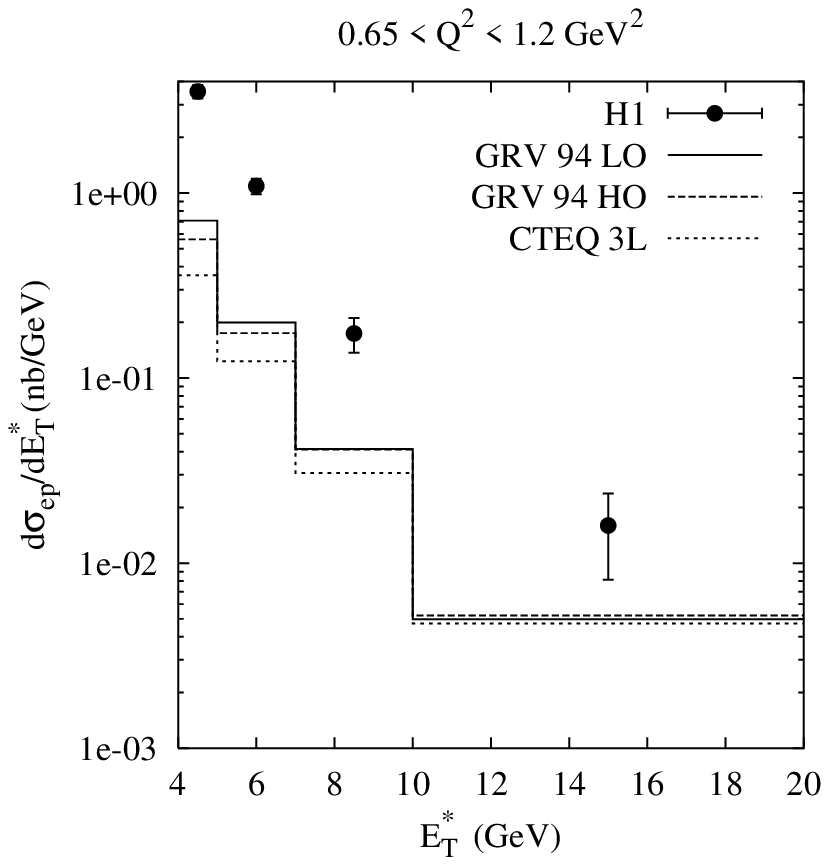,width=78mm}\hspace{-0.5cm}
	   \psfig{figure=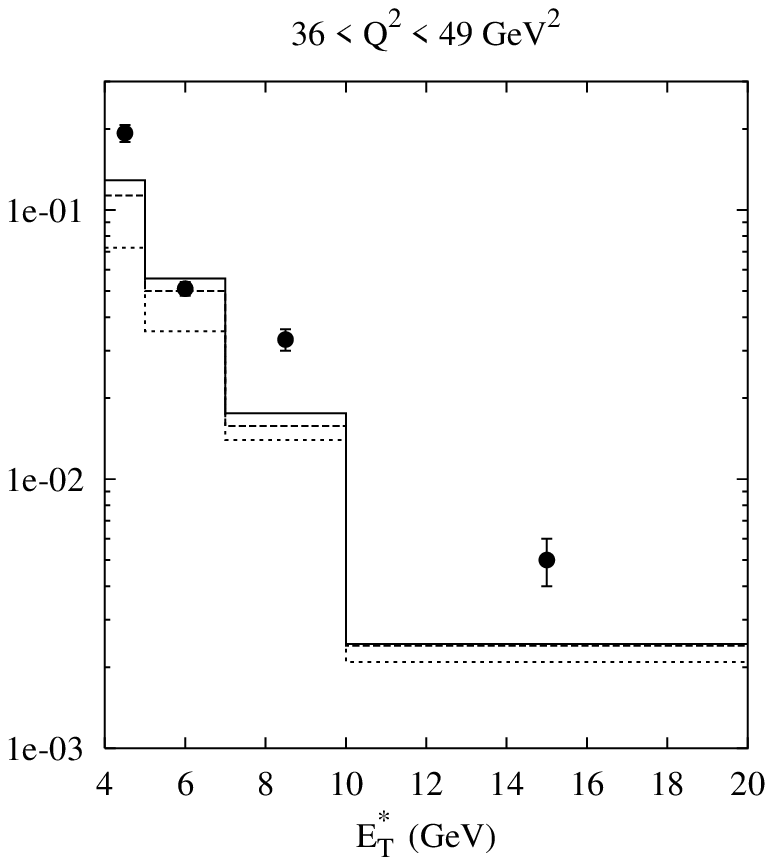,width=78mm}}
     \mbox{ }
     \mbox{\psfig{figure=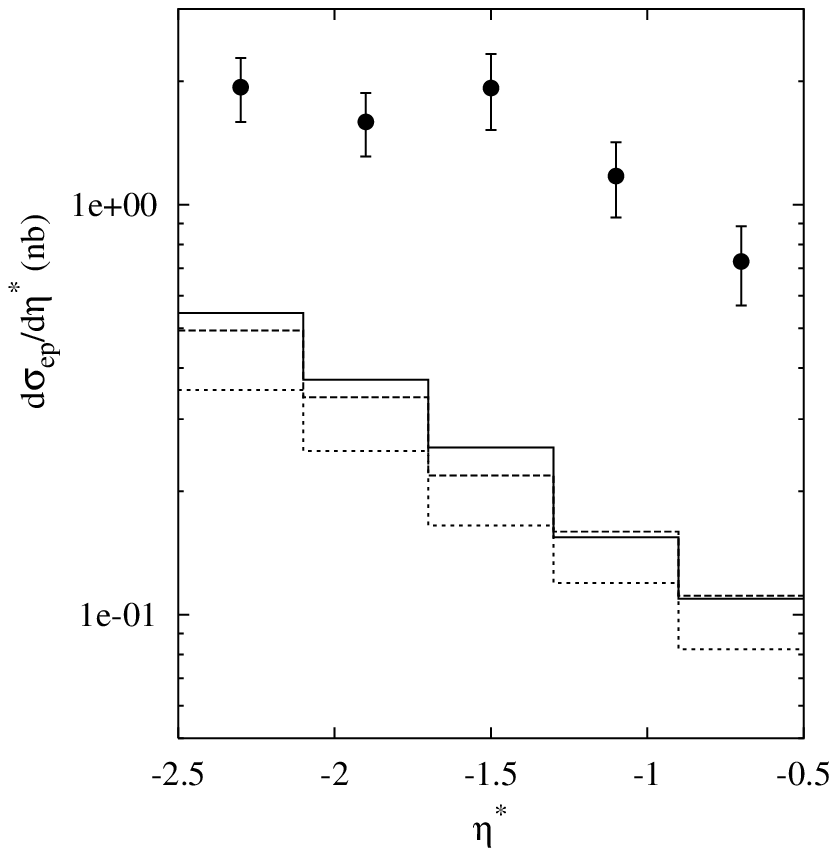,width=78mm}\hspace{-0.5cm}
	   \psfig{figure=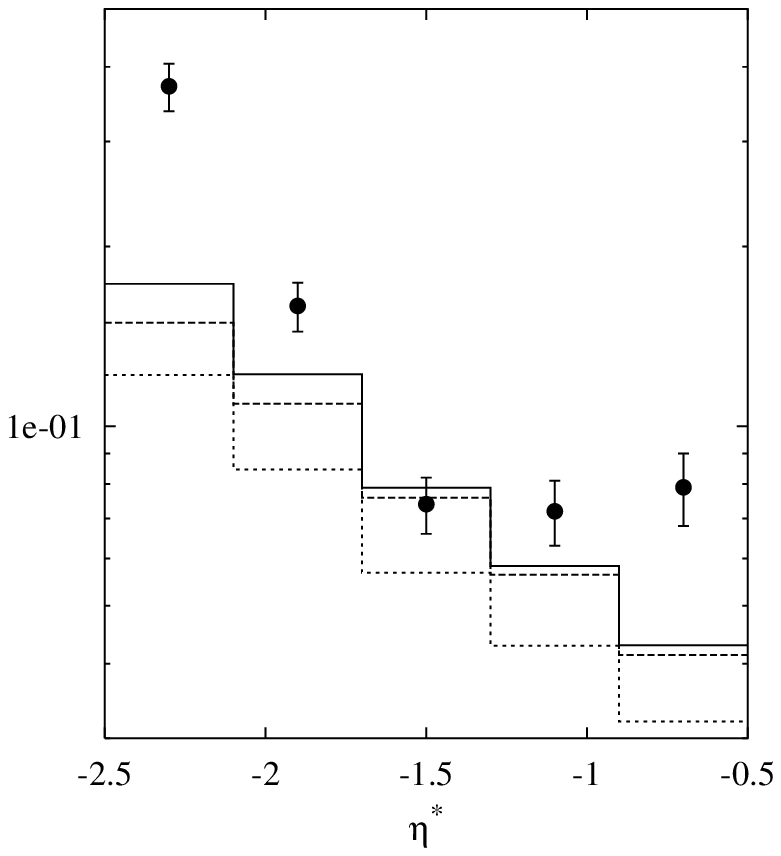,width=78mm}}
   \end{center}
\captive{The differential jet cross sections $\d\sigma_{\e\p}/\d E^*_{\perp}$
and  $\d\sigma_{\e\p}/\d \eta^*$ for jets with $-2.5<\eta^*<-0.5$ and 
$0.3<y<0.6$. Only direct events are used to show the sensitivity to the 
proton parton distribution.
\label{fig:ET_eta-dir}}
\end{figure}

Using CTEQ~3L instead of GRV~94~LO as the proton parton distribution 
reduces the result in some $\ET^*$ and $\eta^*$ bins by half, 
Fig.~\ref{fig:ET_eta-dir}. The GRV~94~HO parton distribution give a slightly 
lower result (as compared to GRV~94~LO). The differences mainly come from 
the gluon distributions, that are not yet so well constrained from data. 
In the modeling of the parton distributions, it is a deceptive accident that 
the more well-known proton parton distribution gives a larger uncertainty 
than the photon one. It offers a simple example that also phenomenology of 
other areas may directly influence the interpretation of photon data. 

\subsubsection{Inclusive $\gast \gast$ Jet Cross Sections}

\begin{figure} [!htb]
   \begin{center}     
   \mbox{\psfig{figure=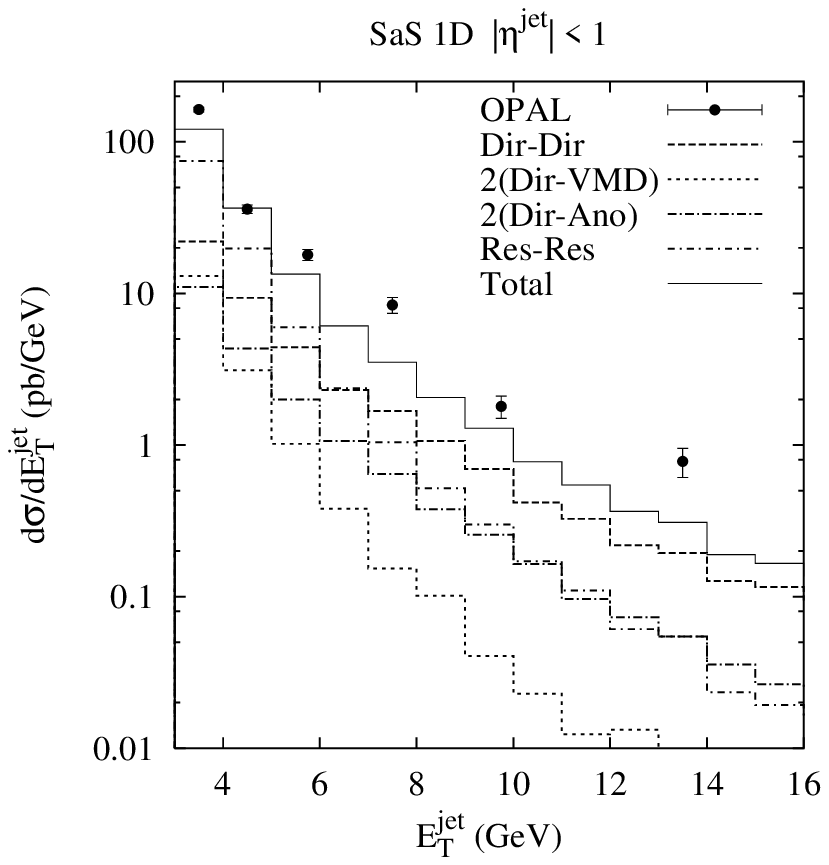,width=78mm}\hspace{-0.5cm}
	\psfig{figure=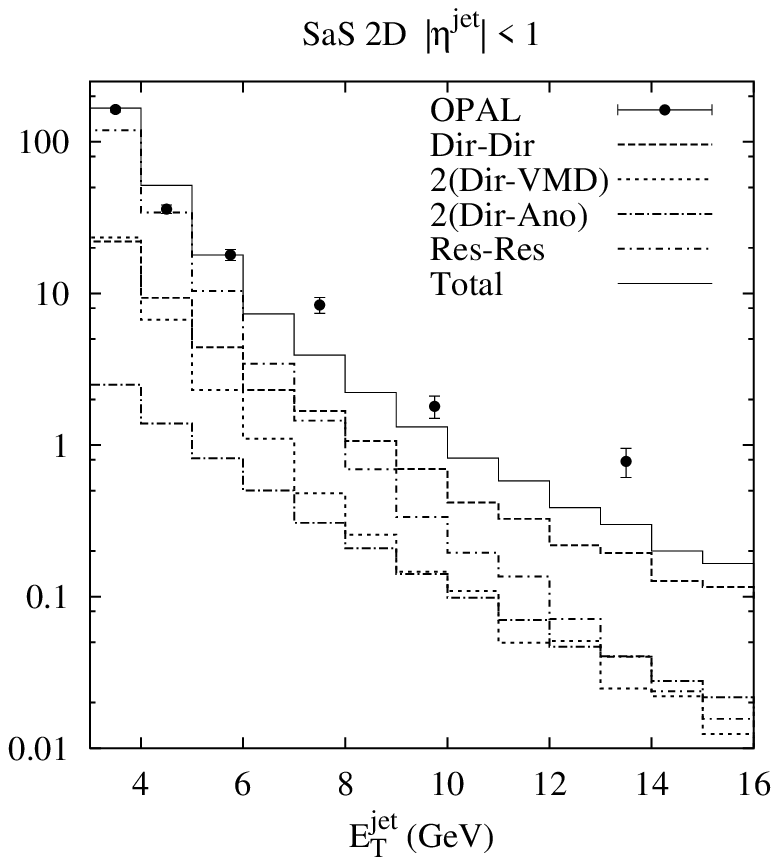,width=78mm}}  
   \end{center}
\captive{The inclusive one--jet cross section as a function of 
$\ET^\mathrm{jet}$ at the centre of mass energies 
$\sqrt{s_{\e^+\e^-}}=130$ and 136~GeV, $|\eta^\mathrm{jet}|<1$. 
The events were generated at a fixed energy of 
$\sqrt{s_{\e^+\e^-}}=133$~GeV, 
$Q^2_\mathrm{max}=0.8$~GeV$^2$, $\pTmin^\mathrm{parton}=1.5$~GeV 
and $\pTmin^\mathrm{MI}=1.4$~GeV.
\label{fig:ET1-jet}}
\end{figure}

\begin{figure} [!htb]
   \begin{center}     
   \mbox{\psfig{figure=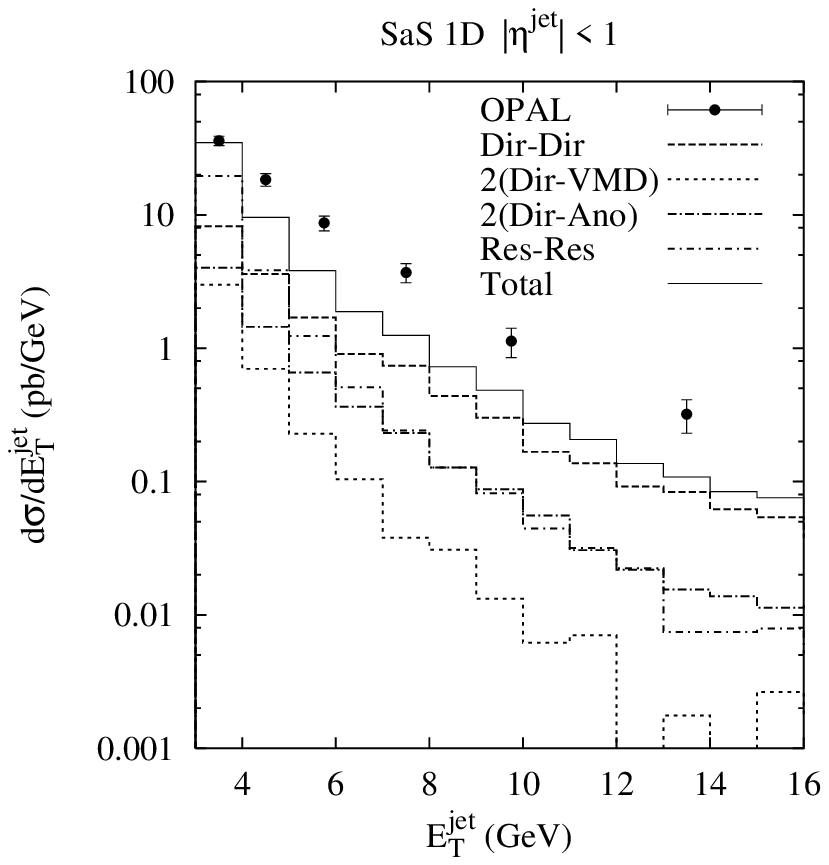,width=78mm}\hspace{-0.5cm}
	\psfig{figure=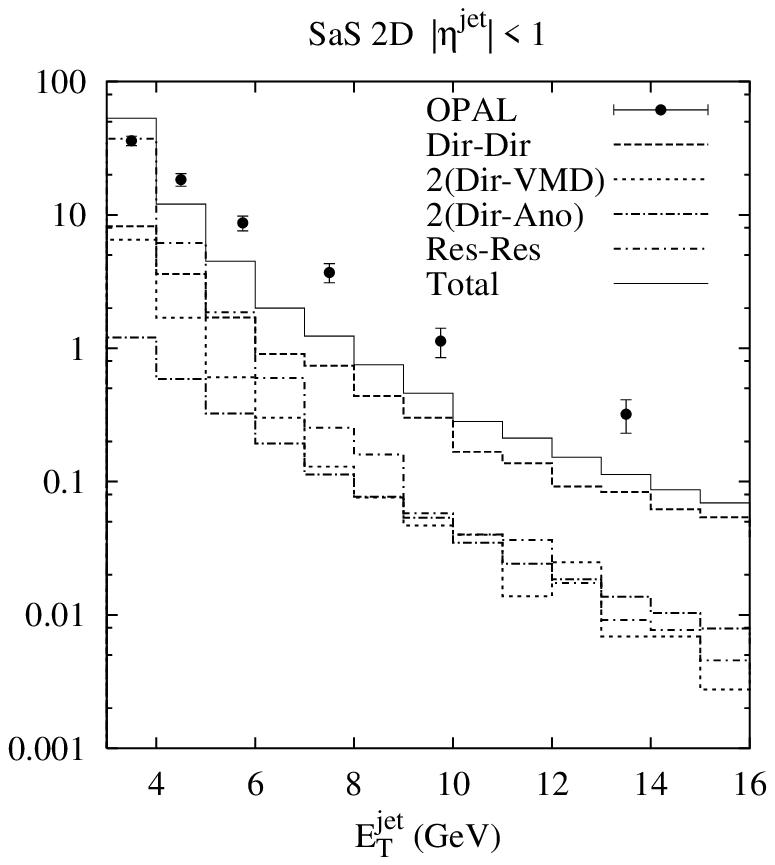,width=78mm}}  
   \end{center}
\captive{The inclusive two--jet cross section as a function of 
$\ET^\mathrm{jet}$; for more information see Fig.~\ref{fig:ET1-jet}.
\label{fig:ET2-jet}}
\end{figure}

\begin{figure} [!htb]
   \begin{center}     
   \mbox{\psfig{figure=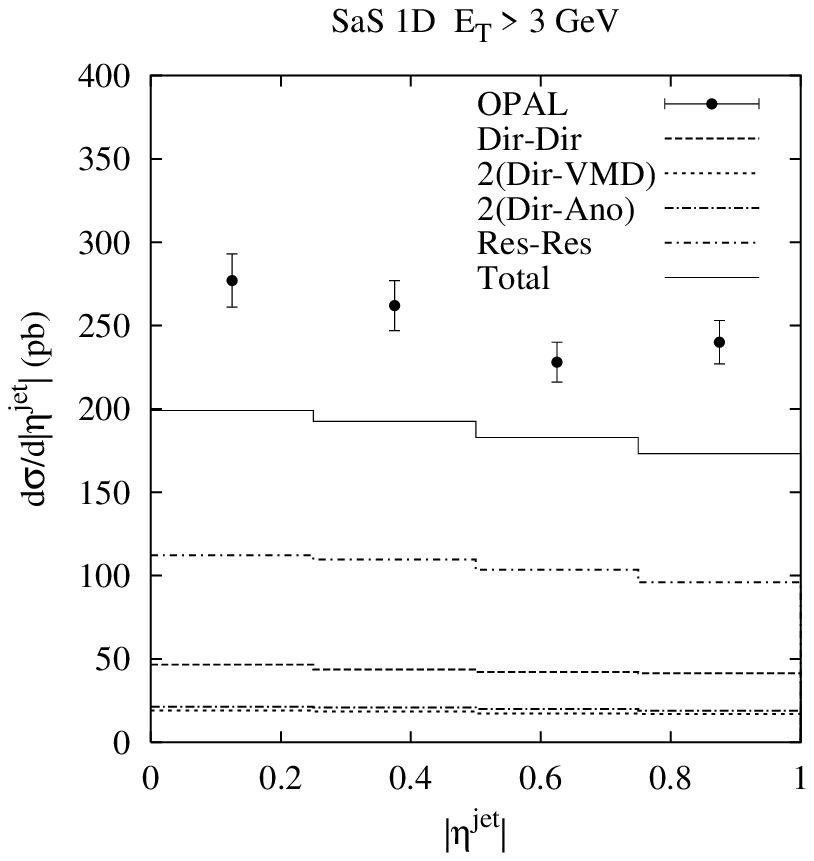,width=78mm}\hspace{-0.5cm}
	\psfig{figure=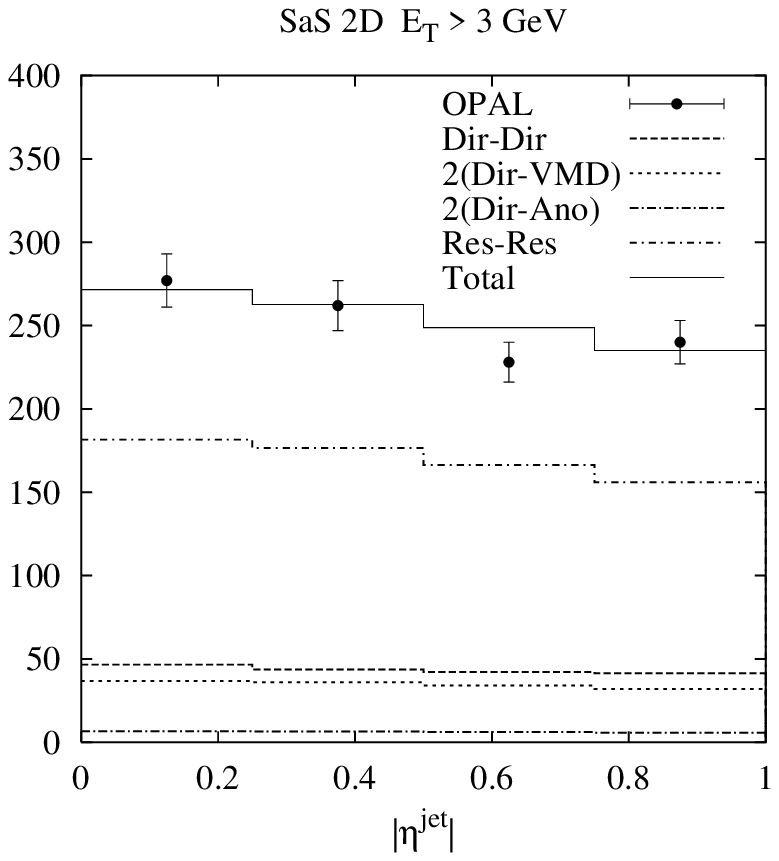,width=78mm}}  
   \end{center}
\captive{The inclusive one--jet cross section as a function of 
$\eta^\mathrm{jet}$ at
the centre of mass energies $\sqrt{s_{\e^+\e^-}}=130$ and 136~GeV, 
$\ET^\mathrm{jet}>3$~GeV. 
The events were generated at a fixed energy of 
$\sqrt{s_{\e^+\e^-}}=133$~GeV, 
$Q^2_\mathrm{max}=0.8$~GeV$^2$, $\pTmin^\mathrm{parton}=1.5$~GeV 
and $\pTmin^\mathrm{MI}=1.4$~GeV.
\label{fig:eta1-jet}}
\end{figure}

\begin{figure} [!htb]
   \begin{center}     
   \mbox{\psfig{figure=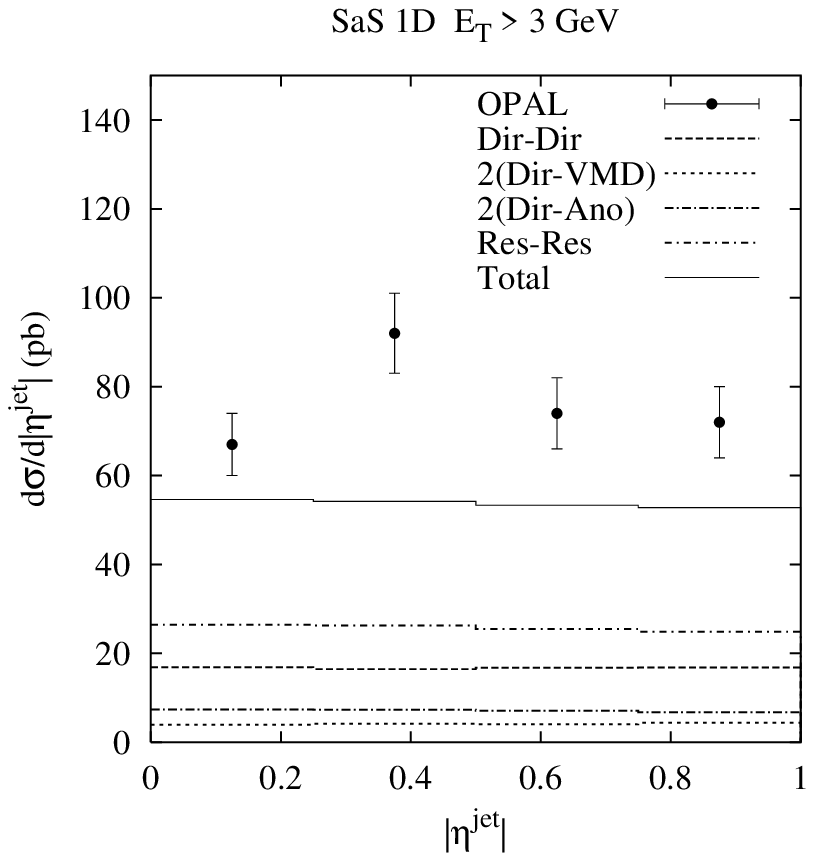,width=78mm}\hspace{-0.5cm}
	\psfig{figure=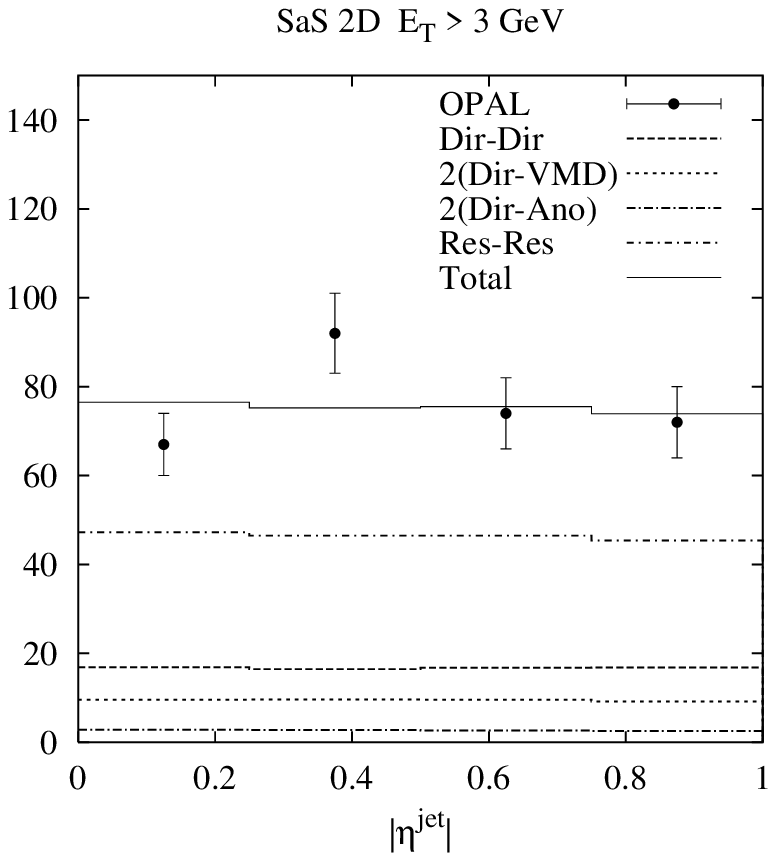,width=78mm}}  
   \end{center}
\captive{The inclusive two--jet cross section as a function of 
$\eta^\mathrm{jet}$; for more information see Fig.~\ref{fig:eta1-jet}.
\label{fig:eta2-jet}}
\end{figure}

The OPAL collaboration has measured inclusive one--jet and two--jet cross 
sections in the range $|\eta^\mathrm{jet}|<1$ and requiring 
$\ET^\mathrm{jet}$ to be 
larger than 3~GeV~\cite{OPAL}. The centre of mass energies were 130 and 
136~GeV. In the analysis, a cone jet algorithm was used with a cone radius 
$R=1$ and $E_{\perp,\mathrm{min}}^\mathrm{jet}=2$~GeV; differences in the 
application of this algorithm may affect the results (we used the PXCONE 
algorithm from HzTool). The two--jet cross sections were obtained by 
measuring events with at least two jets, and then using only the two hardest 
jets. In Figs.~\ref{fig:ET1-jet}--\ref{fig:eta2-jet}, the inclusive 
jet cross sections as a function of $\ET^\mathrm{jet}$ or 
$\eta^\mathrm{jet}$ are shown, with events generated at 
$\sqrt{s_{\e\e}}=133$~GeV. The anti--tagging conditions at this energy 
corresponds to a $Q^2_\mathrm{max}=0.8$~GeV$^2$. Due to migration effects 
$\pTmin^\mathrm{parton}$ was chosen to 1.5~GeV, and 
$\pTmin^\mathrm{MI}=1.4$~GeV. The symmetric cuts used gives no difference 
between the two single--resolved contributions, instead the direct--VMD 
(VMD--direct) and direct--anomalous (anomalous--direct) contributions is
shown together with the direct and double--resolved ones.

At low $\ET^\mathrm{jet}$ the double--resolved events are dominating 
and at larger $\ET^\mathrm{jet}$ it is the direct processes since more 
energy goes into the hard scattering in the latter case. For 
single--resolved events, the SaS~1D VMD component dies out much quicker 
with increasing $\ET^\mathrm{jet}$ than the SaS~2D one which is comparable 
with the direct--anomalous events at high $\ET^\mathrm{jet}$. For both 
cases, at high $\ET^\mathrm{jet}$, the direct--anomalous components give 
the same order of magnitude contribution to the cross section as the 
double--resolved events. The biggest 
difference between the two parton distributions can be seen at low 
$\ET^\mathrm{jet}$ and for the $|\eta^\mathrm{jet}|$ distributions, where the 
double--resolved events dominates; it is a reflection of the difference in
normalization among the contributions. For the SaS~2D case, this kinematical 
region makes the VMD component more important than the anomalous one; as a 
consequence multiple interactions play an important role. The double--resolved 
contribution for SaS~2D without multiple interaction is half of the one 
shown here. Clearly, for the SaS~1D case the opposite is true: the 
importance of the components are reversed. In the region of high 
$\ET^\mathrm{jet}$, where the direct events dominate, the model is 
under\-shooting data. On the other hand, there is nice agreement with data for 
the $|\eta^\mathrm{jet}|$ distribution when using SaS~2D.

\subsubsection{Forward Jets in $\e\p$}
\label{secfwdep}

Jet cross sections as a function of Bjorken-$x$, $x_{\mathrm{Bj}}$, for 
forward jet production (in the proton direction) have been measured at 
HERA~\cite{fwdjet}. The objective is to probe the dynamics of the QCD cascade 
at small $x_{\mathrm{Bj}}$. The forward jet is restricted in polar angle 
w.r.t. the proton and the transverse momenta $\pT^\mathrm{jet}$ should 
be of the same order as the virtuality of the photon, suppressing an 
evolution in transverse momenta. If the jet has a large energy fraction of 
the proton, there will be a big difference in $x$ between the jet and the 
photon vertex, $x_{\mathrm{Bj}} \ll x_{\mathrm{jet}}$, allowing an evolution 
in $x$. The above restrictions will not eliminate the possibility of having a 
resolved photon, although the large $Q^2$ values are not in favour of it. 

\begin{figure} [!htb]
   \begin{center}     
   \mbox{\psfig{figure=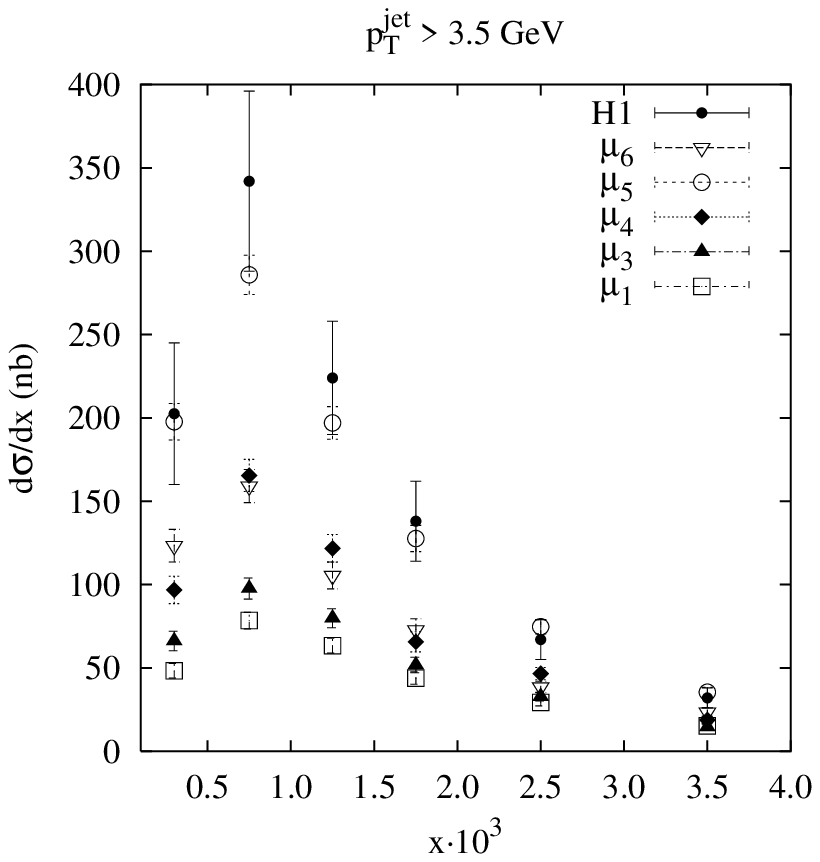,width=78mm}\hspace{-0.5cm}
	\psfig{figure=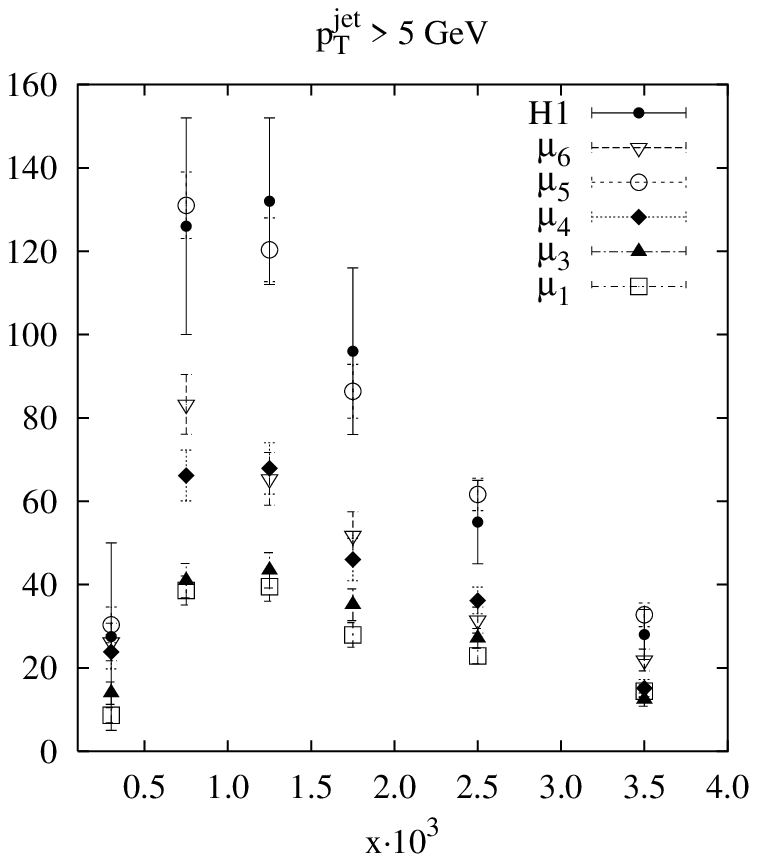,width=78mm}}  
   \end{center}
\captive{Forward jet cross section as a function of $x$ compared with 
H1~data (with statistical and systematic uncertainties added in quadrature). 
Five different scales are shown at two 
different $\pT^\mathrm{jet}$ cuts, 3.5 and 5~GeV. 
$x_\mathrm{jet}>0.035$, $0.5<(\pT^\mathrm{jet})^2/Q^2<2$ and 
$7^{\circ}< \theta_\mathrm{jet} < 20^{\circ}$. 
\label{fig:fwd}}
\end{figure}

\begin{figure} [!htb]
   \begin{center}     
   \mbox{\psfig{figure=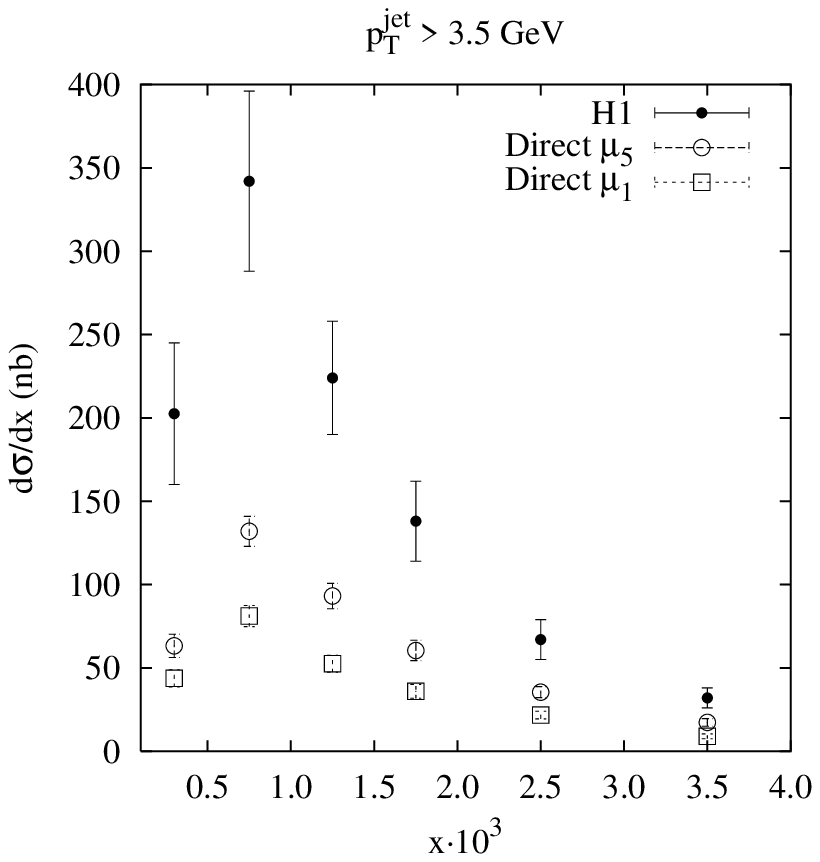,width=78mm}\hspace{-0.5cm}
	\psfig{figure=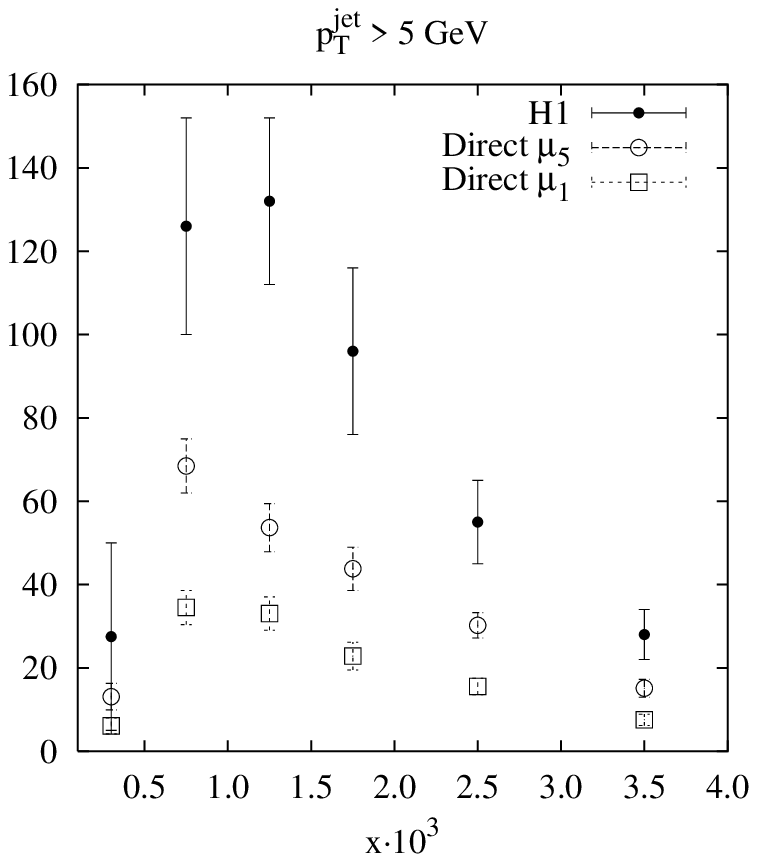,width=78mm}}  
   \end{center}
\captive{Same as in Fig.~\ref{fig:fwd} but only the direct component is 
compared with data for two different scale choices.
\label{fig:fwd-dir}}
\end{figure}

The HzTool routines were used to obtain the results in 
Fig.~\ref{fig:fwd}. Five different scales $\mu_i$ are shown. A larger 
forward jet cross section is obtained with a stronger $Q^2$ dependence, 
with the scale $\mu_5^2=\pT^2+Q^2$ in best agreement with data, in general
agreement with the conclusions in~\cite{LeifHannes}%
\footnote{However, we note a not complete agreement in the description of the
inclusive jet cross sections. In~\cite{LeifHannes} a different option of the
SaS extension to virtual photons was used than here, but we have checked that
this cannot be the main reason for the discrepancy.}. 
The constraint 
$(\pT^\mathrm{jet})^2 \simeq Q^2$ gives a large enhancement for the 
$\mu_4$ and $\mu_5$ scales, which are combinations of 
$(\pT^\mathrm{parton})^2$ and $Q^2$, whereas $\mu_2$ and $\mu_3$ have a 
$Q^2$ dependence that are scaled down by $\shat$. The choice of scale does 
not only affect the resolved photon contribution but also the direct photon, 
arising from the scale dependence in the proton parton distribution, as seen 
in Fig.~\ref{fig:fwd-dir}. 
As a check for the direct component, a simple comparison was made between 
{\sc Pythia}, using the $\mu_1=\pT$ scale and the direct processes, 
and LEPTO  \cite{lepto}, without soft colour interactions, giving similar 
results. The rather large $Q^2$ values, 
$Q^2 \simeq (\pT^\mathrm{jet})^2$, suppresses VMD photons and favours 
the SaS~1D distribution which is the one used here, though the difference 
is small.

Note that the $\mu_6$ scale undershoots the forward jet cross section 
data and overshoots the inclusive jet distributions at low $Q^2$, 
so it is not a real alternative. As a further check, with more data 
accumulated and analyzed, the $(\pT^\mathrm{jet})^2/Q^2$ interval 
could be split into several subranges which hopefully would help to 
discriminate between scale choices. 

\subsubsection{Importance of Longitudinal Resolved Photons}

In this section we will study the importance of longitudinal resolved 
photons. The different $R$-factors described in section~\ref{pdf} will 
be used to provide some estimates. 
Of those distributions studied so far, we will concentrate on the 
$\d\sigma_{\e \p}/\d \ET^*$ distributions for inclusive jets and 
the $\d\sigma_{\e \p}/\d x$ for
forward jet cross sections. A sensible $Q^2$--dependent scale choice, 
$\mu_3$, together with the SaS~1D distribution will be used throughout. 

\begin{figure} [!htb]
   \begin{center}    
     \mbox{\psfig{figure=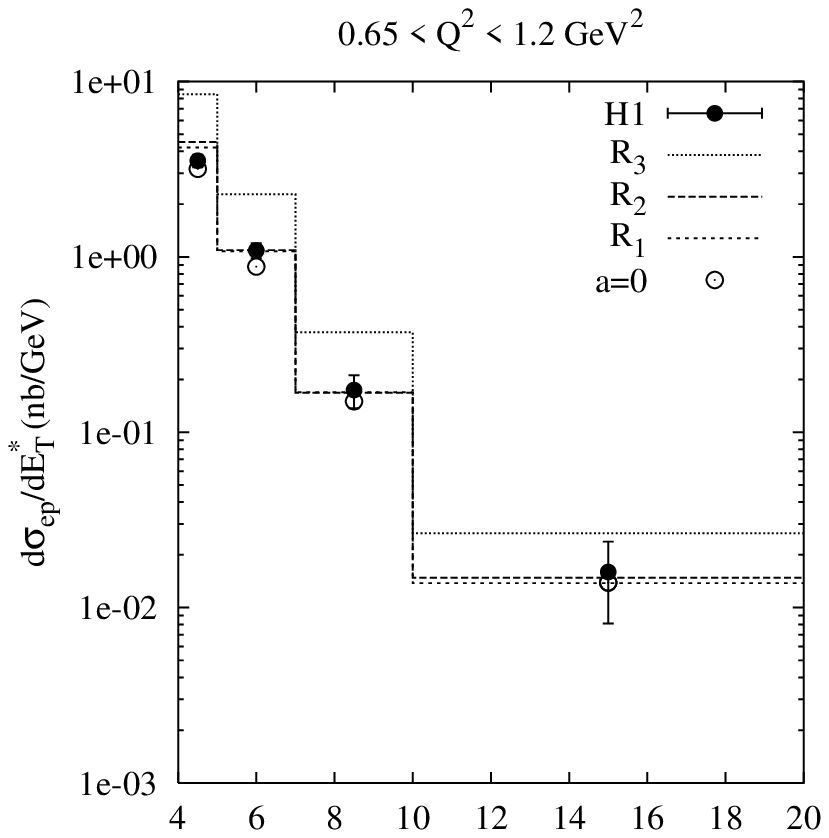,width=78mm}\hspace{-0.5cm}
	   \psfig{figure=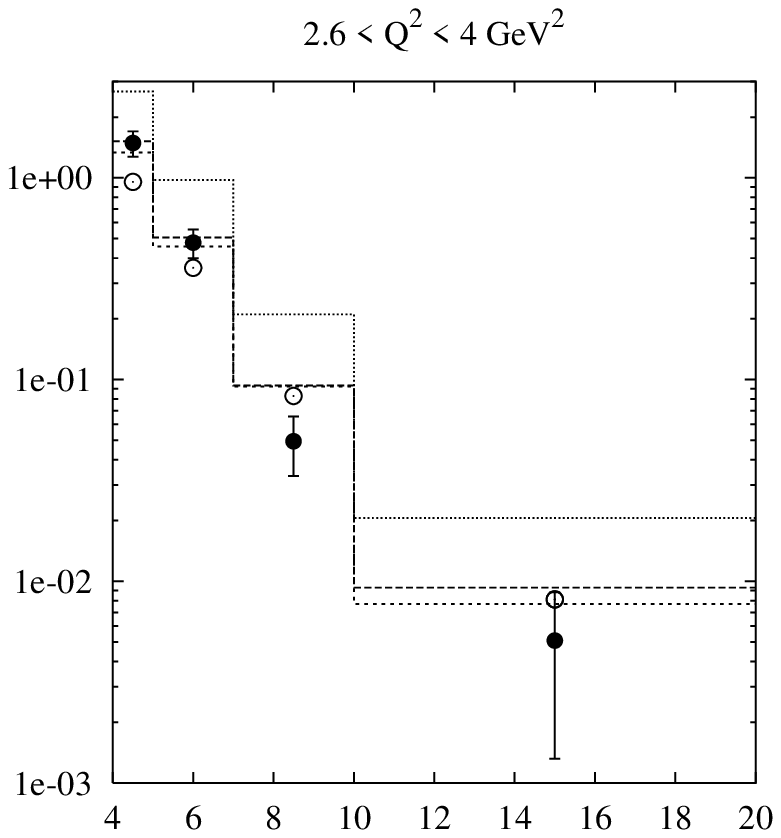,width=78mm}}
     \mbox{ }
     \mbox{\psfig{figure=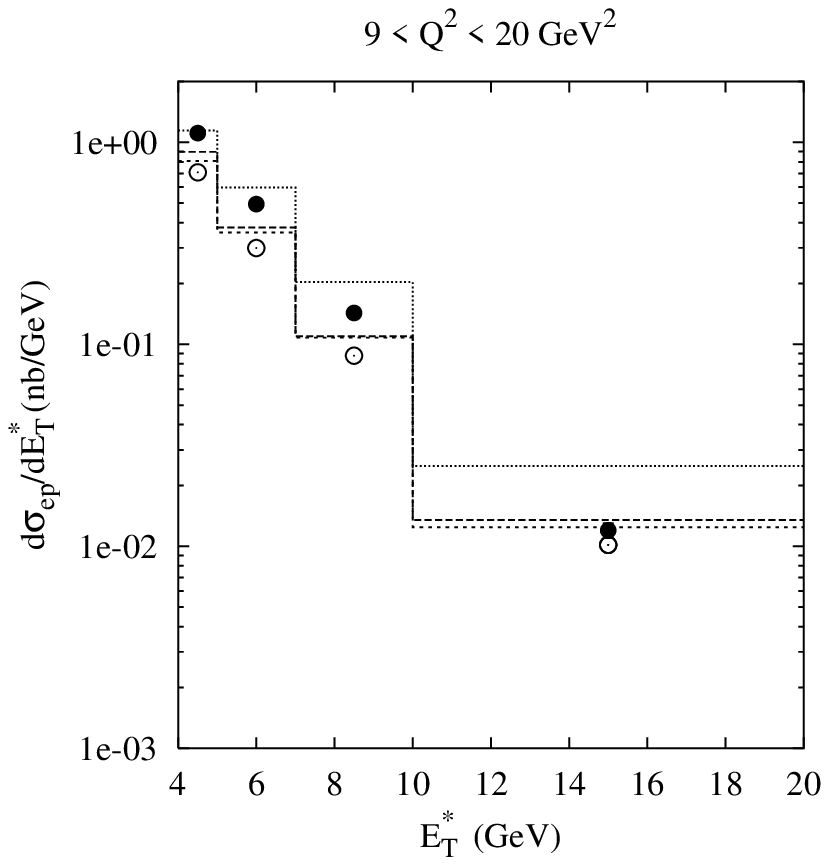,width=78mm}\hspace{-0.5cm}
	   \psfig{figure=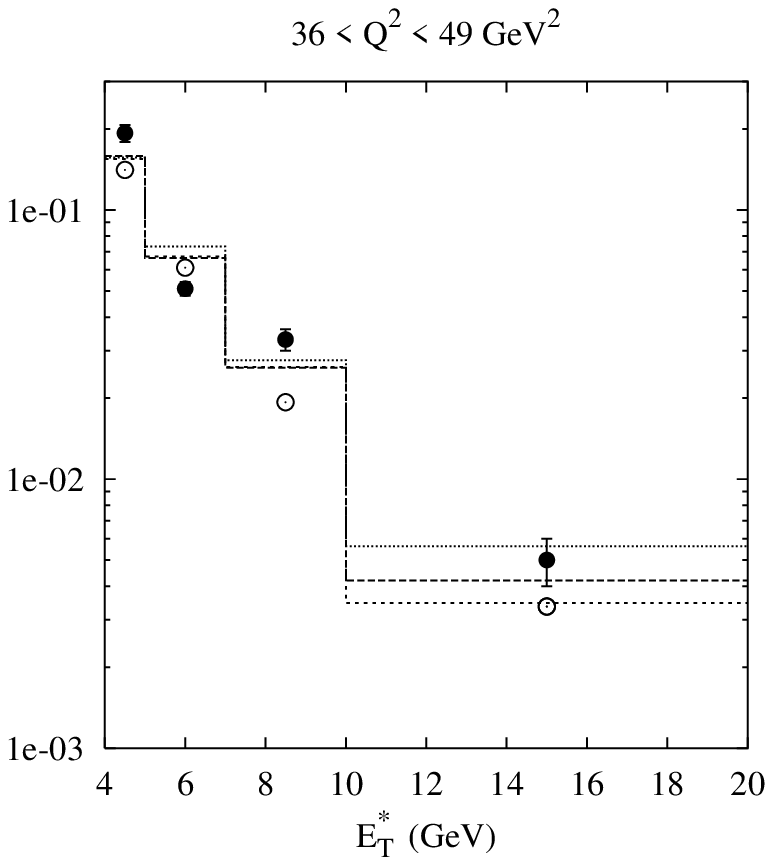,width=78mm}}
   \end{center}
\captive{The differential jet cross section $\d\sigma_{\e\p}/\d E^*_{\perp}$ 
for jets with $-2.5<\eta^*<-0.5$ and $0.3<y<0.6$. 
\label{fig:ETL}}
\end{figure}

With $a=1$ the different alternatives are shown in Fig.~\ref{fig:ETL} 
for the $\d\sigma_{\e \p}/\d \ET^*$ distributions together 
with the result from pure transverse photons, i.e. $a=0$. The importance 
of the resolved contributions decreases with increasing $Q^2$, see 
Fig.~\ref{fig:ET}, which makes the asymptotic behaviour less crucial. 
The onset of longitudinal photons governed by the $R_1$ and $R_2$ 
alternatives are favoured whereas the $R_3$ one overshoots data in the 
context of the other model choices made here. 

\begin{figure} [!htb]
   \begin{center}     
   \mbox{\psfig{figure=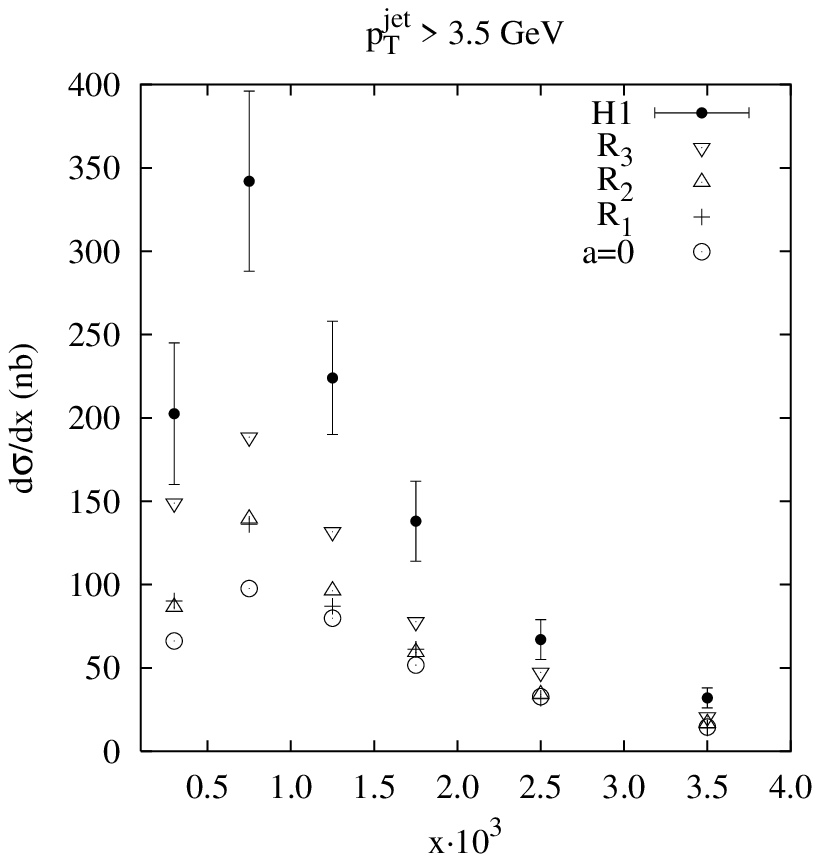,width=78mm}\hspace{-0.5cm}
	\psfig{figure=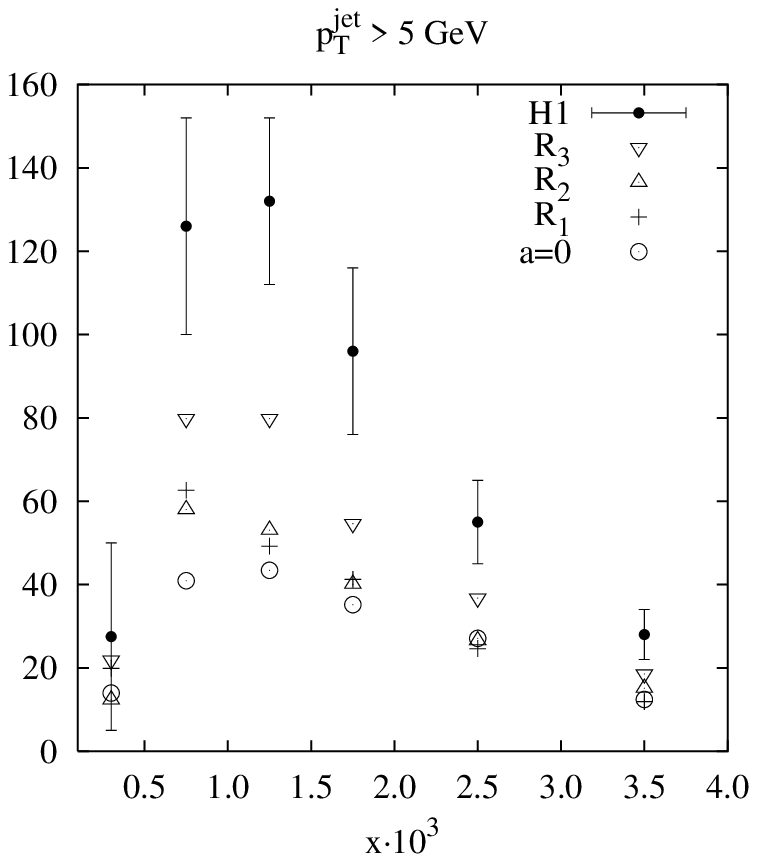,width=78mm}}  
   \end{center}
\captive{Forward jet cross section as a function of $x$.
\label{fig:fwdL}}
\end{figure}

In Fig.~\ref{fig:fwdL} the same alternatives are shown for the forward jet 
cross sections. With this scale choice, $\mu_3$, none of the longitudinal 
resolved components (together with the direct contribution) are sufficient 
to describe the forward jet cross section. The resolved contribution with 
$R_3$ is about the same as the one obtained with the scale 
$\mu_5^2=\pT^2+Q^2$ (without longitudinal contribution); 
the difference in the total results originates from the difference in the
direct contributions, see Fig.~\ref{fig:fwd-dir}. With $R_1$ and $a=1$, the
$\mu_5$ scale (not shown) overshoots the data, but undershoots in combination 
with $\mu_4$. The $\mu_4$ scale in combination with $R_3$ and $a=1$ is in nice 
agreement with data. 

The above study indicates, as expected, that longitudinal resolved photons
are important for detailed descriptions of various distributions. It cannot
by itself explain the forward jet cross section, but may give a significant 
contribution. Combined with other effects, for example, different scale 
choices, parton distributions, underlying events, etc. it could give a 
reasonable description. The model(s) so far does not take into account the 
difference in $x$ distribution or the $k^2$ scale (of the 
$\gast \rightarrow \q\qbar$ fluctuations) between transverse and longitudinal 
photons. As long as the distributions under study allow a large interval in 
$x$ the average description may be reasonable. In a more sophisticated 
treatment these aspects have to be considered in more detail. 

\subsection{Example of Further Tests: Forward Jets in $\e^+\e^-$}

With the experience of forward jets at HERA, we suggest a similar study at
LEP. The optimal kinematical and forward jet constraints have to be set by
each collaboration itself; the study here will give the order of magnitude 
for the cross section and point out uncertainties in the model.

Comparing with forward jets at HERA, one of the leptons will play the role 
of the proton. Some of the constraints will be taken over directly, 
kinematic cuts:
\begin{eqnarray}
   y & >& 0.1,~~E'_e>11~\mathrm{GeV},
\label{eq:fwd-kin}
\end{eqnarray}
and jet selection:
\begin{eqnarray}
 \pT^\mathrm{jet} & > & 3.5~\mathrm{GeV}\\
 x_\mathrm{jet}=\frac{E_\mathrm{jet}}{E_\e} & > & 0.035 \\
                    0.5 & < & \frac{(\pT^\mathrm{jet})^2}{Q^2}<2\\
                      3^{\circ} & < & \theta_\mathrm{jet} < 20^{\circ}
\label{eq:fwd-jet}
\end{eqnarray}
To fulfill the jet selection one of the leptons has to be tagged in order to
know the virtuality of the photon. To obtain a reasonable number of events the 
other lepton is not tagged, imposing a $Q^2_\mathrm{max}$, here chosen to 
1.5~GeV$^2$. With a centre of mass energy of 200~GeV, the smallest accessible 
$x_\mathrm{Bj}=\frac{Q^2}{y s}$ is around $10^{-4}$, where $Q^2$ and $y$ is 
calculated from the tagged electron, omitting the virtuality of the other 
photon. In a more sophisticated treatment also double--tagged events are 
analyzed; then one of the photons plays the role of a proton and the forward 
jet should be defined with respect to one of the photons. 

\begin{figure} [!htb]
   \begin{center}     
   \mbox{\psfig{figure=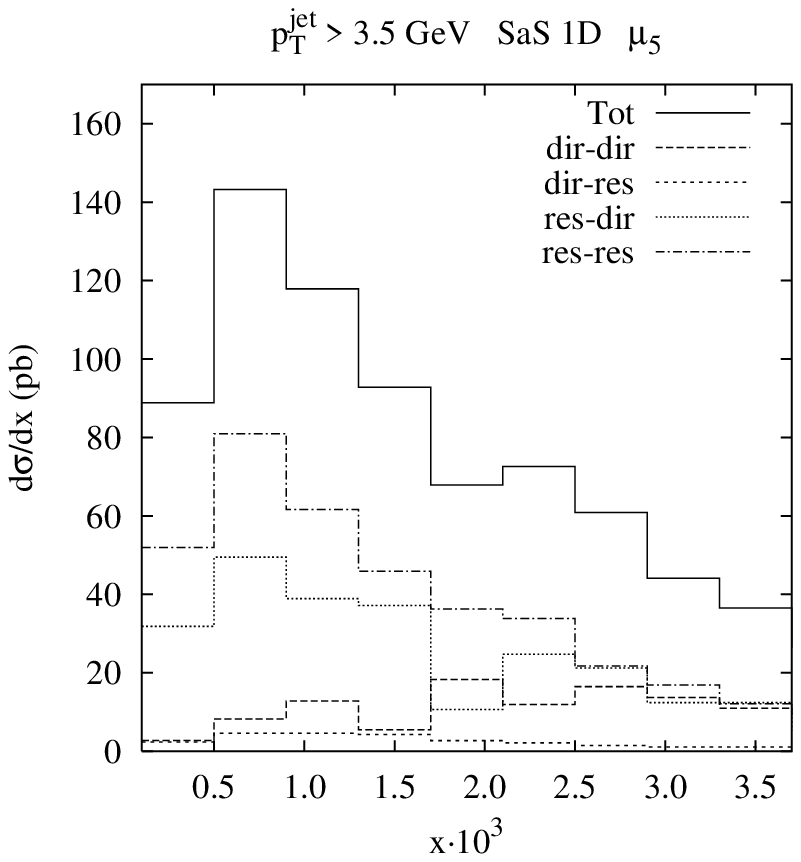,width=78mm}\hspace{-0.5cm}
	 \psfig{figure=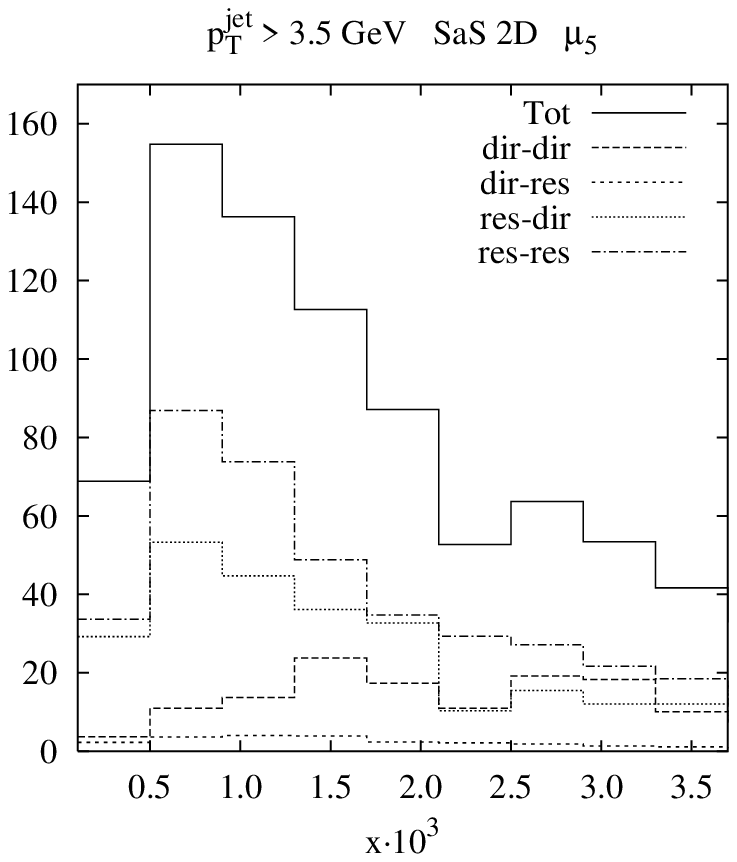,width=78mm}}
   \mbox{\psfig{figure=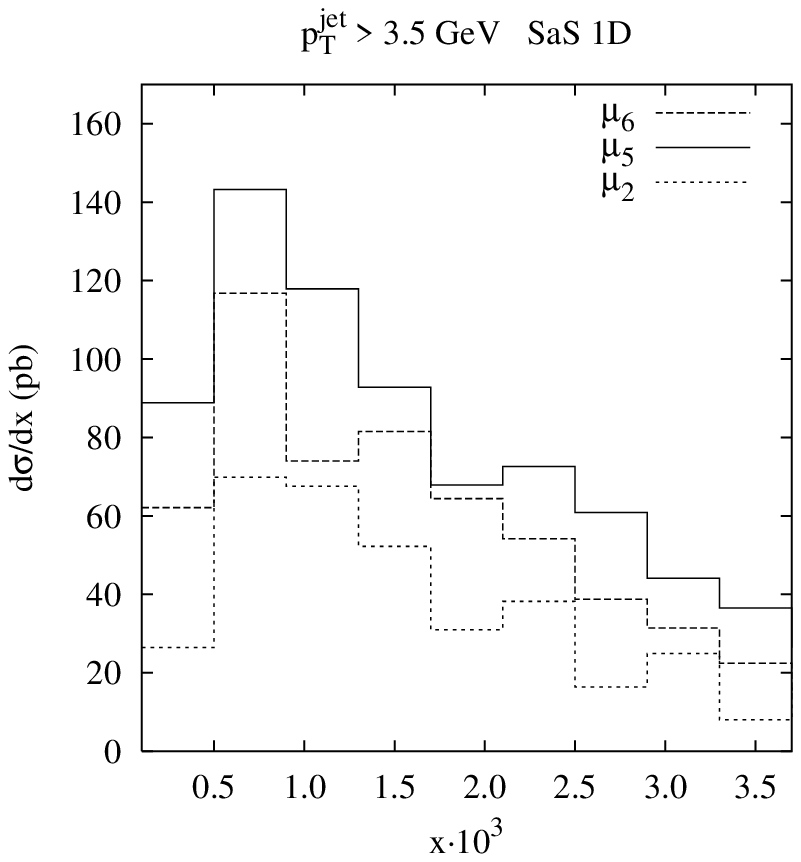,width=78mm}\hspace{-0.5cm}
	 \psfig{figure=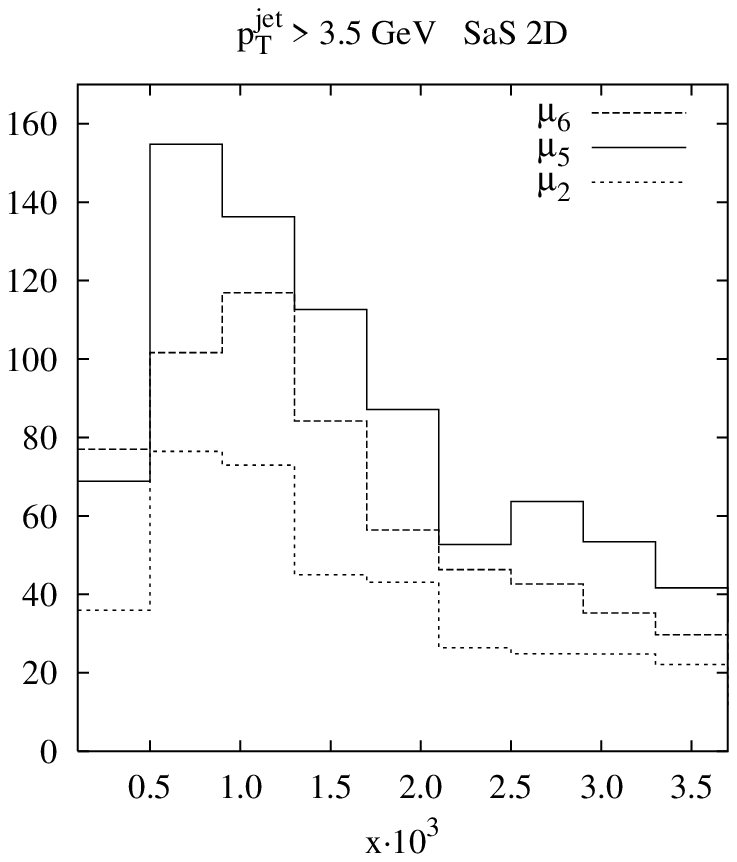,width=78mm}}  
   \end{center}
\captive{Forward jet cross section as a function of $x$. 
$\pT^\mathrm{jet}>3.5$~GeV,  
$x_\mathrm{jet}>0.035$, $0.5<(\pT^\mathrm{jet})^2/Q^2<2$ and 
$3^{\circ}< \theta_\mathrm{jet} < 20^{\circ}$. 
\label{fig:fwdee3.5}}
\end{figure}

\begin{figure} [!htb]
   \begin{center}     
   \mbox{\psfig{figure=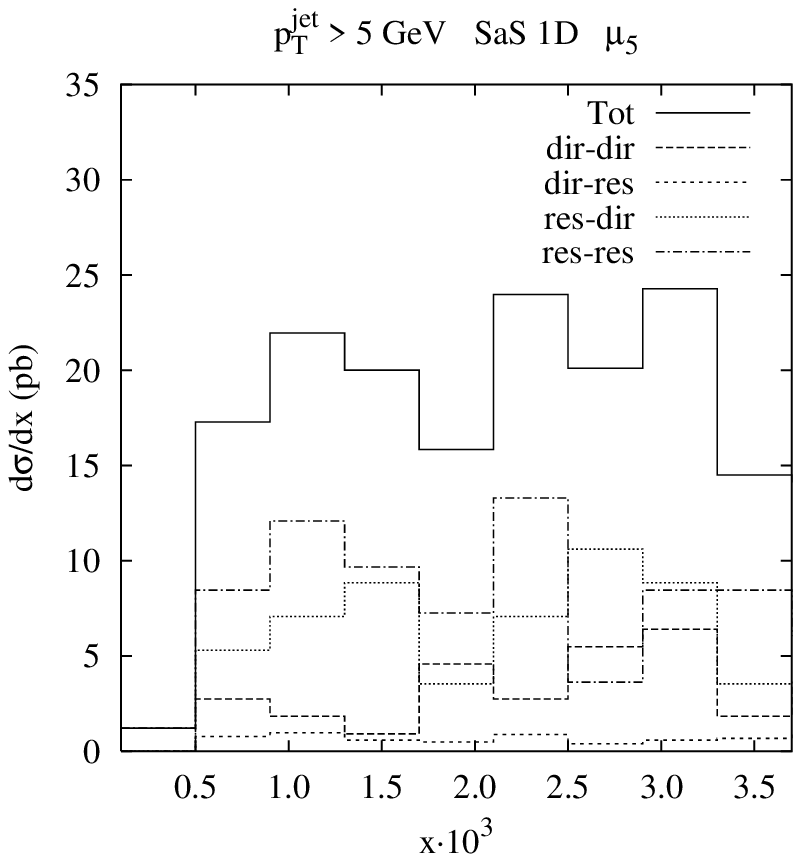,width=78mm}\hspace{-0.5cm}
	 \psfig{figure=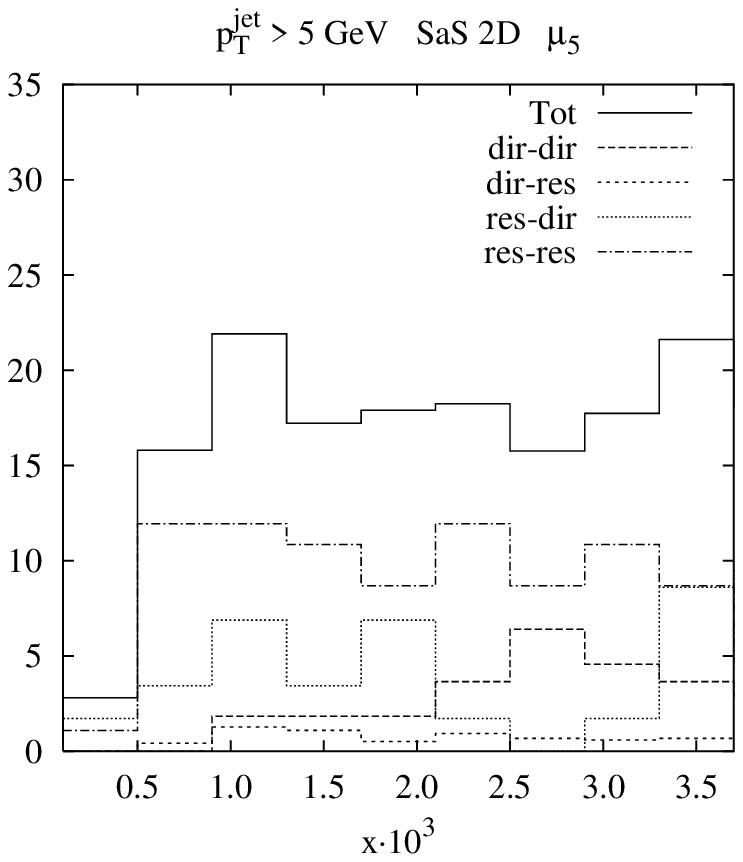,width=78mm}}  
   \end{center}
\captive{Forward jet cross section as a function of $x$. 
$\pT^\mathrm{jet}>5$~GeV,  
$x_\mathrm{jet}>0.035$, $0.5<(\pT^\mathrm{jet})^2/Q^2<2$ and 
$3^{\circ}< \theta_\mathrm{jet} < 20^{\circ}$. 
\label{fig:fwdee5}}
\end{figure}

A cone jet algorithm with cone radius $R=1$ and 
$E_{\perp,\mathrm{min}}^\mathrm{jet}=2$~GeV 
is used for jet finding. As for the case at HERA, the $\mu_5$ scale gives the 
largest forward jet cross section, about twice as large as with the $\mu_2$ 
scale, Fig.~\ref{fig:fwdee3.5}. Most of the differences arise from the 
double--resolved events. Double--resolved and single--resolved events, where 
the resolved photon give rise to the forward jet, dominate the forward jet 
cross section, Fig.~\ref{fig:fwdee3.5} and~\ref{fig:fwdee5}. At low $x$, for 
the $\mu_5$ scale, the double--resolved contribution is close to an order of 
magnitude larger than the direct one. For the $\mu_2$ scale it is about a 
factor of four. As for the case at HERA, the rise of the forward jet 
cross section at small $x$ is dominated by resolved photons. A study like 
this at LEP could be an important cross check for the understanding of 
resolved photons and that of small-$x$ dynamics. 

\section{Summary and Outlook}

The field of photon physics is rapidly expanding, not least by the impact 
of new data from HERA and LEP. The prospects of building a Linear Collider, 
with its objective of high-precision measurements and to search for possible 
new physics, requires an accurate description of photon processes. 
The plan here is to have a complete description of the main physics aspects 
in $\gamma\p$ and $\gamma\gamma$ collisions, which will allow important cross 
checks to test universality of certain model assumptions. As a step forward, 
we have in this article concentrated on those that are of importance for the 
production of jets by virtual photons, and are absent in the real-photon case. 
While we believe in the basic machinery developed and presented here, we have 
to acknowledge the many unknowns --- scale choices, parton distribution sets 
(also those of the proton), longitudinal contributions, underlying events, 
etc. --- that all give non-negligible effects. To make a simultaneous  
detailed tuning of all these aspects was not the aim here, but rather to 
point out model dependences that arise from a virtual photon. 

When $Q^2$ is not small, naively only the direct component needs to be 
treated, but in practice a rather large contribution arises from resolved 
photons. For example, for high-$Q^2$ studies like forward jet cross sections, 
Fig.~\ref{fig:fwd}--\ref{fig:fwd-dir}, or inclusive differential jet cross 
sections, Fig.~\ref{fig:ET}--\ref{fig:inc_gp}. Resolved longitudinal 
photons are poorly understood and the model presented here can be used to 
estimate their importance and get a reasonable global description. 
Longitudinal effects are in most cases small but of importance for 
fine--tuning. 

In the study of the dijet angular distributions in photoproduction the 
relative amount of direct and resolved events is not so well described by the
model. In the future, this could be improved by using lepton-inside-lepton 
structure functions. 

The inclusive $\gast\gast$ one-jet and two-jet cross sections are well 
described except for the high $\ET^\mathrm{jet}$ region of the 
$\ET^\mathrm{jet}$ distribution. In this region, the direct events are 
dominating. Currently, owing to the lesser flexibility in the modeling of the 
direct component, we do not see any simple way to improve the model. The 
factorized ansatz made for the photon flux is expected to be valid in this 
kinematical range; interference terms are suppressed by 
$Q_1^2 Q_2^2/W_{\gast\gast}^2$. However, differences in the application of 
the cone jet algorithm may affect the results. 

The forward jet cross section presented by H1~\cite{fwdjet} is well described
by an ordinary parton shower prescription including the possibility of having
resolved photons. The criteria that the $\pT^\mathrm{jet}$ should be of
the same order as $Q^2$, makes the scale choice crucial and $\mu_5^2=\pT^2+Q^2$
is favoured by data, as concluded in~\cite{LeifHannes}. With this experience 
we predict the forward jet cross section to be obtained at LEP. With more 
data accumulated and analyzed, the $(p_{\perp}^\mathrm{jet})^2/Q^2$ interval 
could be split into several subranges, which hopefully would help to 
discriminate between different scale choices.

Multiple interactions for the anomalous component are not yet included, and is
not expected to be of same importance as in the VMD case. However, for 
low $k^2$ fluctuations it may be important, especially for SaS~1D, and need 
to be investigated. 

After this study of jet production by virtual photons it is natural to extend
the modeling to low--$\pT$ events. Clearly, a smooth transition from 
perturbative to non-perturbative physics is wanted. One idea is to make use of
a parameterization of the total $\gamma\p$ and $\gamma\gamma$ cross section in 
terms of a pomeron and a reggeon exchange. Starting from the real-photon case,
dipole dampening factors are introduced for the generalization to virtual 
photons. For example, the $\gamma\p$ cross section is divided into a VMD, an 
anomalous, a direct and a DIS $\gast\q \rightarrow \q$ process part. In the 
limit $Q^2 \rightarrow 0$, the first three event classes remain. On the 
contrary, when $Q^2$ increases from zero to high $Q^2$; the resolved 
processes dies out (as given by the dipole factors), the direct also drops 
and finally only the DIS process remain. At intermediate $Q^2$ values, the 
direct processes and the DIS (+parton showers) process overlap, since, in 
some regions of phase space, they are equally valid descriptions of the same 
physics. It thus becomes necessary to avoid double-counting, e.g. by 
introducing Sudakov style form factors for the DIS process, suppressing those 
parton configurations covered by the direct processes. We intend to return to 
this issue in a future publication.

\subsubsection*{Acknowledgements}

We acknowledge helpful conversations with, among others, Jon Butterworth, 
Jiri Ch\'yla, Gerhard Schuler, Hannes Jung, Leif J\"onsson, Ralph Engel and 
Tancredi Carli.


\begin{thebibliography}{99}

\bibitem{xgobs}
ZEUS Collaboration, M. Derrick et al., \Journal{\PLB}{322}{287}{1994};\\
H1 Collaboration, T. Ahmed et al., \Journal{\NPB}{445}{195}{1995}.

\bibitem{sasevt}
G.A. Schuler and T. Sj\"ostrand, \\
\Journal{\NPB}{407}{539}{1993}; \Journal{\ZPC}{73}{677}{1997}.

\bibitem{saspdf}
G.A. Schuler and T. Sj\"ostrand, \\
\Journal{\ZPC}{68}{607}{1995}; 
\Journal{\PLB}{376}{193}{1996}.
%\JournalPLB{376}{193}{1996}.

\bibitem{pythia} 
T. Sj\"ostrand, \Journal{\CPC}{82}{74}{1994};\\
http://www.thep.lu.se/$\sim\,$torbjorn/Pythia.html.

\bibitem{herwig}
G.~Marchesini, B.R.~Webber, G.~Abbiendi, I.G.~Knowles, M.H.~Seymour and 
L.~Stanco, \Journal{\CPC}{67}{465}{1992}.

\bibitem{ldc}
H. Kharraziha and L. L\"onnblad, \Journal{\JHEP}{03}{006}{1998}.

\bibitem{lepto}
G. Ingelman, A. Edin and J. Rathsman, \Journal{\CPC}{101}{108}{1997}.

\bibitem{phojet}
R. Engel and J. Ranft, \Journal{\PRD}{54}{4244}{1996}.

\bibitem{rapgap}
H. Jung, \Journal{\CPC}{86}{147}{1995}.

\bibitem{nlome}
M. Gl\"uck, E. Reya and M. Stratmann, \Journal{\PRD}{54}{5515}{1996};\\
D. de Florian, C. Garcia Canal and R. Sassot, \Journal{\ZPC}{75}{265}{1997};\\
M. Klasen, G. Kramer and B. P\"otter, \Journal{\EPJC}{1}{261}{1998}.

\bibitem{weiz}
C.F. von Weisz\"acker, \Journal{\ZP}{88}{612}{1934}.

\bibitem{will}
E.J. Williams, \Journal{\PRV}{45}{729}{1934}.

\bibitem{klaWW}
M. Klasen, G. Kramer and S.G. Salesch, \Journal{\ZPC}{68}{113}{1995}.

\bibitem{EPA}
V.M. Budnev, I.F. Ginzburg, G.V. Meledin and V.G. Serbo, \\
\Journal{\PRP}{15}{181}{1975}.

\bibitem{GS}
G.A. Schuler, \Journal{\CPC}{108}{279}{1998}.

\bibitem{gammaflux}
G. Bonneau, M. Gourdin and F. Martin, \Journal{\NPB}{54}{573}{1973}.

\bibitem{Qfourred}
S. Frixione, M.L. Mangano, P. Nason and G. Ridolfi,
\Journal{\PLB}{319}{339}{1993}.
%\JournalPLB{319}{339}{1993}.

\bibitem{Baier}
V.N.~Baier, E.A.~Kuraev, V.S.~Fadin and V.A.~Khoze, 
\Journal{\PRP}{78}{293}{1981}.

\bibitem{siggap}
G. Altarelli and G. Martinelli, %\Journal{\PLB}{76}{89}{1978};\\
\JournalPLB{76}{89}{1978};\\
A. Mend\'ez, \Journal{\NPB}{145}{199}{1978};\\
R. Peccei and R. R\"uckl, \Journal{\NPB}{162}{125}{1980};\\
Ch. Rumpf, G. Kramer and J. Willrodt, \Journal{\ZPC}{7}{337}{1981}.

\bibitem{sigpp}
B.L. Combridge, J. Kripfganz and J. Ranft, 
%\Journal{\PLB}{70}{234}{1977};\\
\JournalPLB{70}{234}{1977};\\
R. Cutler and D. Sivers, \Journal{\PRD}{17}{196}{1978}.

\bibitem{backwards}
T. Sj\"ostrand, %\Journal{\PLB}{157}{321}{1985};\\
\JournalPLB{157}{321}{1985};\\
M. Bengtsson, T. Sj\"{o}strand and M. van Zijl, 
\Journal{\ZPC}{32}{67}{1986}. 

\bibitem{grspdf}
M. Gl\"uck, E. Reya and M. Stratmann, \Journal{\PRD}{51}{3220}{1995}.

\bibitem{dgpdf}
F.M. Borzumati and G.A. Schuler, \Journal{\ZPC}{58}{139}{1993};\\
M. Drees and R.M. Godbole, \Journal{\PRD}{50}{3124}{1994}.

\bibitem{LeifHannes}
H. Jung, L. J\"onsson and H. Kuster, 
DESY~98-051 (hep-ph/9805396); \\
DESY~99-028 (hep-ph/9903306), to appear in \EPJC.

\bibitem{AGIS} 
B.~Andersson, G.~Gustafson, G.~Ingelman and T.~Sj\"ostrand,\\
\Journal{\PRP}{97}{31}{1983}.

\bibitem{cdfcoher}
CDF Collaboration, F. Abe et al., \Journal{\PRD}{50}{5562}{1994}.

\bibitem{largekT}
S. Frixione, M.L. Mangano, P. Nason and G. Ridolfi,
\Journal{\NPB}{431}{453}{1994};\\
L. Apanasevich et al., \Journal{\PRD}{59}{074007}{1999};\\
%hep-ph/9808467;\\
G. Miu and T. Sj\"ostrand, \Journal{\PLB}{449}{313}{1999}.
%\JournalPLB{449}{313}{1999}.

\bibitem{multint}
T. Sj\"ostrand and M. van Zijl, \Journal{\PRD}{36}{2019}{1987}.

\bibitem{multintHERA}
H1 Collaboration, S. Aid et al., \Journal{\ZPC}{70}{17}{1996}. 

\bibitem{xgpm}
L. L\"onnblad and M. Seymour (convenors), \\
{\it $\gamma\gamma$ Event Generators}, in ``Physics at LEP2'', CERN~96-01, \\
Eds.~G.~Altarelli, T.~Sj\"ostrand and F.~Zwirner, Vol.~2~(1996)~187.

\bibitem{dijetZEUS}
ZEUS Collaboration, M. Derrick et al., \Journal{\PLB}{384}{401}{1996}.
%\JournalPLB{384}{401}{1996}.

\bibitem{hz}
J.~Bromley et al.,\\
HzTool --- A Package for Monte Carlo Generator -- Data Comparison at HERA,\\ 
http://dice2.desy.de/$\sim\,$h01rtc/hztool.html

\bibitem{lowQ2H1}
H1 Collaboration, C.~Adloff et al., \Journal{\PLB}{415}{418}{1997}.
%\JournalPLB{415}{418}{1997}.

\bibitem{grvpdf}
M. Gl\"uck, E. Reya and A. Vogt, \\
\Journal{\PRD}{46}{1973}{1992}; \Journal{\PRD}{45}{3986}{1992}.

\bibitem{OPAL}
OPAL Collaboration, K.~Ackerstaff et al., \Journal{\ZPC}{73}{433}{1997}.

\bibitem{fwdjet}
H1 Collaboration, C.~Adloff et al., \Journal{\NPB}{538}{3}{1998};\\
H1 Collaboration, S.~Aid et al., \Journal{\PLB}{356}{118}{1995};\\
%\JournalPLB{356}{118}{1995};\\
ZEUS Collaboration, J.~Breitweg et al., \Journal{\EPJC}{6}{239}{1998}.

\end{thebibliography}
\end{document}